\documentclass[twocolumn]{aastex62}

\usepackage{lineno,hyperref}
\modulolinenumbers[5]

\usepackage{graphicx}
\usepackage{amssymb}
\usepackage{amsmath}
\usepackage{color}
\usepackage{tabularx}

\received{2018 February 23}
\accepted{2018 August 17}
\published{2018 September 28}
\submitjournal{\pasp}

\shorttitle{Astrophysics with \NH}
\shortauthors{Zemcov et al.}

\newcommand{\nw}{nW m$^{-2}$ sr$^{-1}$}
\newcommand{\NH}{\textit{New Horizons}}

\newcolumntype{b}{X}
\newcolumntype{s}{p{1.07in}}

\begin{document}

\title{Astrophysics with \NH: Making the Most of a
 Generational Opportunity}

\correspondingauthor{Michael Zemcov}
\email{zemcov@cfd.rit.edu}

\author[0000-0001-8253-1451]{Michael Zemcov}
\affiliation{Center for Detectors, School of Physics and Astronomy,
              Rochester Institute of Technology, Rochester, NY 14623,
              USA}
\affiliation{Jet Propulsion Laboratory, 4800 Oak Grove Drive, Pasadena,
              CA 91109, USA}

\author{Iair Arcavi}
\affiliation{Einstein Fellow at the Department of Physics, University of California, Santa Barbara, CA 93106-9530, USA}
\affiliation{Las Cumbres Observatory, 6740 Cortona Drive, Suite 102, Goleta, CA 93117-5575, USA}

\author{Richard Arendt}
\affiliation{CRESST II/UMBC Observational Cosmology Laboratory, Code 665, Goddard Space Flight Center, 8800 Greenbelt Road, Greenbelt, MD 20771, USA}

\author{Etienne Bachelet}
\affiliation{Las Cumbres Observatory, 6740 Cortona Dr Ste 102, Goleta, CA 93117-5575, USA}

\author{Ranga Ram Chary}
\affiliation{U.~S.~Planck Data Center, California Institute of Technology, 1200 East California Boulevard, Pasadena, CA 91125, USA}

\author{Asantha Cooray}
\affiliation{Department of Physics and Astronomy, University of
              California, Irvine, CA 92697, USA}

\author{Diana Dragomir}
\affiliation{NASA Hubble Fellow, Massachusetts Institute of Technology, Cambridge, MA 02139, USA}

\author{Richard Conn Henry}
\affiliation{Henry A. Rowland Department of Physics and Astronomy, The Johns Hopkins University, Baltimore, MD 21218, USA}

\author{Carey Lisse} 
\affiliation{Planetary Exploration Group, Space Department, Johns
              Hopkins University Applied Physics Laboratory, 11100
              Johns Hopkins Road, Laurel, MD 20723, USA}

\author{Shuji Matsuura}
\affiliation{School of Science and Technology, Kwansei Gakuin
              University, Sanda, Hyogo 669-1337, Japan}
\affiliation{Department of Space Astronomy and Astrophysics, the Institute
of Space and Astronautical Science, Japan Aerospace Exploration
Agency, Sagamihara, Kanagawa 252-5210, Japan}

\author{Jayant Murthy}
\affiliation{Indian Institute of Astrophysics, Bengaluru 560 034, India}

\author{Chi Nguyen}
\affiliation{Center for Detectors, School of Physics and Astronomy,
              Rochester Institute of Technology, Rochester, NY 14623,
              USA}

\author{Andrew R.~Poppe}
\affiliation{Space Sciences Laboratory, University of California at
              Berkeley, Berkeley, CA, 94720, USA}

\author{Rachel Street} 
\affiliation{Las Cumbres Observatory, 6740 Cortona Dr Ste 102, Goleta, CA 93117-5575, USA}

\author{Michael Werner}
\affiliation{Jet Propulsion Laboratory, 4800 Oak Grove Dr., Pasadena,
              CA 91109, USA}

\begin{abstract}
  The outer solar system provides a unique, quiet vantage point from
  which to observe the universe around us, where measurements could
  enable several niche astrophysical science cases that are too
  difficult to perform near Earth.  NASA's \NH\ mission comprises an
  instrument package that provides imaging capability from ultraviolet (UV) to
  near-infrared (near-IR) wavelengths with moderate spectral resolution located beyond
  the orbit of Pluto.  A carefully designed survey with \NH\ can
  optimize the use of expendable propellant and the limited data
  telemetry bandwidth to allow several measurements, including a
  detailed understanding of the cosmic extragalactic background light;
  studies of the local and extragalactic UV background; measurements
  of the properties of dust and ice in the outer solar system;
  confirmation and characterization of transiting exoplanets;
  determinations of the mass of dark objects using gravitational
  microlensing; and rapid follow-up of transient events.  \NH\ is
  currently in an extended mission designed to focus on Kuiper Belt
  science that will conclude in 2021.  The astrophysics
  community has a unique, generational opportunity to use this mission
  for astronomical observation at heliocentric distances beyond 50 au
  in the next decade.  In this paper, we discuss the potential science
  cases for such an extended mission, and provide an initial
  assessment of the most important operational requirements and
  observation strategies it would require.  We conclude that \NH\ is
  capable of transformative science, and that it would make a valuable
  and unique asset for astrophysical science that is unlikely to be
  replicated in the near future.
\end{abstract}


\section{Introduction}
\label{intro}

Astronomical observations have been performed from a wide range of
locations, including the surface of the Earth, from atmospheric
platforms, and in space from orbit, as well as further afield
like the Earth's Lagrange points and from Earth-trailing orbits around the sun.  Very
occasionally, humans have sent instruments to the outer edge of the
solar system that are capable of astronomical observation
\citep{Weinberg1974,Broadfoot1977,Weaver2008}.  These instruments have
been used to make astronomical measurements, including studies of the
decrease in the light from interplanetary dust (IPD) with heliocentric
distance \citep{Hanner1974, Matsumoto2018}; Lyman-$\alpha$ emission from the interplanetary medium \citep{Gladstone2013}; the diffuse light from the
Galaxy \citep{Toller1987, Gordon1998}; the brightness of the cosmic
optical background (COB; \citealt{Toller1983,Matsuoka2011,Zemcov2017}) and the cosmic UV background (CUB; \citealt{Holberg1986,Murthy1991,Murthy1999,Edelstein2000}); and the UV
emission from specific objects \citep{Holberg1985}, including studies
of their spectral features \citep{Murthyvoy1993,Murthy2001}.  

Over the years, a number of missions to the outer solar system
including instrumentation expressly designed to obtain astrophysical
measurements have been considered (\textit{e.g.}
\citealt{Mather1996, Bock2012, Matsuura2014, Stone2015}, among others).  However, these
missions are costly and difficult endeavors, and require positive
funding environments.  A more modest strategy is to take advantage of
missions during their cruise phases when they are activated for system
checks and calibration campaigns.  This strategy maximizes science
return by taking advantage of existing assets at only a modest
increase in mission risk and complexity.  

\begin{figure*}[ht]
\centering 
\includegraphics*[width=\textwidth]{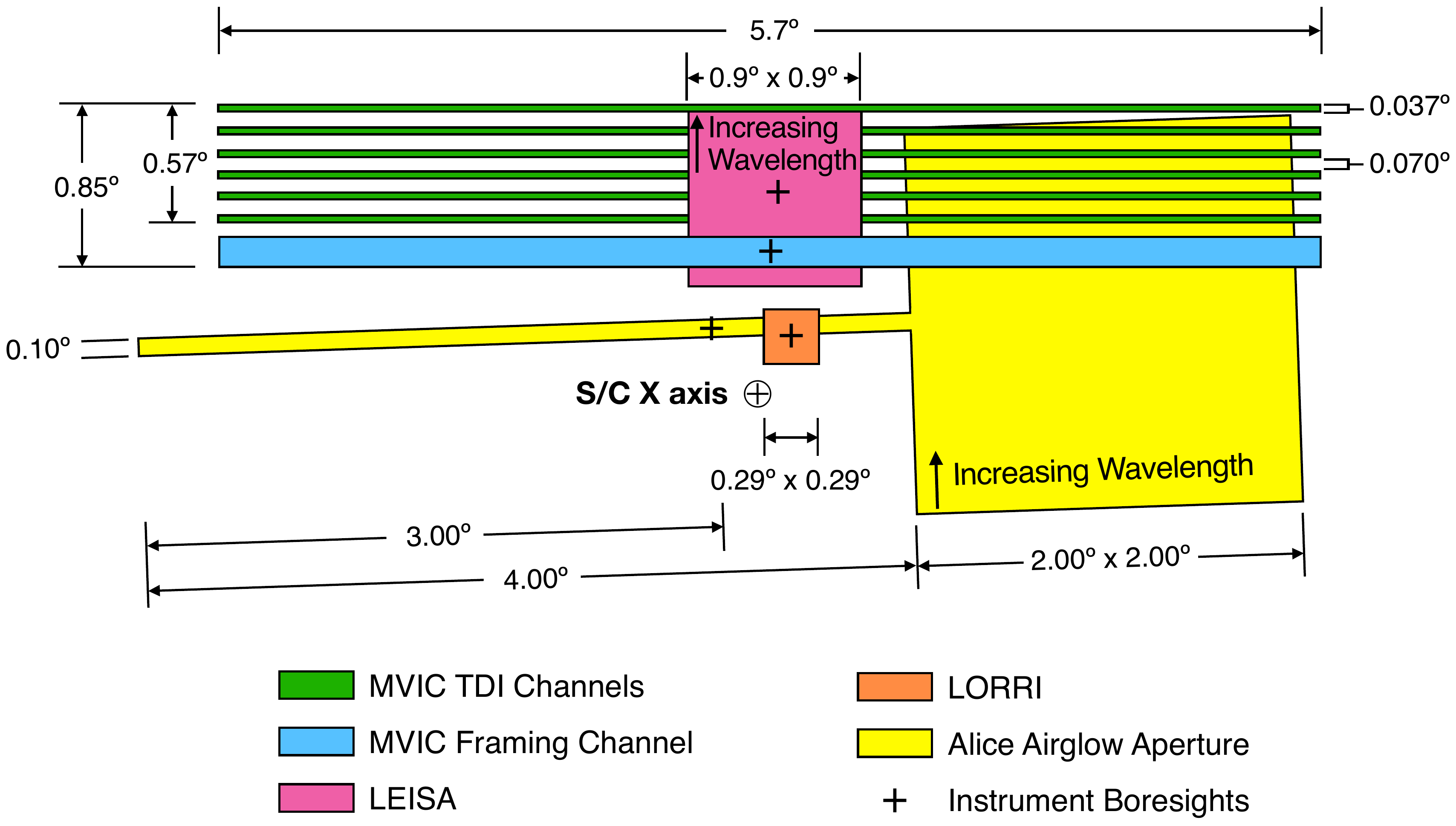}
\caption{Layout of the focal plane of the imaging instruments on
  \NH\ \citep{Weaver2008}.  LORRI has broad bandwidth, but has a
  relatively small footprint on the sky.  MVIC observes in several colors in
  thin, long strips, while LEISA and Alice have relatively large
  fields of view.  The instrument parameters are summarized in Table
  \ref{tab:instruments}. \label{fig:fov} }
\end{figure*}

NASA's \NH\ mission \citep{Stern2008,Weaver2008} recently
performed the first detailed reconnaissance of the Pluto-Charon
system.  \NH\ is currently in an extended mission designed to conduct a close fly-by investigation of the Kuiper Belt Object (KBO) 2014 MU69; perform unique observations of approximately two to three dozen other KBOs and Centaurs; measure the heliospheric plasma, dust, and neutral gas environment out to a heliocentric distance of 50 au. This first extended mission phase is scheduled to conclude in the spring of 2021 \citep{Stern2018}.  \NH\
includes as part of its instrument package the Long Range
Reconnaissance Imager (LORRI; \citealt{Cheng2008}), the Multispectral
Visible Imaging Camera (MVIC), the Linear Etalon Imaging Spectral
Array (LEISA; \citealt{Reuter2008}) and an ultraviolet (UV) spectrograph Alice
\citep{Stern2008}.  In addition to their planetary imaging functions,
these NH instruments can double as sensitive astronomical
instruments working from the UV well into near-infrared (near-IR) wavelengths (see Figure \ref{fig:fov}).  

\NH\ has generated a rich archival data set for both planetary studies and 
astronomy that is currently being analyzed.  However, the spacecraft
itself has operational capability significantly beyond its current
mission, and could operate well into the heliopause.  A tantalizing
possibility is to use the \NH\ instruments for an
extended mission for astrophysics where purposely designed
observations can be performed.  This would help maximize the science
return from the mission, and would take advantage of this unique
resource.  Such an opportunity will not arise again in the foreseeable
future.  

In this paper, we outline the astrophysical studies that
could be performed with the \NH\ instrument suite,
focusing on measurements that require the exceptionally low foreground
emission from the outer solar system, or the $50{-}100 \,$au
separation from Earth to the spacecraft.  These include measurements
of the diffuse UV/optical/near-IR backgrounds away from the obscuring
effects of the sun's immediate environment, and careful photometry science including exoplanet
transits and microlensing that require an exceptionally stable
platform.  These concepts could be used to inform future science and uses of the \NH\ mission.  In Section \ref{S:science} we review the primary science
cases that benefit from access to the outer solar system.  We assess
the sensitivity and stability of the instruments using pre-flight
estimates and in-flight data in Section \ref{S:sensitivity}.  In
Section \ref{S:ops} we outline the operational requirements of these
types of measurements, and describe a hypothetical astrophysical survey.
Finally, Section \ref{S:conclusion} gives some concluding remarks and
an outlook for the future.  

The analysis of data and plotting in this paper are performed using custom routines in {\sc python} and {\sc matlab}; we use functionality from the {\sc scipy} \citep{SciPy}, {\sc numpy} \citep{NumPy}, {\sc ipython} \citep{Perez2007}, and {\sc matplotlib} \citep{Hunter2007} libraries.  As a note to assist readers unaccustomed to working with astrophysical surface brightness, throughout this paper we employ a convention for diffuse surface brightness given by $\lambda I_{\lambda}$ (sometimes called ``intensity''), in which the specific intensity $I_{\lambda}$ (which carries units power per unit area per unit solid angle per unit wavelength, \textit{e.g.}~\nw $\mu$m$^{-1}$) is multiplied by each $\lambda$.  This definition is consistent with that used elsewhere in the astronomical community, but differs from some other fields, and is used because it succinctly describes the power one would measure with a detector with a narrow bandpass at a given wavelength.  The quantity $\lambda I_{\lambda}$ is equal and equivalent to $\nu I_{\nu}$, although $I_{\nu}$ itself carries units of power per unit area per unit solid angle per unit frequency (see \textit{e.g.} \citealt{Rybicki1986} for more details and derivations).  Similarly, we choose a convention where we reference images to flux per 2.6 pixel$^{2}$ beam rather than per pixel, which only affects how diffuse surface brightnesses are calculated from raw data.

\section{Astrophysical Science from the Outer Solar System}
\label{S:science}

\subsection{Measurement of the Extragalactic Background Light}

The formation of stars and galaxies in the universe is accompanied by
the release of photons from both gravitational and nuclear mechanisms
\citep{Hauser2001, Cooray2016}.  A cosmic background radiation in the
UV, optical, and IR parts of the electromagnetic spectrum is therefore
an expected relic of structure formation processes, and measurements
of these backgrounds provide insights into those processes.
Practically speaking, the extragalactic background light (EBL) at
optical/near-IR wavelengths is thought to be dominated by photons released by
nucleosynthesis in stars, and constraints of this stellar emission
integrated over cosmic history and can yield crucial insights into a
variety of astrophysical phenomena.  Specifically, precise measurement
of the EBL enables a cosmic consistency test wherein the integrated
light from all galaxies, stars, active galactic nuclei (AGN), and
other point sources is compared with the EBL intensity \citep{Tyson1995}.
Any excess component suggests the presence of new, diffuse
sources of emission. Potential discoveries with profound implications
for astronomy include the signature of diffuse recombination from the
epoch of reionization (\textit{e.g.} \citealt{Matsumoto2005}), the presence of intra-halo
light in the diffuse intragalactic medium (\textit{e.g.} \citealt{Zemcov2014}), and
diffuse photons associated with dark matter annihilation and their
products (\textit{e.g.} \citealt{Gong2016}).
 
In the past, direct measurements of the EBL have been complicated by
the presence of bright local foregrounds, including the Zodiacal light
(ZL), diffuse galactic light (DGL), and the integrated starlight (ISL)
arising from extended telescope response and faint stars
\citep{Leinert1998}.  Despite a great deal of interest, direct
photometric measurement of the EBL has proved to be challenging,
largely because the atmosphere and ZL are factors of $\sim 100$
brighter than the signal of interest.  Though
some progress has been made in accounting for these foregrounds in the
optical \citep{Bernstein2007,Mattila2012} and into the near-IR
\citep{Gorjian2000, Wright2001, Cambresy2001, Wright2004,
  Matsumoto2005, Levenson2007, Tsumura2013, Sano2015, Matsuura2017}, small errors in this
accountancy propagate to large errors on the inferred COB.  As a result of
misestimation of the foregrounds, the systematic errors of current
photometric measurements of the EBL exceed the integrated light from
all galaxies outside of our own (known as the integrated galactic
light (IGL)) by a factors of at least several (\textit{e.g.} \citealt{Mattila2003, Mattila2006}). It is desirable to
measure the EBL from vantage points where the ZL is not an appreciable
component of the diffuse sky brightness, such as the outer solar
system or above the ecliptic plane \citep{Cooray2009}.

\begin{figure*}[ht]
\centering
\includegraphics*[width=\textwidth]{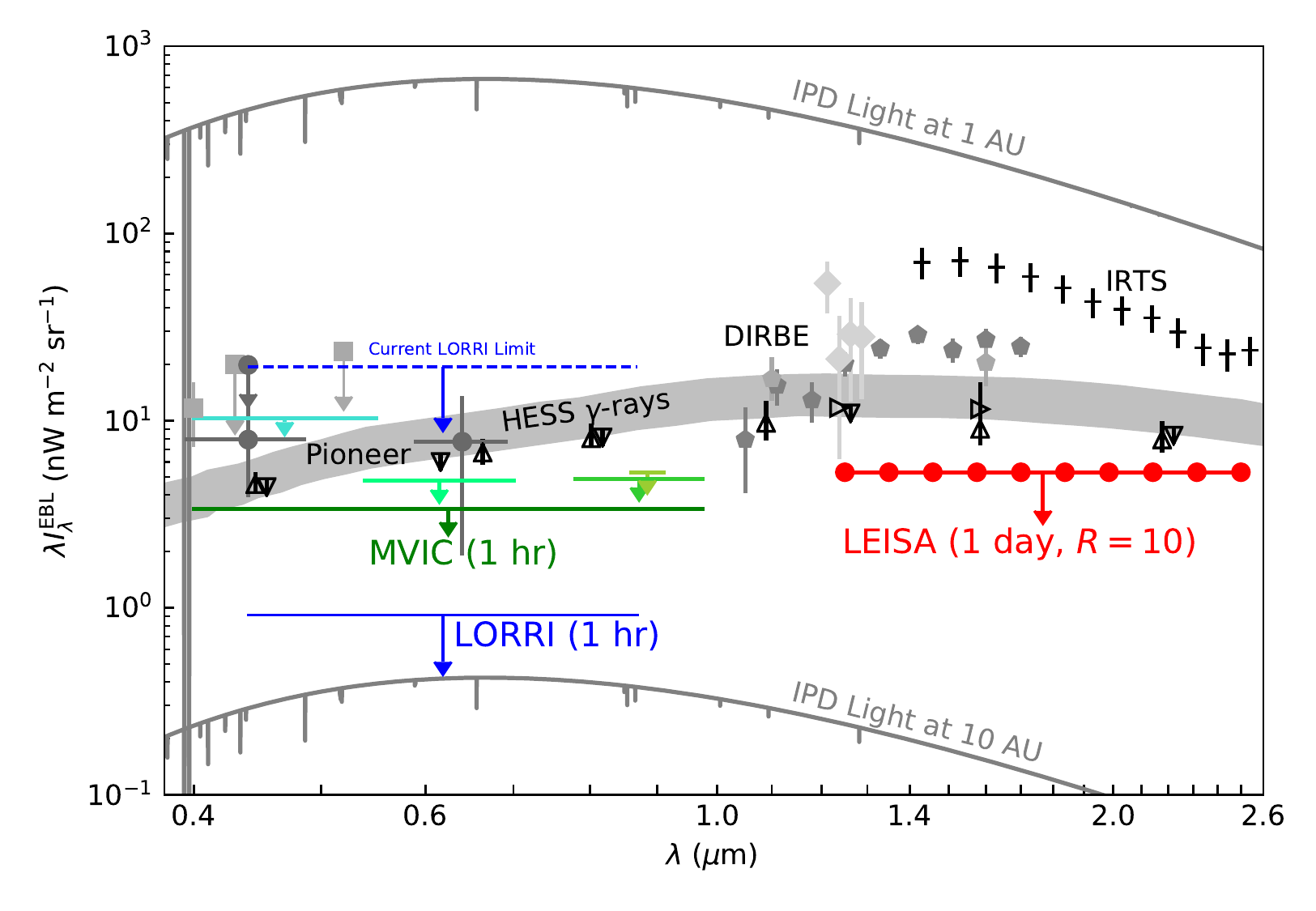}
\caption{Measurements of the EBL surface brightness
  $\lambda I_{\lambda}^{\rm EBL}$ in the optical and near-IR,
  including existing direct photometric constraints on the EBL (filled
  symbols) and the integrated galactic light (IGL; open symbols).  We
  show the expected sensitivity of LORRI in $t_{\rm int} = 1 \,$hour
  of integration time (blue limit), MVIC in $t_{\rm int} = 1 \,$hour
  (green limits, one for each band), and LEISA in
  $t_{\rm int} = 1 \,$day (red limits), as well as the existing
  $2 \sigma$ upper limit from LORRI (blue dashed line; \citealt{Zemcov2017}).
  We show direct measurements of the EBL from observations using the
  ``dark cloud'' method (squares; \citealt{Mattila2017}),
  \textit{Pioneer} 10/11 measurements (circles;
  \citealt{Toller1983,Matsuoka2011}), CIBER (pentagons;
  \citealt{Zemcov2014, Matsuura2017}), combinations of DIRBE and 2MASS
  (diamonds;
  \citealt{Wright2001,Cambresy2001,Wright2004,Levenson2007}), and IRTS
  (plus symbols; \citealt{Matsumoto2005}).  The shaded region
  indicates the HESS $\gamma$-ray constraints on the extragalactic
  background light \citep{HESS2013}.  The IGL points are compiled from
  the Hubble Deep Field (downward open triangles; \citealt{Madau2000})
  and the Subaru Deep Field (upward open triangles and sideways
  pointing triangles;\citealt{Totani2001, Keenan2010}) in the
  optical/near-IR.  The diffuse galactic light is a foreground associated with dust in the Milky Way galaxy that reaches a minimum of about 5 \nw, but can be subtracted using various means.  Ultimately, even modest integration
  times could permit definitive measurements of the brightness of the
  EBL over 3 octaves in frequency. \label{fig:cob} }
\end{figure*}

The surface brightness of the IPD light is thought to drop with heliocentric
distance roughly as $r^{-3}$ to levels significantly below the EBL by
the orbit of Saturn (see Section \ref{sS:EKB}).  As a result, an EBL
measurement from the outer solar system observing out of the plane of
the ecliptic should not suffer from strong IPD light contamination.  Indeed,
data from the early NASA probes \textit{Pioneer} 10 and 11 have been
used to measure both the decrease in the IPD light with heliocentric
distance \citep{Hanner1974}, the diffuse light from the Galaxy
\citep{Toller1987, Gordon1998}, and the brightness of the COB itself
\citep{Toller1983, Matsuoka2011} in two bands spanning $390{-}500 \,$nm
and $600{-}720 \,$nm over heliocentric distances ranging from 1 to 5.3
AU \citep{Weinberg1974}.  Due to the large field of view and poor
angular resolution of the \textit{Pioneer} photometers, these measurements have
uncertainties dominated by errors associated with subtracting galactic
components.  However, an instrument with fine angular resolution can
easily mask stars to the level that their emission is negligible, and
over modest fields of view tracers of galactic dust can be used to
measure a correlation with the DGL component that can then be
regressed from image.  This suggests that a $10 \,$cm-class telescope
in the outer solar system, coupled with a current understanding of the
galactic emission components, would be ideal for measuring the EBL.

The \NH\ mission includes an instrument suite that is
well suited to measurement of the EBL.  Figure \ref{fig:cob} shows the sensitivities of these instruments as compared to current measurements of the optical and near-IR backgrounds.  LORRI is a Newtonian telescope
with a
$20.8 \,$cm diameter Ritchey-Chr{\'e}tien telescope, an
$0^{\circ}.3 \times 0^{\circ}.3$ instantaneous field of view,
$1^{\prime \prime} \times 1^{\prime \prime}$ pixels, sensitivity over
a broad $440{-}870 \,$nm half-sensitivity passband, and (crucially)
real-time dark current monitoring.  The achieved point source
sensitivity of LORRI is $V=17$ ($5 \sigma$) in a $10 \,$s exposure in $4 \times 4$
pixel on-chip ``rebinning'' mode, making it a sensitive astronomical
instrument for which the starlight that challenged earlier \textit{Pioneer}
measurements can be masked out.  LORRI has lately been used to measure
the brightness of the EBL in the optical, yielding an upper limit that
rules out some of the highest previous measurements \citep{Zemcov2017}.
However, that measurement was made on a very limited dataset that was
not optimized for precise measurements of the EBL, and significant
improvements are possible.  In even a limited 4-hour total integration
time with LORRI, uncertainties similar to those on the IGL are
expected.  In fact, the ultimate error from a LORRI measurement is
likely limited by our knowledge of the DGL and ISL foregrounds, rather
than the intrinsic sensitivity of the instrument.

Similarly, MVIC is a broadband imaging instrument, but provides significantly
more spectral information than LORRI.  Compared to LORRI, each band
has a long, thin field of view.  This is not necessarily problematic
for an EBL measurement, but the smaller aperture and narrower
bandpass of the MVIC channels cause a factor of $\sim 10$ per-pixel sensitivity penalty for
measuring the average sky brightness compared with
LORRI.  However, averaging over the array will help, and MVIC observations remain a promising way to gain crucial
spectral information on the shape of the EBL spectrum throughout the
optical, which would provide compelling information compared with a single LORRI data point over a similar wavelength range.

LEISA would make simultaneously the most interesting and challenging
measurement of the EBL.  The near-IR $1{-}3 \, \mu$m background has
proved very difficult to measure from Earth, and is very interesting
as the light from the earliest galaxies will be redshifted into this
range.  LEISA provides detailed spectral information that could be
used to search for \textit{e.g.} the spectral bump expected from Lyman
emission from the galaxies that reionized the universe
\citep{Cooray2004}.  However, LEISA has a relatively small aperture and
$R=240$ spectral resolution, making the per-pixel sensitivity
poor.  Significant integration time would be required to make a
constraining measurement of the EBL.

\subsection{Measurement of the Ultraviolet Background}

A detailed accounting of the cosmic ultraviolet background can provide information on a variety of astrophysical processes in the local interstellar medium (ISM) and other galaxies, including line emission and fluorescence from interstellar gas, high-energy light scattered by dust, the possibility of a massive hot halo of our own galaxy, and constraints on an extragalactic component interesting for reasons similar to those of the EBL discussed above.  Despite the science impact, the interpretation of actual observations of the cosmic ultraviolet background has been controversial for more than a quarter of a century (see \textit{e.g.} \citealt{Bowyer1991} and \citealt{Henry1991} for contrasting viewpoints). This is primarily because it is challenging to separate the different components of the diffuse emission, particularly with imaging surveys. Any spacecraft in low Earth orbit (\textit{e.g.~GALEX}) will be affected by airglow, while any spacecraft observing within the inner Solar System will be affected by the Lyman lines from interplanetary hydrogen and ZL at longer wavelengths. Even if we can account for these foregrounds, distant sources of astrophysical emission are difficult to separate without spectral diagnostics \citep{Murthy2009}.

Almost all of our knowledge of the diffuse UV background in the spectral region longer than 1300 \AA\ has come from {\it GALEX} broadband data. With only imaging data available, \citet{Murthy2016} found that most of the diffuse radiation was due to scattered starlight, albeit with an offset of unknown origin \citep{Hamden2013,Henry2015,Akshaya2018,RCH2018}. However, the different components are impossible to separate photometrically, and the resulting backgrounds are highly model-dependent. In principle, spectroscopic observations from interplanetary space would allow a separation of the components.  \citet{Murthy2001} used observations from the {\it Voyager} ultraviolet spectrographs (UVS) to observe that the diffuse background at shorter wavelengths ($\lambda < 1200$ \AA) is patchy, and demonstrates a poor correlation with the diffuse background in the near-ultraviolet (NUV).  The conclusion of this work is that although the two {\it Voyager} spacecraft observed far from the Sun, thereby avoiding airglow, they were still affected by the interplanetary HI lines. Instrumental scattering from interplanetary Ly $\alpha$ was the source of signal in many regions of the sky and affected the entire spectrum. This was compounded by the relatively low spectral resolution of 27 \AA\ , so that the Ly $\alpha$ (1215 \AA) and Ly $\beta$ (1027 \AA) lines were spread through much of the wavelength range between 900 --- 1200 \AA. This problem is exacerbated in the extraction of the background line emission \citep{Murthyvoy1993, Murthy2001}. 

The advantage of \NH\ is that observations will be made from outside the orbit of Pluto, more than 50 au away. Although instrumental scattering of interplanetary Ly $\alpha$ is still a problem, the magnitude of the line drops by an order of magnitude from 1 au to 50 au \citep{Murthy2001}.  The $9 $ \AA\ resolution of Alice is well suited to search for diffuse emission from the Galaxy, both continuum and in lines.  This is because both the foreground scattered from the interplanetary HI lines is minimized through observations from the outer solar system, and that the spectral shape of the astrophysical emission components can be used to decompose the emission. For example, emission from the Lyman and Werner bands of molecular hydrogen will extend throughout the UV in regions of high density, while diffuse OVI (1032/1038 \AA) emission will track the hot gas \citep{Dixon2006}. These can be used to understand the local ISM through observations of different parts of the sky.  Finally, the dust scattered starlight should correlate with the positions of the emitting O and B stars. Residuals should be due to extragalactic emission at high latitudes or to a previously unknown emissive component at low galactic latitudes. 

\subsection{Edgeworth-Kuiper Belt Dust}
\label{sS:EKB}

Interplanetary dust particles are generated by several
sources including comets, asteroids, and Edgeworth-Kuiper Belt (EKB) objects
and range in size from $\sim$0.1 $\mu$m up to 1 mm. After ejection
from their parent bodies, IPD particles diffuse through the solar system as
they are affected by a variety of forces such as gravitation,
Poynting-Robertson drag, solar radiation pressure, and solar wind drag
\citep[\textit{e.g.}][]{Burns_1979, Gustafson_1994}. As these grains encounter
planetary systems, they have significant impacts on a wide range of
planetary processes, such as the alteration of atmospheric
photochemistry \citep[\textit{e.g.}][]{Moses_1992, Feuchtgruber_1997,
  Moses_2000b, Frankland_2016, Moses_2017}, the injection of metallic
species into planetary magnetospheres \citep{Christon_2015}; the
spatial and compositional evolution of Saturn's main ring system
\citep[\textit{e.g.}][]{Durisen_1989, Cuzzi_1998, Estrada_2015}; and the
production of impact ejecta clouds and/or rings from airless bodies,
like planetary satellites \citep[\textit{e.g.}][]{Verbiscer_2009,
  Hedman_2009, Poppe_2011b}. An accurate understanding of the
size, density, and velocity distributions of IPD
throughout the solar system is critical for studies across a broad
range of planetary science.  The main scientific goal of \NH\ remote sensing measurements of the IPD light would be to discern the makeup of the circumsolar dust cloud and its sourcing planetesimals at $r > 45 \,$au distances. Doing this will help inform comparison of our cloud to the debris disks found around other stars, help us understand the collisions and sublimations of small planetesimals in the outer solar system, and help us understand the exogenous delivery of material of dust from the cloud to solar system bodies (\textit{e.g.} Earth's meteors, Pluto's Haze).  

\begin{figure*}[htb]
	\includegraphics*[width=\textwidth]{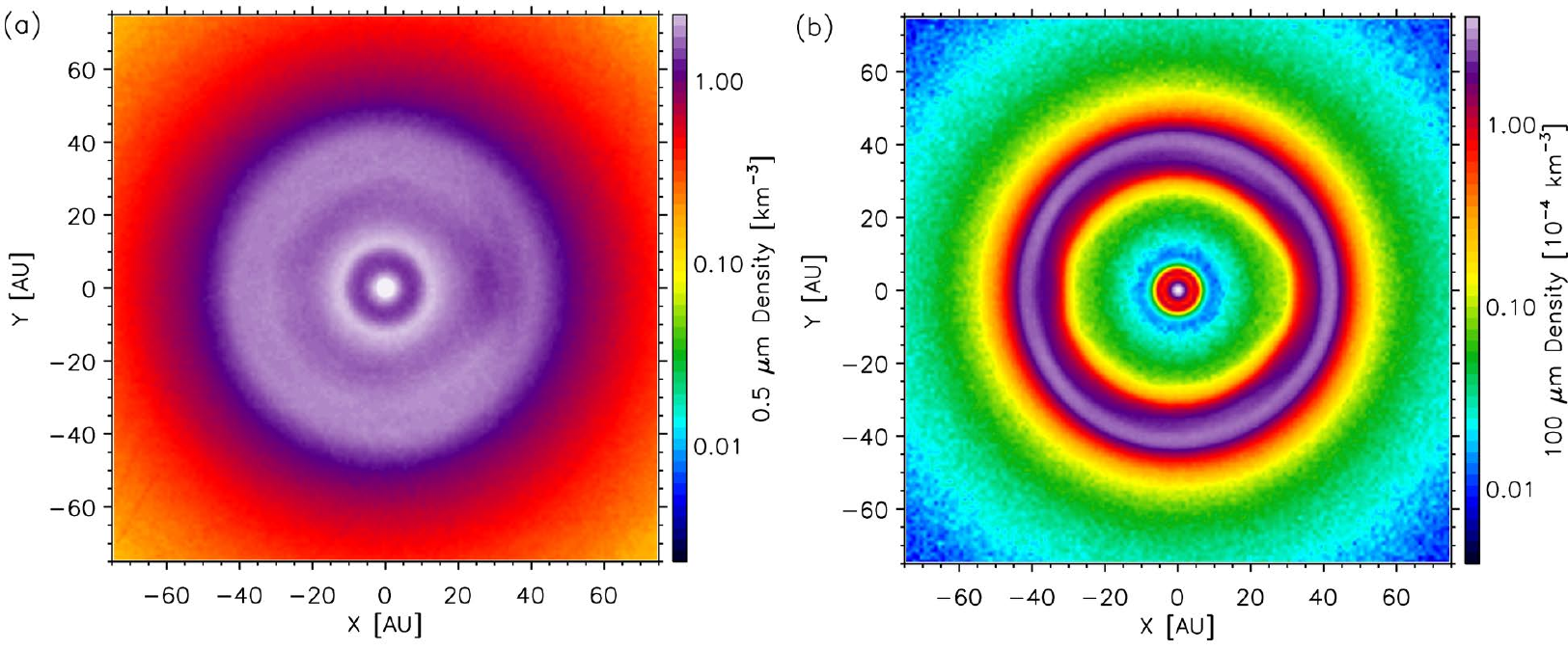}
	\caption{Model for the density distribution of (a) 0.5
          $\mu$m and (b) 100 $\mu$m interplanetary dust particles in
          the ecliptic plane based on that presented in \citet{Poppe_2016}.  Though the surface
          brightness of the light reflected by dust drops as
          $1/r_{\odot}^{2}$, a density enhancement is expected in the
          outer solar system near 40 au where LORRI is capable of making measurements in reflection.}
	\label{fig:density}
\end{figure*}

Our knowledge of the IPD distribution in the inner
solar system is fairly robust, with recent model-data comparisons
concluding that a significant fraction of the IPD particle distribution near 1
au originates from dust emission from Jupiter-family comets with minor
contributions from asteroidal and Oort Cloud cometary dust
\citep{Nesvorny_2011}. The three-dimensional morphology of the inner
solar system IPD particle distribution has been mapped in detail via infrared, optical, and
spectroscopic imaging \citep[\textit{e.g.}][]{Liou_1995, Hahn_2002,
  Ipatov_2008}. In contrast, knowledge of the IPD particle distribution in the
outer solar system is much more limited. \textit{In situ} measurements of outer
solar system IPD densities have been taken by spacecraft such as
\textit{Pioneer} 10 and 11 \citep{Humes_1980}, \textit{Ulysses} \citep{Grun_Uly_1995},
\textit{Galileo} \citep{Grun_Gal_1995}, \textit{Cassini} \citep{Altobelli_2007}, \textit{Voyager}
1 and 2 \citep{Gurnett_1997}, and the \NH\ Student Dust
Counter \citep{Poppe_2010a, Szalay_2013}. Despite producing valuable
results, these measurements have only provided information on grains
with radii between $\sim0.5$ - 10 $\mu$m, whereas the peak in the
IPD mass flux is expected to be near $\sim$100 - 200
$\mu$m. IPD spatial distributions are believed to be a strong function
of grain size; for example, Figure \ref{fig:density} shows the 0.5
$\mu$m and 100 $\mu$m IPD density (including
contributions from Jupiter-family comets, Oort Cloud comets, and
EKB objects) from recent modeling efforts \citep{Poppe_2016}. Furthermore, since the
\textit{Voyagers} have significant out-of-ecliptic trajectories and \textit{Pioneer}
10/11 meteoroid detectors ceased operating inside Uranus' orbit, only the \NH\ Student
Dust Counter \citep{Horanyi2008} has probed the EKB region itself, which
is the primary source of IPD particles in the outer solar system 
\citep[\textit{e.g.}][]{Stern_1996, Liou_1996, Vitense_2010, Vitense_2012,
  Poppe_2016}. Finally, model-data comparisons have constrained the
overall production rate of dust from the EKB and
other cometary sources \citep{Han_2011, Poppe_2016}; however, these limits are only based on measurements of grains $0.5 {-} 10 \, \mu$m in radius and are uncertain within an order of magnitude. We require additional observations and/or constraints on the density of IPD in the outer solar system, especially those that address grains with radii from 10 to several hundred $\mu$m.

\begin{figure*}[htb]
\centering
	\includegraphics*[width=\textwidth]{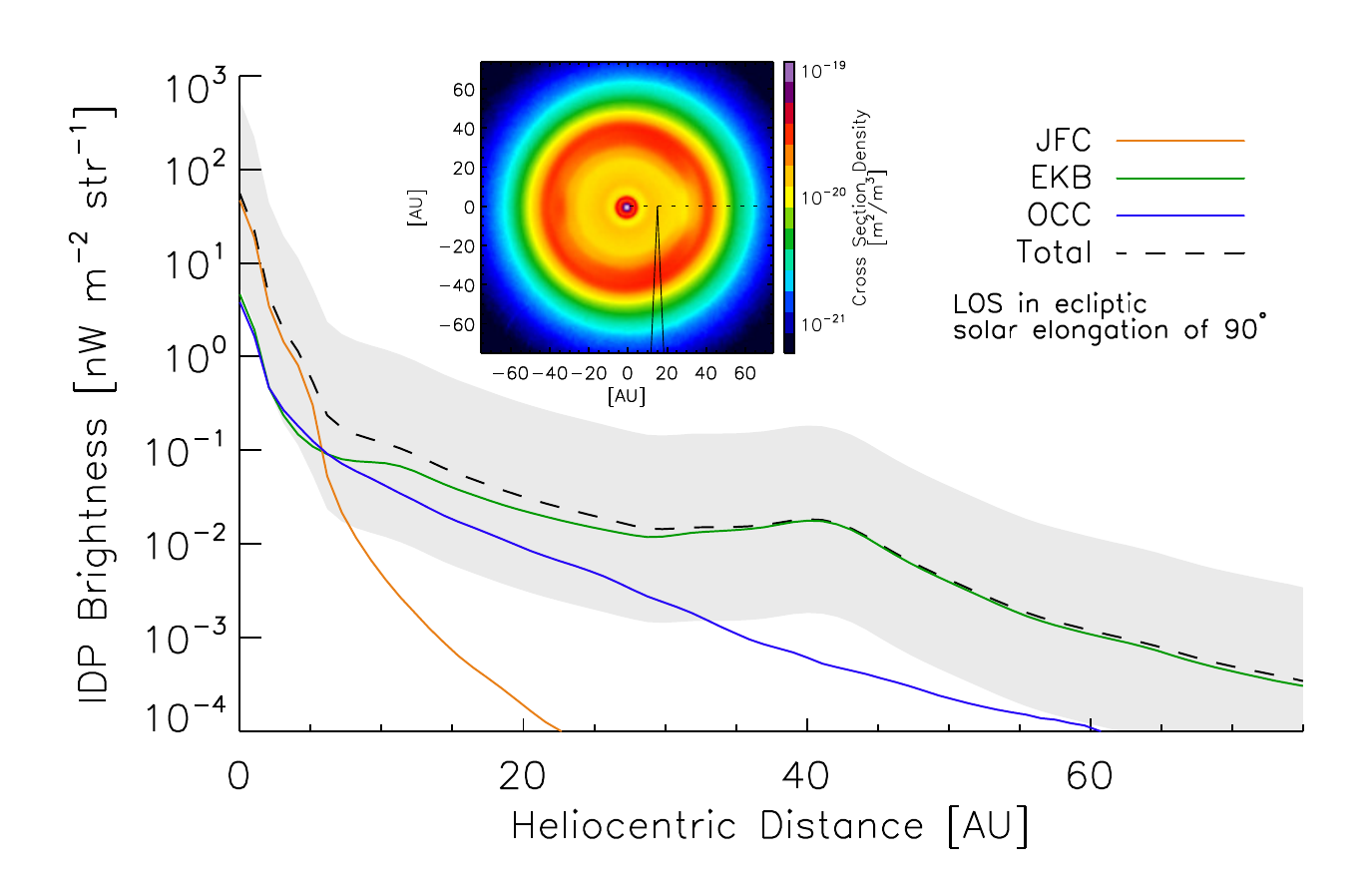}
	\caption{Estimated IPD surface brightness in $\lambda I_{\lambda}$ as a function of
          heliocentric distance for an observer along a radially
          outward trajectory with solar elongation angle of
          90$^{\circ}$. Inset: the IPD cross sectional density
          (m$^2$/m$^{3}$) along with the notional observer.} 
	\label{brightness}
\end{figure*}

Instruments on the \NH\ spacecraft provide a potentially powerful but previously unexplored method of observing the IPD particle density in the outer solar system. Solar light scattered from IPD grains can be observed by \NH\ as a diffuse background and, in combination with appropriate models for the dust density distribution and light scattering characteristics, can be used to place limits on the IPD distribution. To estimate the brightness scattered from the IPD distribution in the outer solar system, we have used the IPD model of \citet{Poppe_2016} which provides three-dimensional IPD densities of Jupiter-family comet, Oort Cloud comet, and EKB dust grains from $0.5 {-} 500 \, \mu$m with 1 au $\times$ 1 au resolution. We assumed the grains are comprised of astrosilicate material and used appropriate optical constants \citep{Jager_2003} to compute the scattering phase function from Mie theory. The differential brightness of solar scattered light from each parcel of IPD density over the LORRI wavelength bandpass was summed along the instantaneous line of sight of a virtual observer representing \NH. Figure \ref{brightness} shows the predicted IPD brightness in nW m$^{-2}$ sr$^{-1}$ as a function of heliocentric distance for an observation at a solar elongation angle of 90$^{\circ}$ (see the inset of Figure \ref{brightness}) along a radially outgoing trajectory (roughly approximating the trajectory of \NH). In the inner solar system ($r < 5 \,$au), scattered IPD light is on the order of $1 {-} 50 \,$\nw\ arising mainly from Jupiter-family comet dust, consistent with \textit{Pioneer} 10 photopolarimetry measurements \citep{Hanner1974}. In the outer solar system, the surface brightness slowly tapers off, averaging approximately $0.01 {-} 1 \,$\nw\ mainly from contributions by Oort Cloud cometary dust and EKB dust. Importantly, the uncertainty for these model predictions is large, as denoted by the shaded region in Figure \ref{brightness}. Measurements of the scattered IPD brightness by LORRI outside of 40 au have the potential to constrain the contributions from Oort Cloud cometary dust and EKB dust to the outer solar system IPD distribution over the summed range of sizes ($0.5 {-} 500 \, \mu$m), representing a powerful new constraint on the outer solar system dust density. 

A previous suggestion of direct detection of light scattering from IPD in the outer solar system comes from work by \citet{Chary_2010}, who inferred the possible presence of high albedo ($a \sim 1$), icy dust between $\sim 20 {-} 80 \,$au based on discrepancies between integrated galaxy light (IGL) and the EBL in the mid-IR. They estimated the IPD brightness at optical wavelengths in the outer solar system to be $\sim 25 \,$\nw, several orders of magnitude higher than predicted by our model. \citet{Chary_2010} theorized that such icy, high albedo dust could be shed from comets at distances far from their perihelia (such activity has been detected in Jupiter-family comets \citealt{Kelley_2013} and could also apply to Oort Cloud comets). If the IPD brightness in the outer solar system is truly this bright, \NH\ will be able to detect it, and this information can be used to add an appropriate icy dust grain composition to the current EKB dust models. Intriguingly, the presence of an icy halo of dust in the outer solar system at unexpectedly high densities may not necessarily conflict with \textit{in situ} measurements by dust detectors \citep{Humes_1980, Poppe_2010a} given that icy grains born in the outer reaches of planetary systems may perhaps migrate outward rather than inwards due to mass loss via photodesorption and/or charged particle sputtering and subsequent ejection via stellar (or solar) radiation pressure (\textit{i.e.}~so-called beta meteoroids; \citealt{Grigorieva_2007}). If \NH\ provides evidence for isotropic, icy dust grains, current IPD dynamics models will be revised by adding an additional component of isotropic, icy dust released from Oort Cloud comets and EKB objects.

\subsection{Transits in Exoplanetary Systems}

Measurements of exoplanet transits can provide a great deal of information about
exoplanetary systems (see \textit{e.g.}~\citealt{Rice2014} for a review).  In the transit
method, the light curves of stars hosting exoplanets are
photometrically monitored for long periods, and occultations of the
star by the planet (or vice versa) are sought.  The duration, shape,
and repetition frequency of the resulting dip in the star's light
curve can yield a great deal of information about the planetary
system.  This is the motivation for instruments like \textit{Kepler}
\citep{Borucki2010}, which photometrically monitored $> 10^{5}$ stars
to search for exoplanets in our galaxy.

Though by now, many thousands of planetary systems have been
identified with this method, its promise has only begun to be
realized.  In addition to wider survey fields, precision photometry
could, in principle, allow detection of structures like
moons around these planets \citep{Heller2017}.  Several methods have
been proposed to search for exo-moons, including transit
time variations (TTV) and transit duration variations (TDV; see
\citealt{Kipping2010} for a review), as well as the orbital sampling
effect (OSE; \citealt{Heller2014}).  All of these methods rely on both
precision photometry of the star, as well as measurements over many
orbits of the moon's parent planet to sample different parts of the
moon's orbital phase.  TTVs are also extremely useful for measuring planets masses in systems with multiple transiting exoplanets.
These measurements require not only extremely stable photometry, but also that observations occur over a long
time baseline to capture the system in different orbital
configurations \citep{Heller2016a, Heller2016b}.

A vantage point in the outer solar system gives a quiet, stable
platform from which to achieve 
$\delta F/F \sim 10^{-5}$ over the long time baselines required for these measurements.  \NH\ offers several possible advantages, including
(\textit{i}) the similarity of the LORRI detector with the
\textit{Kepler} detectors, which have shown remarkable stability
on-orbit \citep{Caldwell2010}; (\textit{ii}) a well-understood point
spread function, which allows accurate modeling of the instrument response
away from the central peak \citep{Morgan2005,Noble2009,Cheng2010}; (\textit{iii}) the lack of ZL
variations, giving an extremely stable, systematic-free background in
measurements of the same field separated by long periods; and
(\textit{iv}) the quiet instrument environment, in which (presumably)
most of the instruments would be in a quiescent state, and
\textit{e.g.} thermal transients from solar heating would be entirely
absent.  

\begin{figure*}[ht!]
\begin{center}
\includegraphics[width=\textwidth]{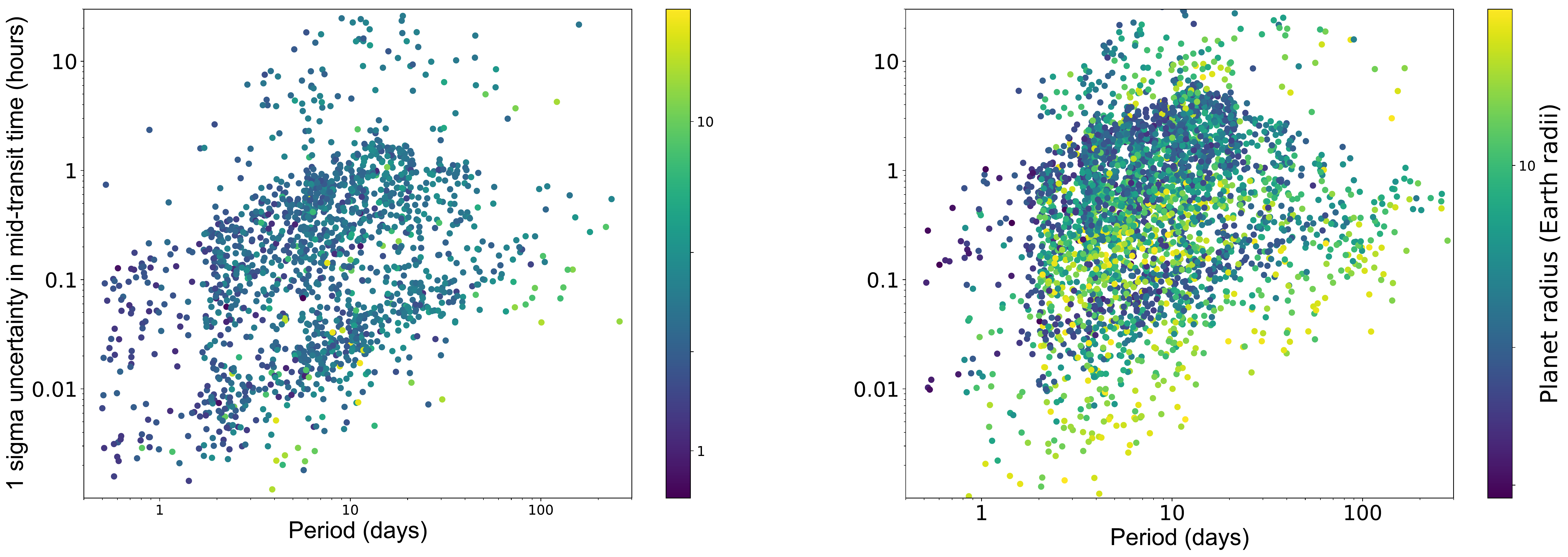}
\caption{\label{fig:tessephemerisloss} Uncertainty in mid-transit time for simulated planets one year after \textit{TESS} observes them.  Left: short cadence (2 minutes). Right: Long cadence (30 minutes). }
\end{center}
\end{figure*}

\textit{TESS} was successfully launched in 2018 April. During its two-year primary mission, \textit{TESS} will carry out a nearly all-sky survey and is expected to discover thousands of new transiting exoplanets around bright stars. These planets will allow a range of follow-up observations, so in this sense, will be much more valuable than \textit{Kepler} and \textit{K2}. However, due to the photometric precision of \textit{TESS} and the relatively short observing baseline (compared to \textit{K2} or \textit{Kepler}), the ephemerides of most \textit{TESS}-discovered planets will become ``stale'' very quickly. Figure \ref{fig:tessephemerisloss} shows the uncertainty in the mid-transit time of all simulated \textit{TESS} planets with two or more transits, one year after the last transit is observed by \textit{TESS} during the primary mission. Out of approximately 1600 two-minute cadence planets expected from \textit{TESS}, 100 will have $1 \sigma$ uncertainties on the mid-transit time greater than one hour, and 60 greater than two hours. For the 30-minute cadence, out of 3000 planets, 1000 will have uncertainties greater than one hour and 500 greater than two hours.

To schedule future observations (new transit photometry for transit timing variation studies, transit and eclipse spectroscopy with the \textit{James Webb Space Telescope} (\textit{JWST}), \textit{etc.}), we will need to recover the ephemerides of these future \textit{TESS} planets. Ground-based resources will not be able to reliably recover transits shallower than $\gtrsim 1$ mmag. Even for deeper transits, recovery of transits from the ground is very challenging if the mid-transit time uncertainties ($1 \sigma$) are greater than $\sim$4 hours, especially when the Earth's diurnal schedule and weather patterns are coupled into the observability window functions. Space-based observatories are needed to avoid ephemeris decay for these planets.

\textit{TESS} is expected to discover approximately 1000 planets with periods longer than fourteen days that show at least two transits in the \textit{TESS} light curves and have a total signal-to-noise ratio (SNR) greater than 7.3 \citep{Sullivan2015}. By also exploiting single-transit events, this yield can be increased by 70\%, with the potential to discover up to 700 {\it additional} planets with periods $>14$ days. Pursuing single-transit planets can make a large impact particularly at periods longer than 200 days; \textit{TESS} is expected to find just two multiply-transiting planets with period longer than 250 days, with single-transit planets increasing this number by up to an order magnitude \citep{Villanueva2018}.

To confirm those single-transit events that correspond to true planets, the usual vetting process will need to be supplemented with extra steps. One of these steps is to capture a second transit. This will happen after an ephemeris has been obtained using radial velocity (RV) monitoring of the system, and constraints from any additional, multi-transiting planets in the system. However, even with these constraints, the uncertainty on the next mid-transit time will generally be between several hours and several days, making it difficult to ensure a transit is captured from the ground. A space-based observatory such as \NH\ could be critical to the confirmation of numerous single-transiting \textit{TESS} planets. 

There are currently three existing or near-term space-based observatories that could be used for the long-term monitoring and recovery of transits: \textit{MOST}, \textit{CHaracterizing ExOPlanets Satellite} (\textit{CHEOPS}), and \textit{Spitzer}. The \textit{MOST} space telescope is not currently funded, and functions only if a user can purchase time. {\it MOST's} photometric precision is also lower than even that of \textit{TESS} (making it difficult to use for shallower single-transit events), and becomes equivalent to that of ground-based facilities for targets fainter than $V$ mag of 11. The European Space Agency is launching CHaracterising ExOPlanets Satellite (\textit{CHEOPS}) in early 2019 to obtain optical transits and phase curves of exoplanets. However, large portions of the \textit{TESS} footprint, particularly toward the ecliptic poles, will not be observable by \textit{CHEOPS}.  Further, \textit{CHEOPS}'s orbit is similar to the \textit{Hubble Space Telescope}'s (\textit{HST}'s), meaning that observations will be periodically interrupted by the Earth so that timing measurements will be more complicated and shorter transit events may be missed. Additionally, the current CHEOPS mission lifetime is only 3.5 years. Finally, \textit{Spitzer} is only funded through fall of 2019, so its usefulness for \textit{TESS} follow-up is limited to a few months at most. Even in combination, these three observatories will not be sufficient to monitor the more than 1000 planets with transits shallower than 3 mmag that \textit{TESS} is expected to discover \citep{Sullivan2015,Barclay2018}.

We note that, while there will likely be a proposal to extend \textit{TESS} past its two-year primary mission, such an extension is far from guaranteed. Even if the mission is approved for a third year extension, the most likely scenario will be to spend the third year reobserving (part of) one of the two ecliptic hemispheres, so we would still need NH for follow-up observations of at least half the \textit{TESS} planets. Further, during an extended mission new planets will be discovered, which will need to be followed up as well (just in time for the potential \NH\ astrophysics mission).  \NH\ would still be a sorely needed resource for observations of additional transits of \textit{TESS}-discovered planets. For the case of single transits, we find that hundreds of single-transit planets will remain unobserved by an extended \textit{TESS} mission \citep{Huang2018}.

To summarize, an extended \NH\ mission that can observe \textit{TESS} planet transits throughout the sky could be critical to the rescue of transit ephemerides for future observations, and to the search and confirmation of new \textit{TESS} planets, thus uniquely enhancing the \textit{TESS}' mission science return. We expect NH, in particular the LORRI detector, will be able to reach the photometric precision required to carry out these observations (see Section \ref{sS:transits} for details).

\subsection{Discovering Intermediate Mass Black Holes and Breaking Mass Degeneracies in Microlensing}

Like exoplanet transits, microlensing of distant stars by foreground massive objects is a time-domain technique wherein photometric monitoring of background stars reveals a distinctive brightening and fading, and where abrupt changes in the light curve can betray the presence of companions to the (normally invisible) lensing body.  Typically, stars in our Galaxy's Bulge are monitored, as this maximizes the number of potential targets per area on the sky. Microlensing is the most effective method for finding exoplanets beyond the snow line of their stars, where the sensitivity of other planet discovery techniques drops off rapidly.  To date, 53 planetary systems detected by microlensing have been published\footnote{\url{https://exoplanetarchive.ipac.caltech.edu}}.  As the technique does not rely on receiving any light from the lens itself, it is uniquely sensitive to any massive body, including compact objects \citep{Wyrzykowski2011} and even free-floating planets \citep{Mroz2017}.  

The mass and distance of a lensing object are degenerate in point source, point lens events, but this can be broken if microlensing parallax can be measured by observing the same event from multiple, widely separated locations \citep{Gould1992, Buchalter1997}.  For events with extremely high magnification, the separation required is as small as the Earth's radius \citep{Gould2009}, but these are rare.  More commonly, parallax is measured either because the event is long enough for the Earth to move in its orbit appreciably during the event \citep[\textit{e.g.~}][]{Muraki2011}, or by obtaining simultaneous light curves from Earth- and space-based observatories such as {\it Spitzer} and {\it K2} \citep[\textit{e.g.~}][]{Dong2007, Yee2015, Street2016, Zhu2017a}.  

\begin{figure}
\begin{centering}
\begin{tabular}{c}
\includegraphics[width=0.48\textwidth]{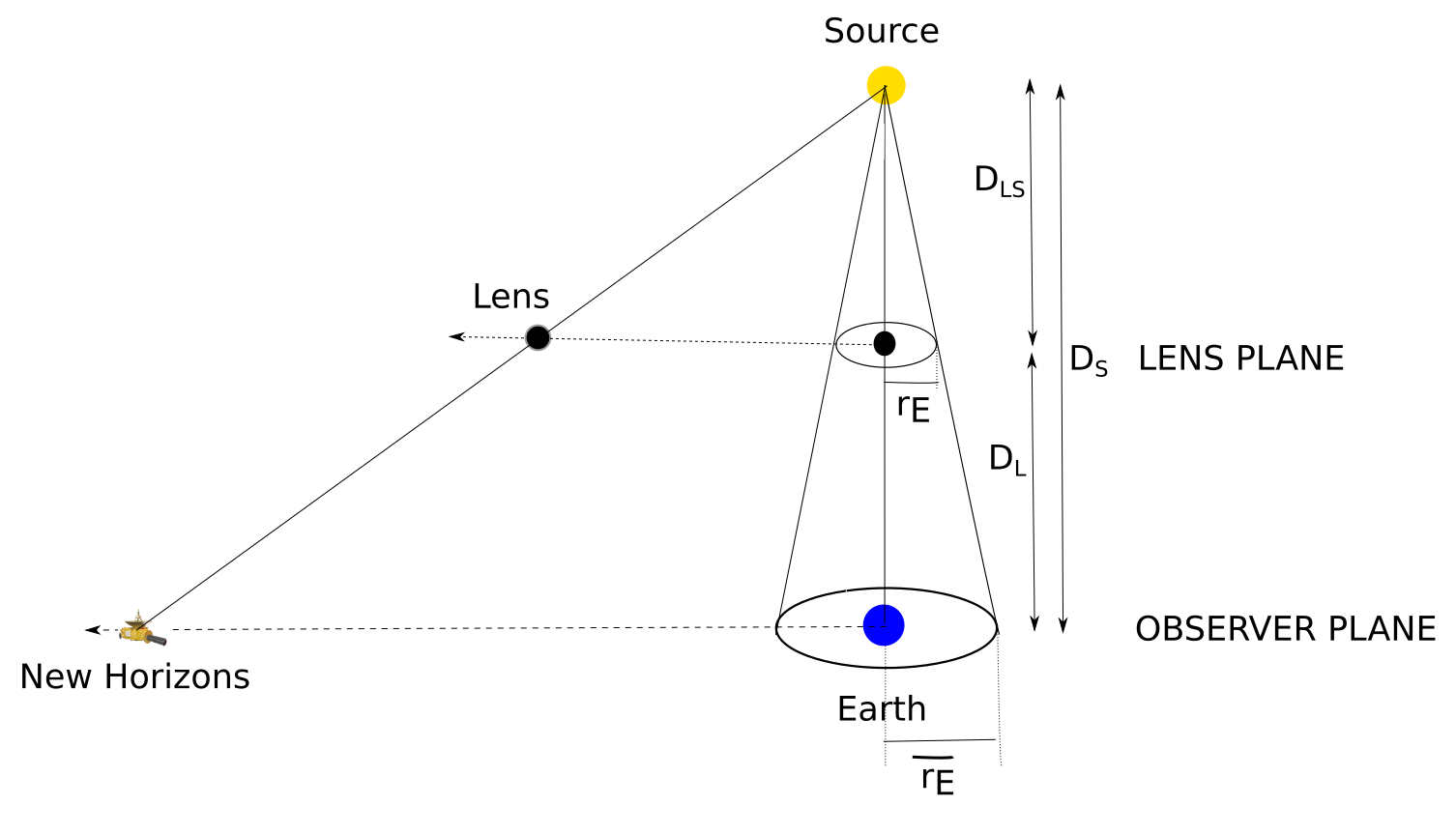}
\end{tabular}
\caption{Schematic diagram of the (simplified) geometry of a lensing event as seen from Earth and {\it New Horizons}, defined such that the Earth-source line is considered to be fixed and the lens moves relative to it.  The lensing object is shown as a black dot at the time of maximum magnification as seen from both Earth and {\it New Horizons}.  \label{fig:lensing_geometry}}
\end{centering}
\end{figure}

In contrast to those missions, {\it New Horizons} is now sufficiently distant from Earth that it will only observe a lensing event simultaneously if the lens is massive.  A lensing object of mass, $M_{L}$, at distance, $D_{L}$, from Earth deflects the light of a source at distance, $D_{S}$, around itself with a characteristic Einstein radius, $r_{E} = \sqrt{\frac{4GM_{L}D}{c^{2}}}$, where $D = \frac{D_{L}D_{LS}}{D_{S}}$ and $D_{LS}$ is the distance between the lens and source (see Fig.~\ref{fig:lensing_geometry}).  To give an illustrative example, for a 1\,$M_{\odot}$ object at 4\,kpc lensing a source at 8\,kpc, $r_{E} = 4.0\,$au.  Projecting this radius to the plane of the observer ($\tilde{r_{E}}$) gives a guide to the region within the solar system from which the event can be seen; any observer within this region will see the object lens the source star at the same time, though the maximum magnification and time of peak will vary as a result of the different closest approach separations observed from different locations.    For this stellar-mass lens example, $\tilde{r_{E}} = 8.1\,$au, beyond which the magnification drops off rapidly.  The component of {\it New Horizons'} separation from Earth perpendicular to the direction of the Galactic Bulge is $\sim 11.9\,$au at time of writing, placing it outside the projected Einstein radius of a stellar-mass lens, meaning that the magnification it would experience while the event is seen lensed from Earth would be undetectably small.  

However, a unique and exciting possibility is to use {\it New Horizons} to observe lensing by stellar-mass black holes.  Compared with the example above, for a 10$M_{\odot}$ black hole, $\tilde{r_{E}}$ = 25.5\,au, which is a good match to \NH' future position.  A number of theories for the formation of these objects have been proposed including primordial objects formed soon after the Big Bang \citep{Carr2016} to the remnants of stellar evolution  \citep{Elbert2018}, but the difficulties of observing them have made these theories hard to test.  Interest in this subject has been renewed as merging binary black holes are one source of the recent detections of gravitational wave events (\textit{e.g.}~\citealt{Abbott2016}).  Microlensing offers a way to measure their masses and binarity and, with an adequate sample, establish a measured mass function which can be compared with predictions. 

{\it New Horizons'} extraordinary velocity, currently 14.0 km s$^{-1}$, is close to half of the average orbital velocity of the Earth ($\sim 30 \,$km s$^{-1}$).  In the course of a typical event lasting $\sim 60 \,$days, the spacecraft moves 7.4$\times 10^{7}$\,km or $\sim 0.5\,$au, on a trajectory where the major component of motion is perpendicular to the Galactic plane.  By comparison, the Earth travels $\sim 1\,$au around its orbit and $\sim 0.3\,$au perpendicular to the Galactic plane within the same time frame.  The trajectory of {\it New Horizons} may produce a significant parallax signature which might be detected from {\it New Horizons} light curves, even without additional data.  It is therefore valuable to explore whether this can distinguished from degeneracies and used to characterize the events. 

Once a lensing event has been seen from Earth, the relative motion of the lens carries it out of alignment with the original observer.  In the case of massive ($\sim$few M$_{\odot}$) lenses, the source will still appear magnified from {\it New Horizons}, with a time offset in the event peak, while the magnification for lower mass lenses will be negligible.  

It is worth noting that for a small fraction of stellar-mass lenses, the relative trajectory of the source could in principle subsequently cross the line of sight to the source from {\it New Horizons}, so that the spacecraft would experience a different lensing event caused by the same lens, after a delay of a few months, for lenses moving with typical relative velocities of $\sim 120\,$km s$^{-1}$.   A small number of events discovered at one observing platform might therefore be followed up from the other.  For events where constraints on parallax can be derived from Earth-bound observations (for instance), one component of the parallax ($\pi_{E} = AU/\tilde{r_{E}} = (\pi_{E,N}, \pi_{E,E})$) is typically measured with far greater precision than the other.  Nevertheless, if the first observer can place some constraints on the event parallax, this information could be used to pre-select targets most likely to exhibit a lensing event from the second platform, and those follow-up observations would provide much tighter constraints on the lens trajectory and event model and hence on the lens' physical parameters.   

Companion objects in any lensing system can cause light curve anomalies that are most likely to occur if the projected separation of the companion from the primary lens at the time of the event happens to coincide with positions of the images of the source created by the primary lens, close to the Einstein radius.  Observations of an event therefore act to probe for companions at specific locations in the plane of the lens, which can be mapped (\textit{e.g.} \citealt{Tsapras2016}).  Naturally, during the delay between the peak of events caused by the same lens as seen from both Earth and {\it New Horizons}, any companion objects move in their orbits around the primary lens.  Observations of a second lensing event therefore effectively probes more of the lens plane, improving our sensitivity to lens companions.  These could also be used to detect binary source stars, as their orbit would subtly change their lens-source-observer alignment between events, and hence affect the observed magnification.  

The probability of microlensing occurring is intrinsically low, but highest in the direction of the dense star fields of the Galactic bulge (\textit{e.g.}~the rate for stellar-mass lensing is $\Gamma $=4.60$\pm$0.25$\times$ 10$^{-5}$ star$^{-1}$ yr$^{-1}$ at $|{b}| \sim$\, -1$\buildrel{\circ}\over{.}$4 and 2$\buildrel{\circ}\over{.}$25 $< l <$ 3$\buildrel{\circ}\over{.}$75 for sources with $I < 20$, see \citealt{SumiPenny2016}).  Microlensing programs therefore necessarily observe in highly crowded fields, and require reasonably high spatial resolution instruments.  For this reason, {\it New Horizons'} LORRI telescope is best suited to this science.  

LORRI offers a spatial resolution of 1$\times$1\,arcsec and a single wide optical passband.  While its pixel scale is somewhat larger than current ground-based optical surveys (\textit{e.g.}~OGLE has 0.26\,arcsec/pixel resolution) it is comparable with some of the telescopes used by ground-based follow-up teams (\textit{e.g.}~MicroFUN\footnote{\url{http://www.astronomy.ohio-state.edu/~microfun/}}).  The larger pixel scale means that the lensed source will suffer somewhat higher blending with nearby stars, but this can be determined by modeling the event light curve provided it is sampled at a range of different magnifications.  LORRI's wide passband is beneficial to harvesting as much light as possible from the relatively faint source stars ($I < 20 \,$mag) and the photometric precision required for microlensing is relaxed compared with transit measurements, typically $\sim$1\%.  Its reasonably wide field of view, which is similar to that of the first-generation microlensing surveys, could be used to monitor multiple events at once.  

In addition to a well-sampled light curve, multi-band photometry is required to determine the spectral type and distance of the source star in a microlensing event.  For massive lenses, this could be obtained from facilities on Earth, but for stellar-mass lenses, the Ralph-MVIC instrument offers five passbands through the optical that could be used for this purpose, though with lower spatial resolution.  While non-optimal in these crowded fields, this resolution is similar to that of {\it Kepler}, which has provided light curves of microlensing events thanks to advanced detrending techniques \citep{Zhu2017b}.  Ralph-MVIC has a brighter limiting magnitude than LORRI (R= 15.3\,mag at current maximum integration time), owing to its smaller aperture, and an asymmetric field of view.  This instrument is therefore better suited to a more targeted strategy, obtaining low-cadence multi-band imaging of selected bright events during their peaks.  

While the rate of black hole lensing is not well established, we can estimate the number of stellar-mass lensing events which {\it New Horizons} could detect from the distribution of baseline source star magnitudes alerted each year by the ground-based surveys.  Of 1834 events found by OGLE in 2017, 824 (44.9\%) had a baseline (\textit{i.e.} unlensed) magnitude $I <$18.6\,mag, LORRI's limiting magnitude in a 30s integration.  Of these, 46 events had a baseline brighter than Ralph-MVIC's limiting magnitude.  This figure underestimates the number of events which could be observable to Ralph-MVIC however, as color observations are primarily required over the peak of an event when the target is brighter.  While ground-based surveys cover a footprint that is much larger than {\it New Horizons} could monitor, they have also shown that events are not uniformly distributed across the Bulge \citep{Poleski2016}; $\sim$41\% are discovered within a central $\sim$3.3$\times$3.3$^{\circ}$ region.  

The most compelling case for microlensing with {\it New Horizons} would be a targeted strategy toward selected events discovered from Earth then observed at low cadence (once every few days/weeks) from the spacecraft, minimizing the downlink overheads (see Section \ref{sS:microlensing} for details).  This cadence would be sufficient to properly constrain the light curves of massive lenses.  

\subsection{Transient Follow-up}

The study of astronomical transients touches on many areas of physics. The explosions of massive stars as supernovae reveal the physics of matter under intense densities and temperatures, and provide insights into shock physics, the origins of the elements, and the sources of extragalactic neutrinos, high-energy particles, and gamma-rays. Rapid follow-up of gravitational wave detections has only begun, but the discovery of the first kilonova is already shedding light on the neutron star equation of state, the physics of their mergers, and the resulting $r$-process nucleosynthesis and its role in producing the heavy elements. 

Many of these phenomena occur with timescales ranging from a few days to a few months. Occasionally, critical phases of these events, or even entire events, could be missed due to the relative positions of the Earth, the Sun, and the event being studied. One notable example is the recent electromagnetic counterpart to the gravitational wave event GW170817 \citep{LIGO170817, LIGOmma}, which has allowed us to constrain the ejecta properties and the associated nucleosynthesis, study the environment of the neutron star merger, and for the first time, use gravitational wave events as ``standard sirens'' \citep{LIGOH0}, providing a completely new probe for cosmology. Had GW170817 occurred just one week later, it would have been unobservable to Earth-based ultraviolet, optical, and infrared telescopes due to sun constraints, so the electromagnetic counterpart would not have been found, and the incredible insights \citep[\textit{e.g.}][and references therein]{Metzger2017} gained from this event would have been lost.

More events need to be observed in order to settle the disputed nature of some of the emission components, and to improve the uncertainties in the inferred cosmological parameters. The expected rate of binary neutron star merger detections is uncertain, due to both order-of-magnitude uncertainties in the intrinsic rates of such mergers and uncertainties in the final sensitivity of the LIGO and Virgo detectors. However, when LIGO and Virgo reach design sensitivity, the event rate could be between a few per year and a few per week \citep{LIGO170817}. Because LIGO and Virgo are not sensitive to the Earth's position relative to the sun and can detect gravitational waves from any position in the sky, a large fraction of these events will be unobservable to any optical, ultraviolet or infrared telescope in existence, except one far from Earth. For approximately half of the year, \NH\ is opposite the sun from Earth, and therefore has exclusive access to large parts of the sky. 

Even a single detection by \NH\ could make the difference between identifying a counterpart candidate and not identifying one, which in itself is an important constraint on the physics of the event. In addition, this would localize the host galaxy, potentially setting interesting constraints on merger environments and hence populations \citep[\textit{e.g.}][]{Blanchard2017,Pan2017}, and allowing a (later) redshift determination for cosmological measurements \citep[\textit{e.g.}][]{LIGOH0}. In addition, a well-timed data point could help constrain the rise, peak or decline emission properties of a kilonova, each of which is critical for discerning competing emission models \citep[\textit{e.g.}][and references therein]{Arcavi2018}.

Several of the instruments that found the electromagnetic counterpart to GW170817 have very similar properties to those of LORRI. Despite the small field of view of LORRI compared with the LIGO and Virgo localization regions (of tens of square degrees), the counterpart was quickly identified by pointing telescopes to a list of known galaxies in the localization region \citep{Nissanke2013, Singer2016, Gehrels2016, Arcavi_Strategy}. This same strategy could be used with \NH\ when gravitational waves from a binary neutron star, or neutron star black hole merger, are detected on the opposite side of the sun from Earth. Even though the data could not be transmitted to us immediately, detecting the counterpart in retrospect, and obtaining even a single flux measurement, would be extremely useful for many of the above science cases. 

Additionally, \NH\ could be used to follow events discovered close to their observability limit, such as was the case with GW170817. In this scenario, the counterpart would be identified by other telescopes, and \NH\ would be used to image it once it can no longer be observed from Earth or Earth orbit. The point source sensitivity of LORRI is adequate to detect the kilonova associated with GW170817 that peaked at $r \sim 17$ \citep{Arcavi2017, Drout2017, Pian2017, Valenti2017}.  The GW170817 kilonova faded rapidly, and would have been below the LORRI detection limit within a few days. Also, had it been more distant, it might have been below the LORRI detection limit for the entire flare. However, it is still not clear how typical the GW170817 kilonova is, and whether future events may have a different peak magnitude and fading time scale. 

Additional high-value and time-critical transients could also be observed by \NH. Even single-epoch flux measurements of particular supernov\ae\ can be critical in bridging observing gaps due to sun constraints.  For example, the nearest superluminous supernova to date, SN\,2017egm, became unobservable due to sun constraints just 2-3 weeks after peak brightness \citep[\textit{e.g.}][]{Nicholl2017,Bose2018}. At a $V$-band magnitude of $\sim15$ it would have been readily observable by \NH. The physical mechanisms responsible for powering superluminous supernovae are still debated \citep[\textit{e.g.}][]{Howell2017}, and even a handful of observations would have been extremely useful to track the light curve over the 2.5 months until it became observable again, constraining its post-peak decline, and informing emission models. There are a handful of similar events each year where \NH\ could fill in such gaps in the data.

\section{Sensitivity and Stability Estimates}
\label{S:sensitivity}

To determine the capability of \NH\ for astrophysical
observations, it is necessary to estimate the sensitivity of the
instrument to both unresolved and resolved emission.  In Table
\ref{tab:instruments} we summarize the parameters of LORRI, Ralph, and
Alice based on published pre-launch and in-flight assessments of their
performance \citep{Morgan2005, Conard2005, Weaver2008, Cheng2008,
  Reuter2008, Stern2008, Noble2009, Cheng2010, Zemcov2017}.  Based on
these parameters, we can derive simple point source and extended
emission sensitivities that take into account the instrument
performance \citep{Bock2013}.  

\renewcommand*{\arraystretch}{1.15}

\begin{table*}[ht]
\centering
\footnotesize
\caption{A Summary of the Characteristics of New Horizons Instruments
  Capable of Astrophysical Observations.} 
\label{tab:instruments}
\begin{tabularx}{\textwidth}{|p{2.1in}|s|s|s|s|}
\hline

Parameter & LORRI & Ralph-MVIC & Ralph-LEISA & ALICE \\ \hline

Instrument type & Single-band imager & Multi-band imager & Imaging spectrometer & Spectrometer \\

Wavelength range$^{a}$ & $440{-}870 \,$nm & $400{-}975 \,$nm & $1.25{-}2.5
\, \mu$m &  $470{-}1880 \,$\AA \\ 

Spectral resolution & $1.2$ & $1.2$ (pan \& framing), $3.2$ (blue), $3.9$ (red), $4.5$ (IR), $17.7$ (CH4) & $240$ & $133$ \\

Spatial resolution (arcsec$^{2}$) & $1.0 \times 1.0$ (or $4.1 \times
4.1^{\rm i}$) & $4.1\times 4.1$ & $12.8 \times 12.8$ & $1000 \times 1000$\\

Number of pixels & $1024 \times 1024$ (or $256 \times 256^{\rm ii}$) &
Framing channel $5024 \times 128$; all others $5024 \times
32$& $256 \times 256$ ($\sim 1$ pixel per spectral element) & $1024
\times 32$ \\

Field of view (sq.~deg.) & $0.29 \times 0.29$ & $5.7 \times 0.037$ &
$0.9 \times 0.9$ & $0.1 \times 4.0$ + $2.0 \times 2.0$ \\

Telescope primary aperture (cm) & $20.8$ & $7.5$ & $7.5$ & $4 \times 4$\\

Point-spread function FWHM (arcsec) & $2$ &  $6$ & $19$ & - \\



Data size (Mb frame$^{-1}$) & $16$ (or $1^{\rm ii}$) & 17.9 &
$1.0$ & $0.5$ \\

Maximum integration time (s) & $30$ & $ 10 $ & $4$ & $3600^{\rm iii}$ \\

Point source sensitivity$^{\rm i}$ & $V=20.5$ in $4 \times 4$ pixel
                                     bins$^{\rm ii}$ for $G$-type star & $R=15.3$ &
                                                                 $J=10.6$,
                                                                 $H=9.8$,
                                                                 $K=8.9$
                                             & - \\

Per-pixel surface brightness sensitivity$^{\rm iv}$ & $2.2 \times
10^{3} \,$\nw & $3.8 \times 10^{4} \,$\nw & $6.0 \times 10^{4} \,$\nw
& $0.4 \,$Rayleigh\\

Characteristic surface brightness sensitivity$^{\rm iv}$ & $10 \,$\nw & $95
\,$\nw\ & $750 \,$\nw\ in $R = 10$ bins & $0.4 \,$\nw\ at $R=133$ \\

\hline
\multicolumn{5}{l}{$^{\rm i}$ Approximate performance; please see \citet{Stern2008} and references therein for details.} \\
\multicolumn{5}{l}{$^{\rm ii}$ Deep observations are typically performed in $4 \times 4$ pixel binning mode to improve sensitivity.} \\
\multicolumn{5}{l}{$^{\rm iii}$ The maximum programmable integration time is actually 65,535 s, but 3600 s is typically considered the}\\
\multicolumn{5}{l}{functional maximum.} \\
\multicolumn{5}{l}{$^{\rm iv}$ $1 \sigma$ $\lambda I_{\lambda}$ at maximum integration time (as discussed in Section \ref{sS:limits}).  Red channel specifications listed for MVIC.} \\
\end{tabularx}
\end{table*}

\renewcommand*{\arraystretch}{1.0}

The surface brightness sensitivity estimate for LORRI listed in Table 
\ref{tab:instruments} is based on in-flight performance that 
the \NH\ team has measured.  \citet{Zemcov2017}
performed a detailed study of the LORRI performance in the context of
astrophysical observations of diffuse surface brightness, and find performance 
figures in agreement with the LORRI team.  In that work, the quoted calibration factors are referenced to a flat-spectrum source, but there is no tension between them and recent calibration work with \NH.  

\begin{figure*}[htb]
\centering
	\includegraphics*[width=\textwidth]{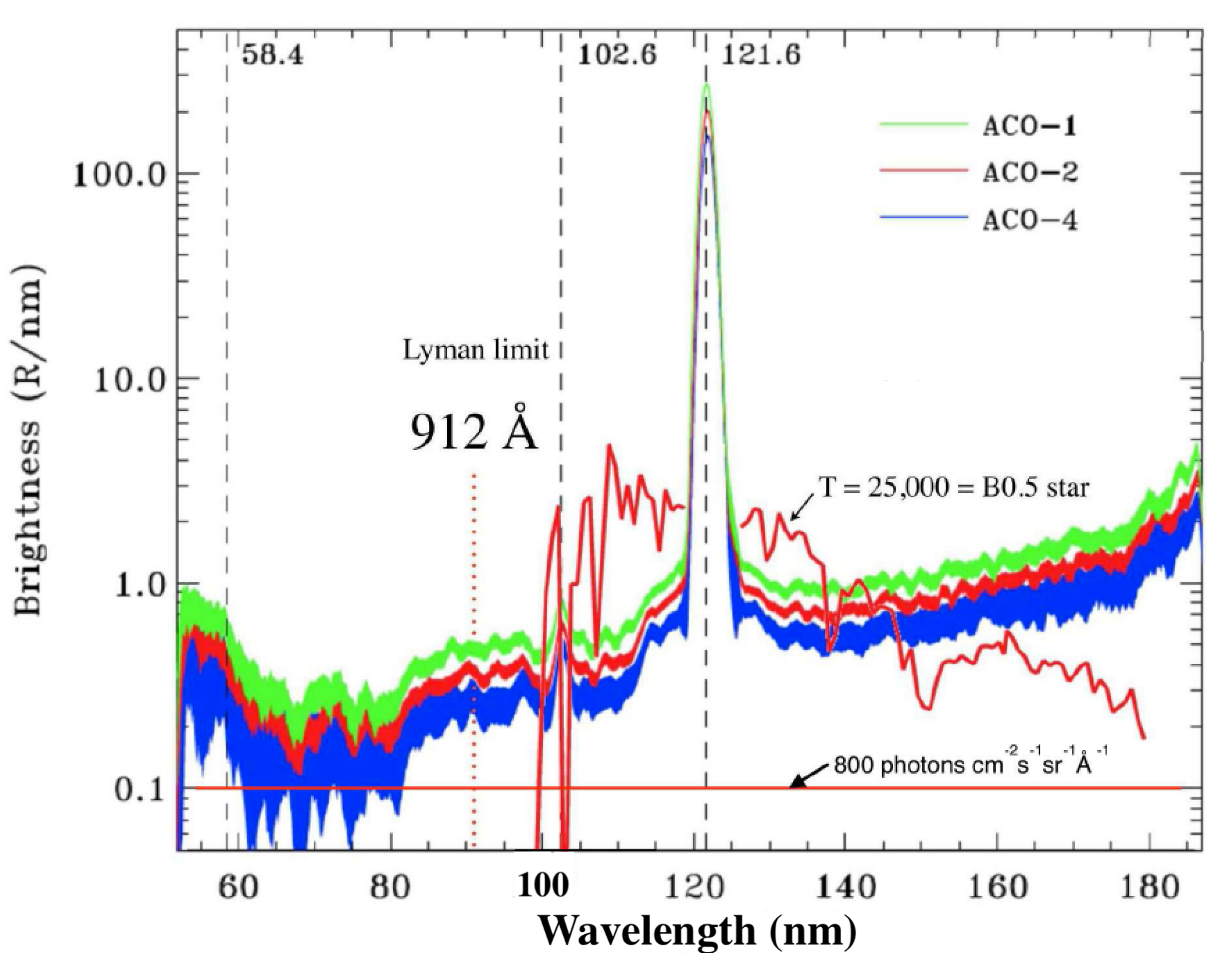}
	\caption{Three published Alice spectra of the interplanetary medium, showing the decline in surface brightness with increasing heliocentric distance observed in the same field \citep{Gladstone2013}. Also appearing is the spectrum of a hot star, with a vertical scale exaggerated by a factor 1.5 to bring out more clearly the stellar spectral features that we hope to detect (or, more dramatically, to fail to detect) in Alice observations of the cosmic background.  A brightness of 0.1 R/nm corresponds to 0.01 R/\AA\ or 800 photons cm$^{-2}$ s$^{-1}$ sr$^{-1}$ \AA$^{-1}$. These spectra are dominated by the light of solar Lyman-$\alpha$ scattering off the interstellar hydrogen that is constantly flowing through the solar system. As \NH\ becomes more distant, this foreground component decreases, as is apparent from these observations that were made many months apart.
} 
	\label{Alice_sensitivity}
\end{figure*}

The sensitivity characteristics of Alice are given in \citet{Stern2008} and summarized in Figure \ref{Alice_sensitivity}.  The instrumental Lyman-$\alpha$ foreground has been declining steadily, and recent unpublished Alice observations show that the instrument is currently at a level such that we can expect to obtain spectra with astrophysical as opposed to instrumental information in coming years (Murthy, private communication).

MVIC is well characterized, and has observed a variety of astronomical objects during cruise phase \citep{Olkin2006,Howett2017}.  As a result, its noise properties and radiometric calibration are quite well understood, and are summarized in Table \ref{tab:instruments}.  As a check of the predictions given in \citet{Reuter2008}, we performed an analysis of the 2006 observations of Asteroid 2002 JF56, and found array standard deviations well matched to the notional noise levels in calibrated data.  As predicted, the effective surface brightness sensitivity is worse than that for LORRI, largely due to the combination of smaller aperture, narrower spectral bandpass, and shorter maximum integration time.

To assess the sensitivity of LEISA, we have studied data taken on the 
star Vega ($\alpha$Lyr) in late 2008.  In this observation, the star was scanned across the dispersive direction of the imaging array with $t_{\rm int} = 0.59 \,$s per resolution element.  The total observation time was $198 \,$s.  The data are calibrated to $I_{\lambda}$ in erg s$^{-1}$ cm$^{-2}$ sr$^{-1}$ \AA$^{-1}$ using the nominal calibration factor for LEISA (derived, in part, from these same data).  In our analysis, we performed aperture photometry of the star image in each frame, using a circular aperture of $r=2 \,$pixels and an outer annular aperture of $2 < r < 4$ pixels.  We then subtract the \textit{HST} CALSPEC Flux Standard for $\alpha$Lyr and compute the standard deviation of the residuals \citep{Bohlin2014} to determine $\sigma(I_{\lambda})$.  From this, we estimate $\sigma(\lambda I_{\lambda})$ to be $< 6.7 \times 10^{4} \,$\nw per pixel in a 4 s integration and assuming the FWHM of the beam is $1.44 \times$ the pixel resolution \citep{Reuter2008}, which is consistent with the published estimate of $6.0 \times 10^{4} \,$\nw per pixel.

\begin{figure*}[htb]
\centering
	\includegraphics*[width=0.53\textwidth]{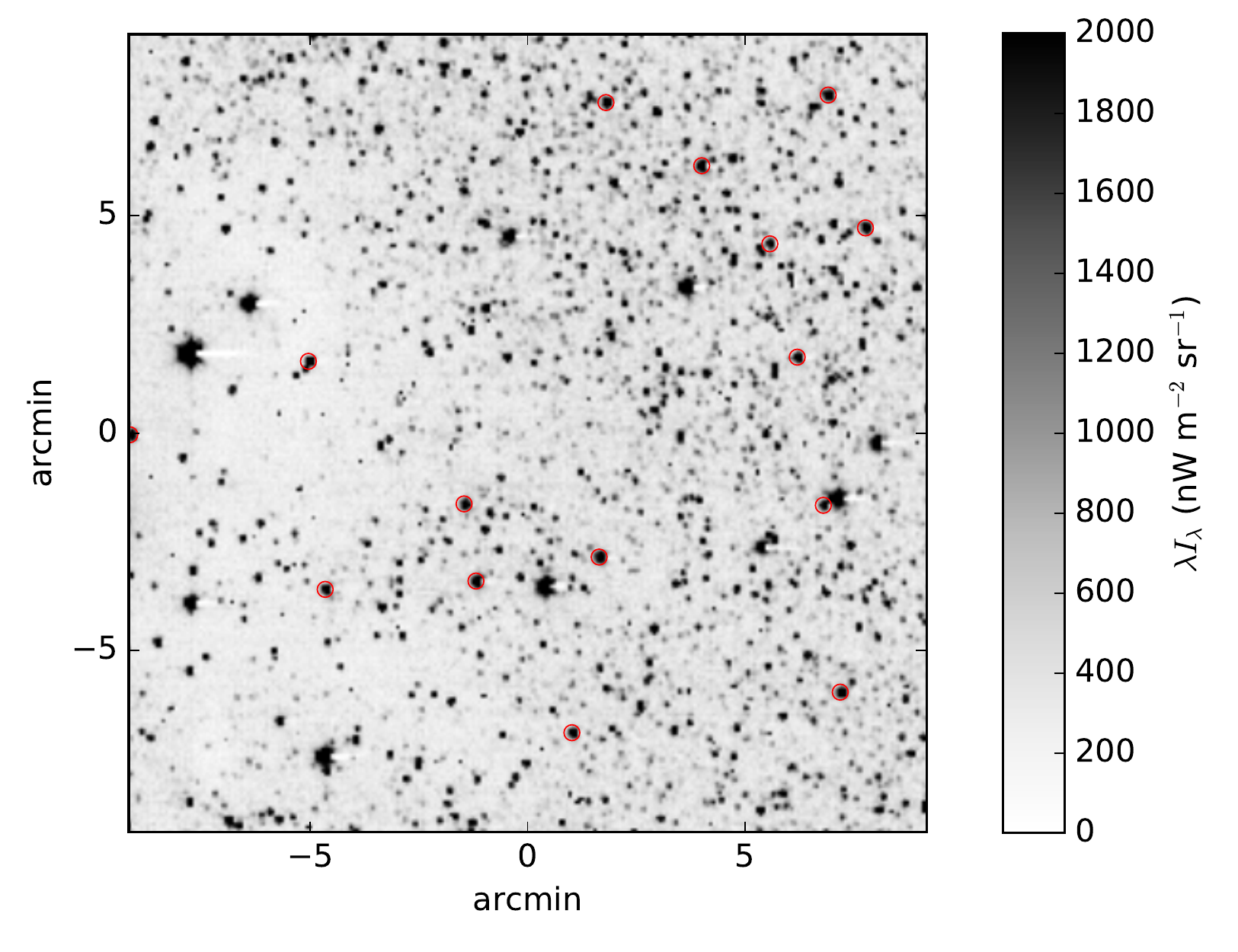}
    \includegraphics*[width=0.46\textwidth]{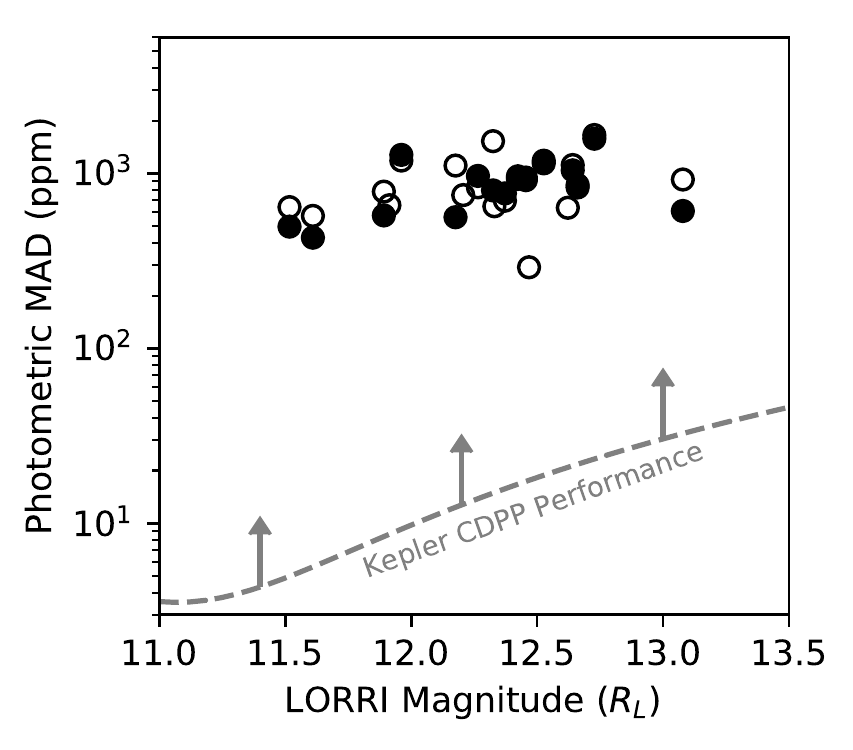}
	\caption{Left panel: an example image of the photometry stability field we study in this work.  The image is calibrated as $\lambda I_{\lambda}$.  Stars selected for the photometry study described here are circled, and lie within $13.1 < R_{\rm L} < 11.3$.  Some of the stars used in the study do not fall into this image, but do fall in others in the set of observations.  Right panel: an assessment of the photometric precision of LORRI based on 191 $t_{\rm int} = 10\,$s observations acquired from summer 2012 to summer 2014 (open circles), and 87 observations over a 5.5 hour period from July 2013 (filled circles).  Both populations are expressed as parts per million in flux.  This measurement is not ideal, as we do not have access to six hours of continuous 10s integrations, but compare our results to the lower limit from the \textit{Kepler} mission, which (in ``long integration'' mode) is more than an order of magnitude more stable at these source fluxes.   That difference in performance can be accounted for by the different aperture size, integration time, pointing control, and data analysis between these observations and \textit{Kepler}'s.
} 
	\label{fig:photom}
\end{figure*}

The instrument most useful for exoplanet investigations is LORRI, where the primary parameter of interest is the photometric stability of the instrument.  To help assess the photometric stability of LORRI, we have used data from the Pluto cruise phase of the \NH\ mission centered on $(\alpha_{\rm J2000}, \delta_{\rm J2000}) = (18^{\rm h} 02^{\rm m}.6,-14^{\circ} 37^{\prime}.8)$.  This field happened to be the position of Pluto as viewed from \NH\ between 2012 and 2014 while the mission was in-bound from about the orbit of Uranus.  Pluto was still a faint object in these images, and many stars are visible in them.  An example image is shown in Figure \ref{fig:photom}.

\begin{figure*}[htb]
\centering
	\includegraphics*[width=\textwidth]{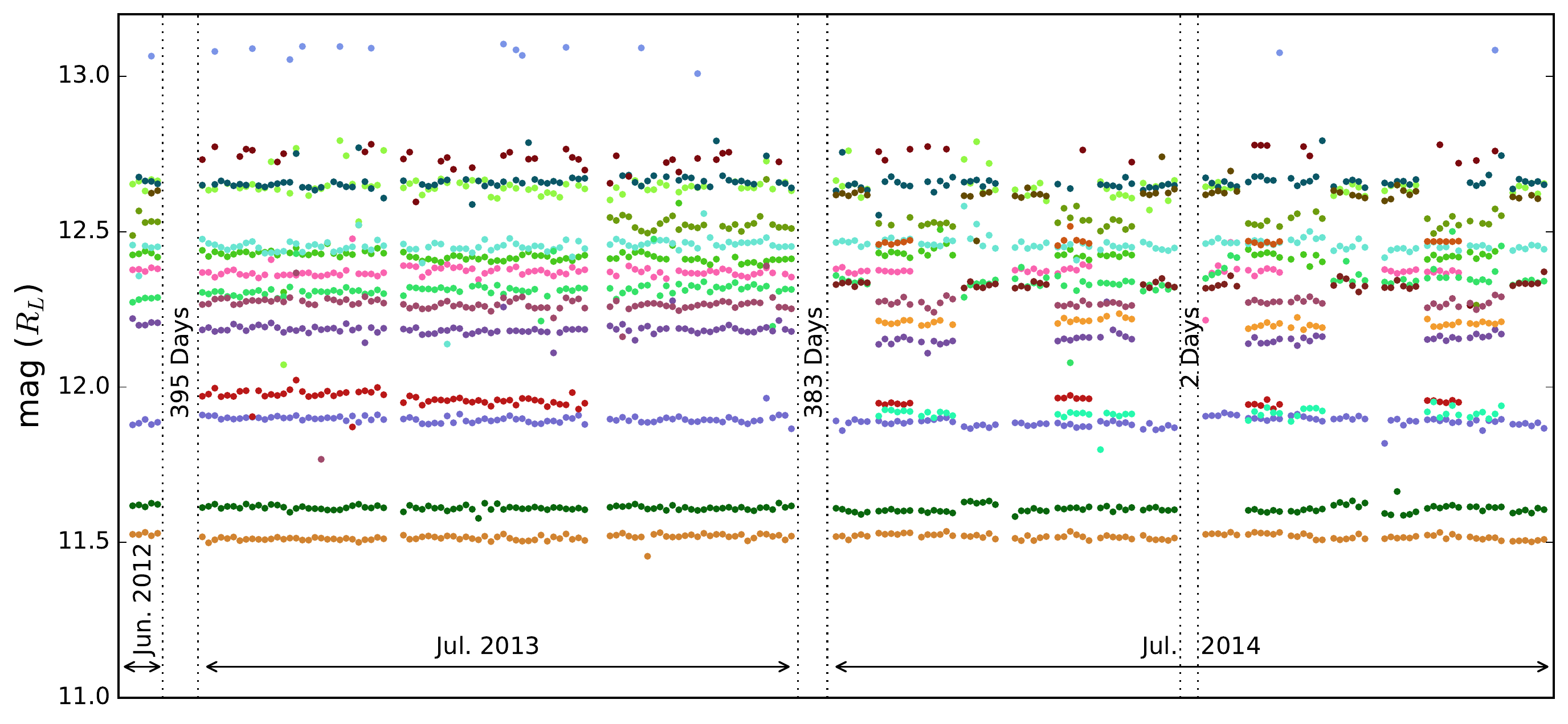}
	\caption{Measured fluxes of sources over three major observation epochs on the same field acquired between 2012 June and 2014 July.  The time axis has been compressed to aid visualization; gaps between points are roughly logarithmic, with the shortest corresponding to tens of seconds and the longest corresponding to hundreds of days.  The largest temporal discontinuities are highlighted by vertical dashed lines.  Because of the changing central position and position angle of the images with time, a given source may or may not be visible in a particular observation.  The variations in these photometric data are used to calculate the photometric precision shown in Figure \ref{fig:photom}, particularly the data set taken in 2013 July which covers about 5.5 hours.  We find no evidence for effects related to waking the instrument after year-long hibernations, and find the stability of the detector is excellent over the period.   
} 
	\label{fig:photomstreams}
\end{figure*}

The data discussed here were reduced and calibrated using the pipeline described in \citet{Zemcov2017}.  As in that work, these observations are ``found data'' that are not ideal for this type of stability characterization, but they do provide an estimate sufficient for our purposes.  The data records consist of 191 $t_{\rm int} = 10 \,$s integrations on the Pluto monitoring field taken from June 2012 to July 2014.  Following calibration, for each field we find $R_{\rm L} > 13.1$ sources and perform photometry on them using {\sc sextractor} in AUTO\_MAG mode.  We cross-identify the sources over images using their positions, and reject $R_{\rm L} < 11.3$ sources as they saturate the detector in this integration time.  Because the field is near the Galactic plane, they suffer from source crowding, giving us a wide sampling of environments.  Also, the position angle and pointing of the images shifts over the course of the observation epoch, so a particular source is not always present in a given image.  Figure \ref{fig:photomstreams} summarizes the photometry measurements for a selection of 20 relatively bright sources over the course of the observations.  Importantly, we see no evidence for turn-on effects after $\sim 1$ year of hibernation, meaning that observations separated by long time intervals do not seem to suffer from transient effects related to power cycling.

To summarize the photometric performance of LORRI, we compute the median absolute deviation (MAD) of the flux measurements for each source.  We perform this calculation for the full set of 191 observations taken over two years, and for a subset of the data taken in 2013 July.  The subset consists of three blocks of contiguous observations, each consisting of $29\times$ 10 s integrations, each spread about 2.5 hours apart.  This subset constrains the behavior of the photometry over a $\sim 6$ hour period, which can be compared against six-hour accuracy measurements for other instruments in the literature.

The MAD results scaled to ppm of flux for the ensemble are summarized in Figure \ref{fig:photom}.  For reference, we compare this to \textit{Kepler's} ``long integration'' photometric accuracy \citep{Jenkins2010, Gilliland2011, Christiansen2012, Vanderburg2014}.  We find the LORRI photometric stability is $1{-}2$ orders of magnitude less accurate at these magnitudes.  This estimate does not probe the photometric precision continuously over the six-hour period during which complex effects we would be unaware of can begin to have an effect, and if these are present they could inflate the variance in the data.  Though not the most accurate conceivable instrument, we conclude that LORRI is capable of precise enough photometry to do interesting science.

\section{Considerations for Survey Operations}
\label{S:ops}

Though the astrophysical science possible from \NH\ is 
compelling, there are practical considerations that limit the
observations possible with the spacecraft.  There are both programmatic and technical complications that would need to be addressed before \NH\ could be used for these observations. In addition to the costs associated with data telemetry, keeping the operations team active, \textit{etc.}, these observations may impose extra stress on the spacecraft that could result in additional risk to the mission.

In this section, we discuss
these technical limitations, their impacts on the science cases, and present a
hypothetical operations scenario that would generate a rich and unique
dataset.  Our predictions are based on publicly available information, and we note that a detailed engineering study of the observations fully considering the spacecraft subsystem performance is beyond the scope of the current work.  Properly designed, the new insights these observations would
lead to are unlikely to be rivaled for the foreseeable future.

\subsection{Attitude Control Considerations}
\label{sS:attitude}

Due to power considerations, \NH\ does not have a
reaction wheel-based pointing system.  Instead, hydrazine thrusters
are used to provide pointing control.  Attitude data from the star tracker
and a laser-ring gyroscope system are input to a feedback loop to set
the pointing within prescribed limits in both absolute astrometry and
drift.  In three-axis mode, a targeted position can be found to
within $1^{\prime}.2$ ($1 \sigma$), and active scans can be controlled to that location
within a typical deadband of $1^{\prime}.7$.  The nominal passive drift rate
once an attitude has been achieved is $5^{\prime \prime}$ s$^{-1}$ \citep{Conard2017}, though analysis of images suggests it is frequently significantly better \citep{Zemcov2017}.
Details of the attitude control system can be found in
\citet{Rogers2006} and \citet{Fountain2008}.  

In addition to the attitude control performance, the propellant
required to point the spacecraft is a limiting factor to the observations
performed during any extended mission.  At this time, the predicted mass of propellant following the end of the KEM mission is 10 kg, as compared with about 40 kg remaining at the end of the primary Pluto fly-by mission \citep{Bushman2017}.  As a benchmark, a change in \NH' spin rate of 5 RPM (the change from the nominal spin rate to zero
RPM for three-axis control mode) requires approximately 0.125 kg of hydrazine \citep{Fountain2008}.  Ultimately, the remaining propellant is likely to be the limiting factor in determining precisely which observations and science cases are possible in an extended mission for astrophysics.

\subsection{Telemetry Considerations}
\label{sS:telemetry}

\begin{figure}[htp]
\centering
	\includegraphics*[width=0.49\textwidth]{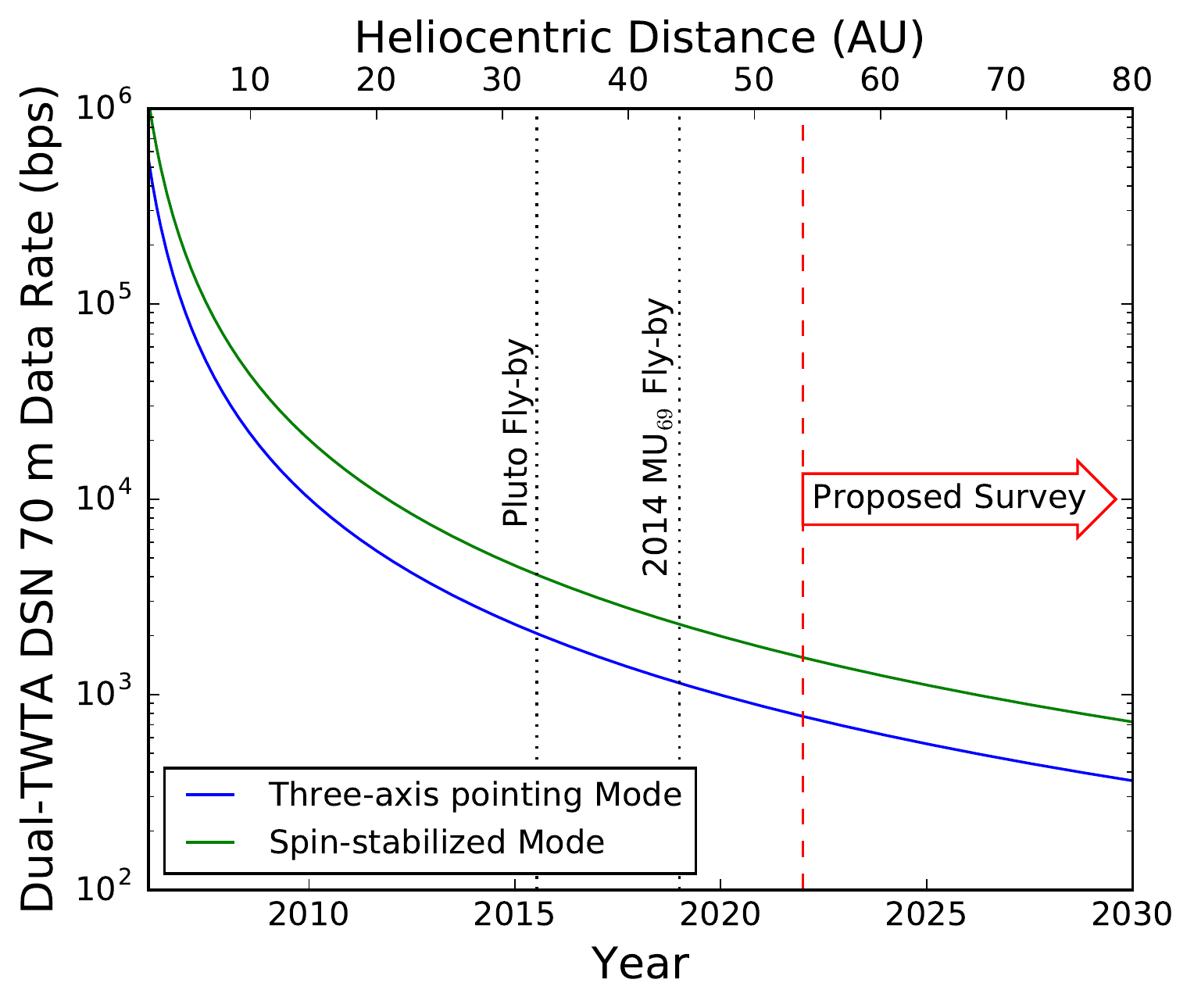}
	\caption{Data downlink rate from \NH\
          versus time and heliocentric distance.  The maximum
          achievable rate given in bits per second (bps) decreases as
          $R_{\odot}^{-2}$ and depends on the attitude control mode
          of the spacecraft.  In this calculation we assume the DSN
          70m dish is used in Dual-TWTA mode (see \citealt{Fountain2008} for details).  }
	\label{fig:datarate}
\end{figure}

Downlinking data from distant instruments has presented a challenge
since the beginning of deep-space missions.  As an example, the 
data acquired
for the prime \NH\ Pluto fly-by mission required only one
week to acquire, but over 16 months to telemeter back to Earth.  The
available bandwidth only decreases with time as the distance to
\NH\ increases.  In Figure \ref{fig:datarate}, we
show the achievable data rate from the beginning of the \NH\ mission until 2030, at which point the spacecraft will be
some 80 au from us \citep{Fountain2008}.

Assuming the proposed astronomical measurements do not occur until
2022 following the 2014 MU$_{69}$ extended mission, we expect a
maximum data rate in three-axis pointing mode (\textit{i.e.} the mode
in which observations will be performed) to be $\sim 900$ bits per
second (bps).  If observational data were telemetered in
spin-stabilized mode this increases to
$\sim 1.8\,$kbps.  The per-frame size of the various \NH\ data products is given in Table \ref{tab:instruments}, and
typically measure in the $\sim 10 \,$Mb range.  Assuming a $50 \,$\%
duty cycle and 30\% compression, even at maximum telemetry speed this corresponds to only
$240 \,$Mb per day, which is approximately 120 mins of LORRI data, or 15
mins of MVIC/LEISA integrations.

\subsection{Other Instrument Limitations}
\label{sS:limits}

In addition to attitude control and telemetry, there are additional
constraints on the instrument hardware to consider.  The first of
these is the maximum allowable integration time for the instruments,
which was set to $30 \,$s for LORRI \citep{Cheng2008}, $10 \,$s for MVIC, and $4 \,$s for LEISA \citep{Reuter2008} at launch.
Alice's maximum integration time of $18 \,$hrs is less restrictive, though long-term pointing drifts may prove problematic here.  
For flux-limited observations, because of the low data downlink rate,
it is desirable to increase the integration time to achieve equal
sensitivity in fewer detector reads.  However, because of \NH'
relatively poor attitude control performance, longer integrations may
suffer from image smearing as source images track along the detector.  Given  typical attitude drift rates, integrations lasting several minutes might offer an advantage.  Such changes are feasible; the \NH\ team is in the process of increasing LORRI's maximum integration time to $60 \,$s (H.~Weaver, private communication 2017).
Optimizing integration times requires a detailed trade study to understand the benefits and costs given the constrained attitude control, telemetry rate,
and particulars of a science case, which we leave to future work.

A second consideration is the optical performance of the instruments.
LORRI's rejection of off-axis light is relatively poor, such that
viewing within a solar elongation angle of $90^{\circ}$ results in
scattered light in the image \citep{Cheng2010}.  Further, there are
optical ghost paths between $0^{\circ}.2$ and $0^{\circ}.35$ from the
optical axis, which leave out-of-focus images of the secondary
mirror on the detector array \citep{Cheng2010}.  In addition, LEISA suffers from a solar light leak with rays sensed by the detector coming from behind the instrument at the $10^{-7}$ level \citep{Reuter2008}.  Though it is
not difficult to work within these restrictions, some science cases
(for example, imaging towards the inner
solar system) are precluded.  Other optical
features that must be considered in the design of
observations requiring high sensitivity and stability may also be present; this is another issue that requires detailed communication with the instrument teams to optimize.

\subsection{Assessment of Science Cases} 

Given the instrument sensitivities and practical considerations
discussed above, here we assess the feasibility of the different
science cases presented in Section \ref{S:science}.  The sky available for observations covers the $2 \pi$ sr away from the sun, which contains the $\Delta \ell = 180^{\circ}$ around the Galactic center, as well as a variety of extragalactic deep fields for EBL measurements.  There are no power limitations nor spacecraft maneuvering constraints associated with pointing the instruments away from the line of sight to the Earth.

\subsubsection{Measurement of EBL} 

Of the \NH\ instruments, the most sensitive
instrument to diffuse emission is LORRI, which has the largest
telescope aperture and widest bandpass.  The expected IGL at LORRI's
wavelength is $\sim 8 \,$\nw.  As a result, if no pixel masking were
required and only uncorrelated random noise were present, it should be
possible to measure the IGL at $S/N \gtrsim 0.5$ in a single 30 s
integration with LORRI.  However, it is necessary to mask some
fraction of pixels that contain bright stars, and the actual noise in
the instrument is not ideal.  As a result, the previous measurement 
\citep{Zemcov2017} reached a statistical error of $7 \,$\nw\ in
$240 \,$s of integration time.  To reach an uncertainty level
comparable to the current uncertainty on IGL in a single field, some
$400 \times$ 30 s integrations would be required.  In Table
\ref{tab:sbsens} we give estimates of the total number of integrations
and total observation time required for this measurement (assuming no
overheads), as well as an estimate of the time required to telemeter
the data.  Assuming a best telemetry rate of $900 \,$bps, 30\% data compression ratio, and 50\%
duty cycle, this data set would require 1.6 days to transmit to Earth.
Though more time-consuming than the actual observations by a factor of
600, this is a relatively inexpensive measurement.  Statistical sensitivity is not likely to limit this
measurement as the CCD dark current stability, astrophysical
foregrounds, and other effects would be relatively large at these low flux
levels. The observation design would therefore rest on acquiring
adequate knowledge of the system performance and foregrounds to be
confident in the measurement.

\begin{table*}[ht]
\centering
\footnotesize
\caption{Surface Brightness Sensitivity Targets and Requirements.} 
\label{tab:sbsens}
\begin{tabular}{|l|s|s|s|s|}
\hline
Instrument & Target Sensitivity (\nw, $1 \sigma$) & Number of Integrations
                                        Required$^{\rm a}$ & Integration
                                                         Time Required
  & Time to Telemeter$^{\rm b}$ \\ \hline

LORRI & 1 & 400 & $200 \,$min & 1.6 days \\

MVIC & 2.5 & 1,500 & $240 \,$min &100 days \\

LEISA (+10\% SIM$^{\rm{c}}$) & 5 (at $R=10$) & 25,000 & $27 \,$hrs & $95$ days \\

\hline
\multicolumn{5}{l}{$^{\rm a}$ Assumes maximum programmable integration time.} \\
\multicolumn{5}{l}{$^{\rm b}$ Assumes a data rate of $900 \,$bps, 30\% compression ratio, and
  50\% data transmission duty cycle.} \\
\multicolumn{5}{l}{$^{\rm c}$ Assumes ``Solar Illumination'' mode (SIM) is used to monitor the dark current every 10 observations.}\\
\end{tabular}
\end{table*}

With a pixel RMS of $>4\times10^{4} \,$\nw\ in 10 s, MVIC would require
$\sim 10^{3}$ integrations to reach a statistically significant EBL
measurement, which in turn would require a few hours to execute.
Information from all five MVIC channels would be telemetered at once,
providing low-resolution spectral information in the optical, which is
an important addition to a LORRI measurement.  However, the data
telemetry for MVIC becomes a real consideration, with the transmission
time for this data set estimated to be a significant fraction of a
year.  MVIC
does have dark (\textit{i.e.} non-illuminated) pixels to allow a
measurement of the detector current in the absence of photons
\citep{Reuter2008}, an important prerequisite to absolute photometric
measurements \citep{Matsuura2017}.

Finally, LEISA would provide a unique measurement of the EBL, as it
covers crucial infrared bands where observations of the emission from
early galaxies are possible.  LEISA's sensitivity is poor, and it
would require at least two days equivalent observation time to execute observations that would yield a
$\sim 3 \sigma$ detection of the EBL per $R=10$ band.  However, we
again encounter a situation where the data telemetry is
time-consuming, largely due to the number of $4 \,$s integrations
required to achieve a useful surface brightness sensitivity.  Though
LEISA does not have built-in dark pixels, the instrument can be used
in ``Solar Illumination'' mode where a small pick-off mirror assembly
couples only $\sim $3,000 pixels to external illumination
\citep{Reuter2008}.  A suitable choice of standard mode observations
interspersed with Solar Illumination mode observations can offer
close to real-time assessment of the dark current in the PICNIC
detector array.  

It is likely that the dominant foreground after star masking will be the DGL.  This signal follows the structure of the Milky Way, and can be slightly fainter than the EBL, but is never zero.  As a result, a typical approach to subtracting the DGL component from the isotropic EBL is to observe a number of fields and correlate against a template for the DGL emission, usually based on maps of thermal emission from dust in the ISM.  The extrapolation to $I_{\rm DGL} = 0$ then provides an estimate for the EBL component.  To demonstrate the isotropy of the measured EBL, it is necessary to repeat this process in
several independent field sets at various ecliptic and galactic latitudes.  For the LORRI
observations this would multiply the single-field estimates by the number of fields to be observed, however, the per-field observation time could be reduced because the error on the EBL would be dominated by the overall fit uncertainties rather than the absolute uncertainty in a single field.  A precise optimization of the observations depends on details we leave for the future, but it is likely that an order of magnitude more LORRI observations than indicated above would allow us to demonstrate isotropy of the signal.  This type of measurement is prohibitively
expensive for the Ralph measurements, but we could reasonably rely on isotropy demonstrated by LORRI alone.  

\subsubsection{Ultraviolet Background Sensitivity}

The nominal sensitivity of Alice is $\sim 1 \,$R per pixel at $R=133$ over a 
32 kpixel detector array in a $600 \,$s integration.  Averaging over pixels, 
we estimate a total background sensitivity of about $0.02 \,$R/nm in this 
integration time, which corresponds to approximately 1,600 photons cm$^{-2}$
s$^{-1}$ sr$^{-1}$ \AA$^{-1}$.  Current estimates for the UV background place 
its surface brightness at $< 100 \,$photons cm$^{-2}$ s$^{-1}$ sr$^{-1}$
\AA$^{-1}$, meaning that we would require $256$ integrations to reach an
interesting sensitivity limit \citep{Henry2015}.  However, the UV background 
is faint compared to the galactic foregrounds, so Alice is expected to yield
interesting new information in only 30 integrations.  Because Alice 
observations are relatively inexpensive to telemeter, we would be able 
to both execute the observations and telemeter the resulting data in an 
equal amount of time.  For 256 observations, we estimate 2 days of 
observation and 2 days of telemetry are required.  This observation is 
inexpensive enough that many fields could be targeted on the sky, as 
long as propellant costs did not become problematic.

\subsubsection{EKB Dust Observations}

The primary regions of interest for IPD light scattering will be those
centered on the ecliptic where the IPD particle density is highest, but also
away from the galactic background, where DGL will contaminate the
images.  Though the very lowest levels of modeled IPD surface
brightness would be too challenging to reach with LORRI, deep
$< 1 \,$\nw\ sensitivities remain a real possibility, and would allow
us to constrain models for the composition and structure of the EKB.
The example calculation of IPD brightness shown in Figure
\ref{brightness} assumes silicate grains; however, we can also input
other dust grain compositions including ice
mantle/silicate cores, carbonaceous, and organic compositions
\citep[\textit{e.g.}][]{Warren_1984, Jenniskens_1993, Quinten_2002,
  Jager_2003}.  The observations would likely be performed as a
function of solar elongation, and repeated over time as the sight line
through the dust cloud changed to help deconvolve the structure
profile.  One interesting possibility to boost the signal is to
measure the EKB analog to the Gegenschein, which is due to reflection of
sunlight from dust in the directly anti-solar direction that boosts the
surface brightness of the local ZL signal by factors $\sim 100$.  Unfortunately, the anti-solar direction for \NH\ lies not far from the Galactic plane $\{\ell,b\} = (16^{\circ},-14^{\circ})$, so the galactic backgrounds may be large.  However, even upper
limits to the EKB dust surface brightness would be unique and useful
in this regard.  We would estimate that $\sim 10$ positions at
different solar elongations, observed every 5 au in heliocentric
radius, would make an excellent data set requiring only two months to telemeter.

\subsubsection{Exoplanet Transits}
\label{sS:transits}

Exoplanet transits are typically studied in relatively bright star systems,
and require photometric accuracy better than 1:1000 over $\sim 1
\,$hour timescales for studies of anomalous features in the light curve.  For a $V=8$ star, we would require a $1 \sigma$
photometric accuracy of $\delta V = 15.5$ to detect the presence of
the planet around \textit{e.g.} HD209458, which is significantly above
LORRI's $t_{\rm int} = 30 \,$s sensitivity.  This bodes well for the
use of LORRI in observing transits.  We do not suggest that LORRI would be appropriate for finding transiting systems, but rather that it could be used to follow up particularly interesting targets with known (or well-constrained) ephemerides.

Due to the need for propellant, it would be far too
expensive to operate LORRI in a constant-monitoring mode as was done
with \textit{e.g.~Kepler}. A more efficient use of the finite resources
would be to target known transiting systems in order to better characterize them.  Both
the transit timing and transit duration methods require precise
measurements of a planet's light curve over many transits to build up
a model of transit timing variations for mass measurements, or of a possible moon's orbit.  
The quiet environment and diagnostic information about the detector would be very useful in
ensuring instrument stability of this time.  
However, LORRI will mostly be used to improve the transit parameters and the uncertainty on the ephemerides of known
transiting exoplanets (including those to be discovered by \textit{TESS}).
All of these observations rely on well-understood and low-error light curves,
which \NH\ is in a unique position to generate.

In terms of operations, the data requirement of, for example, a $30
\,$s observation every five minutes for two hours would generate only 24
frames for telemetry.  The requirement of pointing stability would be
more problematic, as would the active pointing required to keep the
source in the same pixels over time, since this requires active use of
propellant.  This expenditure would have to be traded in the context of the larger mission goals and observations.

\subsubsection{Microlensing}
\label{sS:microlensing}

Microlensing events can magnify a source star by up to several magnitudes over the course of events lasting between $\sim$1\,day to several months.  Single-lens events can be detected with relatively low-cadence imaging, where the frequency required is a function of the Einstein timescale, $t_{E}$, and hence proportional to the lens mass. A sampling rate of once every $0.5{-}3\,$days is needed to detect stellar-mass lenses while black hole lens events require imaging only once every 1-2\,weeks.  However, both stellar and planetary {\em binary} events that comprise $\sim$10\% of the total are characterized by short-lived ($\sim$hours -- days) light curve anomalies which must be sufficiently well sampled to constrain the model.  Typical observations aim for a photometric precision of $<$0.01\,mag, and a cadence of at least four per hour.  We consider the practical implications of several possible observing strategies. 

LORRI's wide field of view suggests a survey strategy where {\it New Horizons} would repeatedly image the region of highest microlensing rate over the course of $>$2\,months.  The overall length of the observations would be determined by the need to measure the lensing light curve both over the peak of the event and at unlensed baseline in order to properly constrain the event magnification and timescale.  Surveying the full $\sim 3^{\circ}.3 \times 3^{\circ}.3$ central Bulge region would require a 11$\times$11 mosaic of LORRI images.  Although in principle it could achieve a cadence of $\sim$4\,hrs, this strategy would be prohibitively expensive on propellant.  Furthermore, it would accumulate data far in excess of the downlink capacity, some $\sim$11.9\,GB/day (noting that bulge observations could not be binned in order to preserve spatial resolution).  Surveying four LORRI field pointings once a day (or conversely, one field every 6\,hrs) has a more practical data rate of 67.2\,MB day$^{-1}$.  The wider footprint would ensure more events are detected ($\sim$22 year$^{-1}$ vs. $\sim$5 year$^{-1}$), while a single pointing would conserve propellant.  Arguably the most practical survey strategy would be to image as large a footprint as possible at a cadence of $\sim$once per week for a total duration of $>$150\,d, with the goal of detecting black hole lenses.  Concurrent observations of the same footprint conducted from Earth could be used to measure the event parallax and determine the physical properties of the lenses. However, it is difficult to estimate the yield of black holes detected this way as the rate is not well known. 

A second possible strategy would take advantage of {\it New Horizons'} unique position to act as an ``early warning system.''  As noted above, some fraction of events observed by {\it New Horizons} may subsequently be observed from Earth in separate lensing events after a delay of $\sim$0.5\,yr.  Were the spacecraft to undertake a very wide angle, but low- ($\sim$1-3 day) cadence survey of a wide region, there would be sufficient time to downlink the data and discover events which could then be intensively followed up from Earth and near-Earth missions.  LORRI could survey a 4$\times$4 grid of pointings, $\sim 1^{\circ}.16 \times 1^{\circ}.16$ each, once every 3 days with a data rate of 89.6\,MB day$^{-1}$.

The final (and probably most practical) option would be the converse: to use {\it New Horizons} to follow-up selected events discovered from Earth and/or by {\it WFIRST}.  Events observed from both Earth and {\it WFIRST} will already have constraints on the lens-source relative trajectory, allowing more stringent target selection, and many will already be known to have planetary or binary signatures.  In this way, {\it New Horizons} could act as a ``force multiplier'' for those surveys, to search for other planetary or stellar companions in the same systems thanks to its distinct line of sight to the event.  This strategy would require higher cadence observations (ideally $<$1\,hr) but over a shorter period during the peak of the event, with lower cadence (every $\sim$3\,days) observations taken before and after the peak to measure the event magnification.  As an example, a similar measurement from EPOXI used only 20 observations over the course of 48 hours to break degeneracies in the system MOA-2009-BLG-266 \citep{Muraki2011}.  This suggests the characterization of a single system could require only tens of observations, which would require much less than a day to downlink.

\subsubsection{Transient Follow-up}

LORRI has the point source sensitivity to reach the flux of some transient sources, and is stable over long time periods.  Measurement of transient events would require a ``fast track'' observation upload scheme.  It is likely that at least several days would pass between the detection of an event and a \NH\ observation taking place, and data would not necessarily be telemetered immediately.  Further, these measurements would only be useful for the period when \NH\ is near the sun as viewed from the Earth and observation from the ground is impossible.  It would be advantageous to optimize the \NH\ observation epochs to coincide with these periods, so that the instrument would already be in a mode to execute astrophysical observations. This requirement may not be compatible with rapid command uplink using the high-gain antenna. As a result of these additional requirements, the use of \NH\ for transient measurements remains somewhat speculative at this point.

\subsection{A Hypothetical Observational Campaign}

In designing an observational campaign, there are three major factors
to consider: (\textit{i}) the time required to telemeter the data back
to Earth, and the storage capacity and reliability of the onboard
data volumes; (\textit{ii}) the need to expend fuel for observations
requiring active pointing control; and (\textit{iii}) optical and
communications restrictions on the attitude of the spacecraft.

As shown in Table \ref{tab:sbsens}, the time to telemeter the data can
easily grow to be prohibitive.  The most cost-efficient instrument in
terms of sensitivity per data volume is LORRI, and we assume that most
of the observations would be performed with it.  Even so, the data
storage considerations impose a survey design similar to the
\textit{New Horizons}' planetary encounters, where an observation
campaign is pre-programmed and executed contiguously, and then later
telemetered to Earth while the spacecraft is in spin-stabilized mode.
This scheme takes advantage of the downlink rate boost of
spin-stabilized mode.  Based on purely data telemetry considerations,
we therefore propose a scheme where observations are performed
roughly annually in a short burst, and then telemetered during a
cruise phase.  This pattern could be repeated for a number of years, and
would ultimately be limited by the fuel required to maneuver the
spacecraft. 

The attitude control system likely limits the lifetime of the
mission.  To conserve the resource, observations that would
not require active pointing control, or at least could be performed
with periodic pointing correction, would be preferable.  Assuming the
nominal post-acquisition drift rate of $5^{\prime \prime}$
sec$^{-1}$, a target centered on the LORRI detector array would drift
off the field of view in $> 1.7 \,$minutes.  This sets a natural
cadence for attitude correction during measurements of point sources
that minimizes fuel consumption.  For deep observations of diffuse
surface brightness, and even more conservative viewing mode would be
to point the telescope on target, and then let it drift for some
specified time before re-pointing.  For observations of emission that
varies smoothly over sub-degree scales (for example, EBL, DGL, or
IPD light), the spacecraft could wander for up to one hour, by which time the center
of the field of view would have drifted by $0^{\circ}.5$.  Point
source emission could easily be masked following the post facto image
registration, and foreground emission requiring image-space
correlation could just use the reconstructed pointing of each image
separately.  The most challenging measurements are those requiring photometric precision, where drift causes a source to wander between pixels that have different relative photoresponse.  These observations are likely to require tighter attitude control than studies of diffuse brightness.  However, if controlled, in this work we have shown that LORRI can perform adequately to allow unique observations of both exoplanet transits and mircolensing.

The third consideration in our survey design are attitude constraints
due to the instruments, communications, or other features of the
spacecraft.  One obvious constraint is for the imaging instruments to
have a solar elongation $> 90^{\circ}$ at all times during an
observation, which constrains the field of regard to $2 \pi \,$sr away 
from the Sun, which will be close to frozen in
celestial coordinates for the duration of the mission.  There are
almost certainly additional constraints for the high-gain antenna and
other systems, and for keeping the solar illumination of the spacecraft
roughly constant to minimize thermal disturbances.

All of these constraints considered, \NH\ is still
capable of generating a rich and unique data set for astrophysical
science.  For the EBL science case, we would measure $5{-}10$
independent fields with LORRI to $\pm 1 \,$\nw\ to show isotropy in
the signal, and at least one field to $\pm 3 \,$\nw\ error with MVIC
and $\pm 4 \,$\nw error with LEISA.  These measurements would require a large number of integrations added together, likely
acquired over different epochs.

A measurement of the UV background with Alice is 
quite tractable, and requires only of order days of integration and 
telemetry time to achieve interesting sensitivities that cannot be 
reached from vantage points near the Earth.  The scientific goals of our proposed Alice observations are served well by any and all observations 
of any regions of the sky.  Particularly valuable will be comparison of 
Alice spectra obtained while pointed toward regions at high galactic 
latitudes compared against the same at low galactic latitudes (where 
starlight scattered from dust is expected to dominate the spectra, 
demonstrating instrumental capability).  Valuable data will be 
obtained on any target that is observed in pointed mode, and the 
pointing stability can drift considerably without harm to the value of 
the data obtained.  To conserve propellant 
and aid in cross-correlation studies, it is likely that the UV 
measurements would be performed concurrently with and on the same 
fields as the EBL measurements.  Obviously, it would be necessary 
to develop a detailed observation plan that would optimize the 
observation strategy.

For the EKB dust measurement, we would measure $\sim 10$ fields placed
at different ecliptic latitudes in each epoch of a multi-year mission.
Because the EKB dust is spatially smooth, these observations would not
require tight pointing control.  Measuring to $\pm 1 \,$\nw\ error
with LORRI in each epoch would allow a detailed probe of the structure
of the EKB.  Spectral information from MVIC and LEISA is likely too
expensive to be considered, but the EBL measurements may permit
interesting constraints on the longer wavelength behavior of the EKB
dust emission.  It is likely these fields and the extragalactic background fields would be designed in a coordinated fashion, since the observational requirements are very similar.

In our envisioned survey we would also observe a subset of known transiting systems
to improve the ephemerides, and monitor transit timing variations if there are multiple transiting planets. The subset of systems to be observed
would be selected based on prioritization of targets for atmospheric characterization with the
\textit{JWST} and large ground-based telescopes. These observations require pointing on a particular target 
for long periods of time (hours to days) with observations at a relatively fast cadence ($\sim 10$ per hour), so
would be expensive in terms of propellant. Assuming a 2-day measurement with a 5 minute cadence of $30 \,$s 
LORRI observations, we would require 576 frames to be telemetered.  Potentially, $\gtrsim 10$ such
observations could be carried out in a year.

Traditional exoplanet microlensing measurements require close to
constant monitoring of fields in the Galactic bulge region to increase
the number of possible targets. A microlensing survey based on this
design would thus require fairly constant sampling of a single target
field for as long a baseline as possible, and active pointing
correction to keep the field of view on target.  Here, we envision a
different approach.  An Earth-based microlensing survey monitoring a
known field could have a real-time event pipeline that triggers on
suspected star-star lensing events.  During \NH\
observation campaigns, these triggers could be passed to the science
team and programmed into the queue with priority.  The $\sim 10$ day
duration of these events gives ample time to design and upload an
observation into the queue.  The light curve of the source would be
monitored for several days, and short-duration microlensing events
indicative of exoplanets could be sought.  

Science cases that require point source photometry would benefit from windowing the image to a region around the target of interest, as this would significantly reduce the telemetry bandwidth requirement.  At the other extreme, onboard co-addition of images could allow an increase in the signal-to-noise ratio of static sources of emission.  Both algorithms would have to take into account the absolute pointing accuracy of the instrument and the drift of the images over time.  Finally, to maximize the available fuel resources, observations would
need to be designed to minimize slew distances on the sky.  Since
observations are planned well in advance (except for microlensing
events), this is not a prohibitive requirement.  Following a $\sim 1$
week long observation campaign each year, \NH\ would
go into spin-stabilized mode and begin transmitting the data to Earth.

\subsection{The Possibility of Science Observations During Spin-Stabilized Operations}

Given the limited propellant budget for pointed observations, one 
possibility of interest is to perform astrophysical observations in 
some form of the spin-stabilized operation.  This would provide the 
benefit of increasing the data telemetry rate while allowing different 
parts of the sky to be surveyed by the instruments.  The primary 
drawback of this scheme is related to the detectors; all of the 
detectors on the \NH\ instrument suite suitable for
astrophysical observations are of the charge integrating type, 
which usually require stable pointing over the course of an integration 
to provide clean images of the sky.  The cost of having 
\NH\ spin during observations is that the 
astrophysical signal would be smeared over multiple pixels, thereby 
complicating image analysis and, in the limit of read-noise limited 
measurements, decreasing the total signal-to-noise ratio on the 
source.  For a purely isotropic signal, or one that is not spatially 
structured on the angular scale of the spin smear, this is not 
problematic.  However, most of the science cases discussed here 
require spatial resolution either to monitor a point-like source, 
or to remove it through masking.  As a result, allowing the 
spacecraft to spin could be problematic.

In \NH' standard spin-stabilized mode the spacecraft 
is spinning around the high-gain antenna's boresight at 5 RPM, which 
corresponds to $30^{\circ}$ minute$^{-1}$.  This is clearly prohibitively 
fast, as it means (for example) that LORRI's field of view is moving one 
full array width every 0.6 s.  Even in unreasonably short integration 
times, images of stars would be smeared.  As a conservative estimate 
for the preferred spin rate, we impose the requirement that, over a full 
$30 \,$s integration, the LORRI image can shift by 0.5 pixels, or 2 arcsec.
This corresponds to a spin rate of $3.1 \times 10^{-6}$ RPM, which is 
clearly a different engineering regime than the current spin-stabilized mode.  Due to their larger pixels and shorter integration times, the other 
instruments could accept relatively faster spin rates, though still 
within an order of magnitude of the LORRI requirement.  Faster spin rates may also be acceptable, with a concomitant loss in scientific capability.  This kind of observation may be enabling for EBL science with MVIC and LEISA, where the required integration times and data volumes are probably prohibitive in pointed mode, but if the observations can be spread over many months they become more tractable.  

We conclude that, though it may be technically challenging to implement, 
it is worth studying the possibility of a spin-stabilized mode with a 
very slow spin rate.  Observing in this mode would not require any 
propellant, would increase the data telemetry rate, and would allow 
maps of large areas of sky to be constructed.  Some of the science 
cases, particularly those related to diffuse emission, could potentially 
benefit from such an observation strategy.

\section{Conclusions}
\label{S:conclusion}

With a fully functioning \NH\ beyond the orbit of Pluto, the astrophysical and planetary communities have a rare opportunity to perform unique science with an instrumentation suite capable of deep and precise observation of the cosmos.  In this paper we have motivated the broad scientific fields such observations can address, as well as studied the performance of the instruments and discussed the various limitations and considerations a future survey with \NH\ would have to address.  We find that \NH\ is well suited to astrophysical observation, and that a carefully designed survey optimizing the expenditure of propellant and telemetry bandwidth while minimizing spacecraft operational risk could provide interesting new insights in astrophysics.  Some data of astrophysical interest is already available in the archive, and the analysis of these is ongoing.  Insights from these will help us design better observations.  Going forward, we suggest a study of the detailed feasibility of astrophysical observations with \NH\, combining the \NH\ instrument and engineering teams with astrophysical experts in the various scientific fields discussed here.  This will permit an accurate assessment of the current capabilities of the instruments and spacecraft and a detailed observation plan to be formulated.  

\acknowledgments

We would like to thank the \NH\ science and instrument teams for their decades of dedicated effort designing,
building and flying such a complex mission, and in particular H.~Weaver for his patience in answering our largely impenetrable queries and his thoughtful input on our work.  
We would also like to thank B.~Crill for his insightful comments that helped improve the study, and our six referees for their incisive thoughts.

The \textit{New Horizons} cruise phase data sets used in this work were obtained from the
Planetary Data System (PDS).  Support for I.A.~was provided by NASA
through the Einstein Fellowship Program, grant PF6-170148.  A.R.P.~was supported by the NASA Planetary Atmospheres program, grant \#NNX13AG55G.  D.D.~acknowledges support provided by NASA through Hubble Fellowship grant HSTHF2-51372.001-A awarded by the Space Telescope Science Institute.

\bibliography{nh_pasp}                

\begin{thebibliography}{}
\expandafter\ifx\csname natexlab\endcsname\relax\def\natexlab#1{#1}\fi
\providecommand{\url}[1]{\href{#1}{#1}}
\providecommand{\dodoi}[1]{doi:~\href{http://doi.org/#1}{\nolinkurl{#1}}}
\providecommand{\doeprint}[1]{\href{http://ascl.net/#1}{\nolinkurl{http://ascl.net/#1}}}
\providecommand{\doarXiv}[1]{\href{https://arxiv.org/abs/#1}{\nolinkurl{https://arxiv.org/abs/#1}}}

\bibitem[{{Abbott} {et~al.}(2016){Abbott}, {Abbott}, {Abbott}, {Abernathy},
  {Acernese}, {Ackley}, {Adams}, {Adams}, {Addesso}, {Adhikari}, \&
  et~al.}]{Abbott2016}
{Abbott}, B.~P., {Abbott}, R., {Abbott}, T.~D., {et~al.} 2016, \apjl, 833, L1,
  \dodoi{10.3847/2041-8205/833/1/L1}

\bibitem[{{Abbott} {et~al.}(2017{\natexlab{a}}){Abbott}, {Abbott}, {Abbott},
  {Acernese}, {Ackley}, {Adams}, {Adams}, {Addesso}, {Adhikari}, {Adya}, \&
  et~al.}]{LIGO170817}
---. 2017{\natexlab{a}}, Physical Review Letters, 119, 161101,
  \dodoi{10.1103/PhysRevLett.119.161101}

\bibitem[{{Abbott} {et~al.}(2017{\natexlab{b}}){Abbott}, {Abbott}, {Abbott},
  {Acernese}, {Ackley}, {Adams}, {Adams}, {Addesso}, {Adhikari}, {Adya}, \&
  et~al.}]{LIGOmma}
---. 2017{\natexlab{b}}, \apjl, 848, L12, \dodoi{10.3847/2041-8213/aa91c9}

\bibitem[{{Abbott} {et~al.}(2017{\natexlab{c}}){Abbott}, {Abbott}, {Abbott},
  {Acernese}, {Ackley}, {Adams}, {Adams}, {Addesso}, {Adhikari}, {Adya}, \&
  et~al.}]{LIGOH0}
---. 2017{\natexlab{c}}, \nat, 551, 85, \dodoi{10.1038/nature24471}

\bibitem[{{Akshaya} {et~al.}(2018){Akshaya}, {Murthy}, {Ravichandran}, {Henry},
  \& {Overduin}}]{Akshaya2018}
{Akshaya}, M.~S., {Murthy}, J., {Ravichandran}, S., {Henry}, R.~C., \&
  {Overduin}, J. 2018, \apj, 858, 101, \dodoi{10.3847/1538-4357/aabcb9}

\bibitem[{Altobelli {et~al.}(2007)Altobelli, Dikarev, Kempf, Srama, Helfert,
  Moragas-Klostermeyer, Roy, \& Gr{\"u}n}]{Altobelli_2007}
Altobelli, N., Dikarev, V., Kempf, S., {et~al.} 2007, J. Geophys. Res., 112

\bibitem[{{Arcavi}(2018)}]{Arcavi2018}
{Arcavi}, I. 2018, \apjl, 855, L23, \dodoi{10.3847/2041-8213/aab267}

\bibitem[{{Arcavi} {et~al.}(2017{\natexlab{a}}){Arcavi}, {McCully},
  {Hosseinzadeh}, {Howell}, {Vasylyev}, {Poznanski}, {Zaltzman}, {Maoz},
  {Singer}, {Valenti}, {Kasen}, {Barnes}, {Piran}, \& {Fong}}]{Arcavi_Strategy}
{Arcavi}, I., {McCully}, C., {Hosseinzadeh}, G., {et~al.} 2017{\natexlab{a}},
  \apjl, 848, L33, \dodoi{10.3847/2041-8213/aa910f}

\bibitem[{{Arcavi} {et~al.}(2017{\natexlab{b}}){Arcavi}, {Hosseinzadeh},
  {Howell}, {McCully}, {Poznanski}, {Kasen}, {Barnes}, {Zaltzman}, {Vasylyev},
  {Maoz}, \& {Valenti}}]{Arcavi2017}
{Arcavi}, I., {Hosseinzadeh}, G., {Howell}, D.~A., {et~al.} 2017{\natexlab{b}},
  \nat, 551, 64, \dodoi{10.1038/nature24291}

\bibitem[{{Barclay} {et~al.}(2018){Barclay}, {Pepper}, \&
  {Quintana}}]{Barclay2018}
{Barclay}, T., {Pepper}, J., \& {Quintana}, E.~V. 2018, ArXiv e-prints.
\newblock \doarXiv{1804.05050}

\bibitem[{{Bernstein}(2007)}]{Bernstein2007}
{Bernstein}, R.~A. 2007, \apj, 666, 663, \dodoi{10.1086/519824}

\bibitem[{{Blanchard} {et~al.}(2017){Blanchard}, {Berger}, {Fong}, {Nicholl},
  {Leja}, {Conroy}, {Alexander}, {Margutti}, {Williams}, {Doctor}, {Chornock},
  {Villar}, {Cowperthwaite}, {Annis}, {Brout}, {Brown}, {Chen}, {Eftekhari},
  {Frieman}, {Holz}, {Metzger}, {Rest}, {Sako}, \&
  {Soares-Santos}}]{Blanchard2017}
{Blanchard}, P.~K., {Berger}, E., {Fong}, W., {et~al.} 2017, \apjl, 848, L22,
  \dodoi{10.3847/2041-8213/aa9055}

\bibitem[{Bock {et~al.}(2012)Bock, Beichman, Cooray, Reach, Chary, Werner, \&
  Zemcov}]{Bock2012}
Bock, J., Beichman, C., Cooray, A., {et~al.} 2012, {SPIE} Newsroom,
  \dodoi{10.1117/2.1201202.004144}

\bibitem[{{Bock} {et~al.}(2013){Bock}, {Sullivan}, {Arai}, {Battle}, {Cooray},
  {Hristov}, {Keating}, {Kim}, {Lam}, {Lee}, {Levenson}, {Mason}, {Matsumoto},
  {Matsuura}, {Mitchell-Wynne}, {Nam}, {Renbarger}, {Smidt}, {Suzuki},
  {Tsumura}, {Wada}, \& {Zemcov}}]{Bock2013}
{Bock}, J., {Sullivan}, I., {Arai}, T., {et~al.} 2013, \apjs, 207, 32,
  \dodoi{10.1088/0067-0049/207/2/32}

\bibitem[{{Bohlin}(2014)}]{Bohlin2014}
{Bohlin}, R.~C. 2014, \aj, 147, 127, \dodoi{10.1088/0004-6256/147/6/127}

\bibitem[{{Borucki} {et~al.}(2010){Borucki}, {Koch}, {Basri}, {Batalha},
  {Brown}, {Caldwell}, {Caldwell}, {Christensen-Dalsgaard}, {Cochran},
  {DeVore}, {Dunham}, {Dupree}, {Gautier}, {Geary}, {Gilliland}, {Gould},
  {Howell}, {Jenkins}, {Kondo}, {Latham}, {Marcy}, {Meibom}, {Kjeldsen},
  {Lissauer}, {Monet}, {Morrison}, {Sasselov}, {Tarter}, {Boss}, {Brownlee},
  {Owen}, {Buzasi}, {Charbonneau}, {Doyle}, {Fortney}, {Ford}, {Holman},
  {Seager}, {Steffen}, {Welsh}, {Rowe}, {Anderson}, {Buchhave}, {Ciardi},
  {Walkowicz}, {Sherry}, {Horch}, {Isaacson}, {Everett}, {Fischer}, {Torres},
  {Johnson}, {Endl}, {MacQueen}, {Bryson}, {Dotson}, {Haas}, {Kolodziejczak},
  {Van Cleve}, {Chandrasekaran}, {Twicken}, {Quintana}, {Clarke}, {Allen},
  {Li}, {Wu}, {Tenenbaum}, {Verner}, {Bruhweiler}, {Barnes}, \&
  {Prsa}}]{Borucki2010}
{Borucki}, W.~J., {Koch}, D., {Basri}, G., {et~al.} 2010, Science, 327, 977,
  \dodoi{10.1126/science.1185402}

\bibitem[{{Bose} {et~al.}(2018){Bose}, {Dong}, {Pastorello}, {Filippenko},
  {Kochanek}, {Mauerhan}, {Romero-Ca{\~n}izales}, {Brink}, {Chen}, {Prieto},
  {Post}, {Ashall}, {Grupe}, {Tomasella}, {Benetti}, {Shappee}, {Stanek},
  {Cai}, {Falco}, {Lundqvist}, {Mattila}, {Mutel}, {Ochner}, {Pooley},
  {Stritzinger}, {Villanueva}, {Zheng}, {Beswick}, {Brown}, {Cappellaro},
  {Davis}, {Fraser}, {de Jaeger}, {Elias-Rosa}, {Gall}, {Gaudi}, {Herczeg},
  {Hestenes}, {Holoien}, {Hosseinzadeh}, {Hsiao}, {Hu}, {Jaejin}, {Jeffers},
  {Koff}, {Kumar}, {Kurtenkov}, {Lau}, {Prentice}, {Reynolds}, {Rudy},
  {Shahbandeh}, {Somero}, {Stassun}, {Thompson}, {Valenti}, {Woo}, \&
  {Yunus}}]{Bose2018}
{Bose}, S., {Dong}, S., {Pastorello}, A., {et~al.} 2018, \apj, 853, 57,
  \dodoi{10.3847/1538-4357/aaa298}

\bibitem[{{Bowyer}(1991)}]{Bowyer1991}
{Bowyer}, S. 1991, \araa, 29, 59, \dodoi{10.1146/annurev.aa.29.090191.000423}

\bibitem[{{Broadfoot} {et~al.}(1977){Broadfoot}, {Sandel}, {Shemansky},
  {Atreya}, {Donahue}, {Moos}, {Bertaux}, {Blamont}, {Ajello}, \&
  {Strobel}}]{Broadfoot1977}
{Broadfoot}, A.~L., {Sandel}, B.~R., {Shemansky}, D.~E., {et~al.} 1977, \ssr,
  21, 183, \dodoi{10.1007/BF00200850}

\bibitem[{{Buchalter} \& {Kamionkowski}(1997)}]{Buchalter1997}
{Buchalter}, A., \& {Kamionkowski}, M. 1997, \apj, 482, 782,
  \dodoi{10.1086/304163}

\bibitem[{Burns {et~al.}(1979)Burns, Lamy, \& Soter}]{Burns_1979}
Burns, J.~A., Lamy, P.~L., \& Soter, S. 1979, Icarus, 40, 1

\bibitem[{Bushman(2017)}]{Bushman2017}
Bushman, S.~S. 2017, Performance of the New Horizons Propulsion System through
  the Pluto Encounter (American Institute of Aeronautics and Astronautics).
\newblock \url{https://doi.org/10.2514/6.2017-4746}

\bibitem[{{Caldwell} {et~al.}(2010){Caldwell}, {Kolodziejczak}, {Van Cleve},
  {Jenkins}, {Gazis}, {Argabright}, {Bachtell}, {Dunham}, {Geary}, {Gilliland},
  {Chandrasekaran}, {Li}, {Tenenbaum}, {Wu}, {Borucki}, {Bryson}, {Dotson},
  {Haas}, \& {Koch}}]{Caldwell2010}
{Caldwell}, D.~A., {Kolodziejczak}, J.~J., {Van Cleve}, J.~E., {et~al.} 2010,
  \apjl, 713, L92, \dodoi{10.1088/2041-8205/713/2/L92}

\bibitem[{{Cambr{\'e}sy} {et~al.}(2001){Cambr{\'e}sy}, {Reach}, {Beichman}, \&
  {Jarrett}}]{Cambresy2001}
{Cambr{\'e}sy}, L., {Reach}, W.~T., {Beichman}, C.~A., \& {Jarrett}, T.~H.
  2001, \apj, 555, 563, \dodoi{10.1086/321470}

\bibitem[{{Carr} {et~al.}(2016){Carr}, {K{\"u}hnel}, \& {Sandstad}}]{Carr2016}
{Carr}, B., {K{\"u}hnel}, F., \& {Sandstad}, M. 2016, \prd, 94, 083504,
  \dodoi{10.1103/PhysRevD.94.083504}

\bibitem[{{Chary} \& {Pope}(2010)}]{Chary_2010}
{Chary}, R.-R., \& {Pope}, A. 2010, ArXiv e-prints.
\newblock \doarXiv{1003.1731}

\bibitem[{{Cheng} {et~al.}(2010){Cheng}, {Conard}, {Weaver}, {Morgan}, \&
  {Noble}}]{Cheng2010}
{Cheng}, A.~F., {Conard}, S.~J., {Weaver}, H.~A., {Morgan}, F., \& {Noble}, M.
  2010, in \procspie, Vol. 7731, Space Telescopes and Instrumentation 2010:
  Optical, Infrared, and Millimeter Wave, 77311A

\bibitem[{{Cheng} {et~al.}(2008){Cheng}, {Weaver}, {Conard}, {Morgan},
  {Barnouin-Jha}, {Boldt}, {Cooper}, {Darlington}, {Grey}, {Hayes},
  {Kosakowski}, {Magee}, {Rossano}, {Sampath}, {Schlemm}, \&
  {Taylor}}]{Cheng2008}
{Cheng}, A.~F., {Weaver}, H.~A., {Conard}, S.~J., {et~al.} 2008, \ssr, 140,
  189, \dodoi{10.1007/s11214-007-9271-6}

\bibitem[{{Christiansen} {et~al.}(2012){Christiansen}, {Jenkins}, {Caldwell},
  {Burke}, {Tenenbaum}, {Seader}, {Thompson}, {Barclay}, {Clarke}, {Li},
  {Smith}, {Stumpe}, {Twicken}, \& {Van Cleve}}]{Christiansen2012}
{Christiansen}, J.~L., {Jenkins}, J.~M., {Caldwell}, D.~A., {et~al.} 2012,
  \pasp, 124, 1279, \dodoi{10.1086/668847}

\bibitem[{Christon {et~al.}(2015)Christon, Hamilton, Plane, Mitchell,
  {DiFabio}, \& Krimigis}]{Christon_2015}
Christon, S.~P., Hamilton, D.~C., Plane, J. M.~C., {et~al.} 2015, J. Geophys.
  Res.: Space Physics, 120, 2720

\bibitem[{{Conard} {et~al.}(2017){Conard}, {Weaver}, {N{\'u}{\~n}ez}, {Taylor},
  {Hayes}, {Cheng}, \& {Rodgers}}]{Conard2017}
{Conard}, S.~J., {Weaver}, H.~A., {N{\'u}{\~n}ez}, J.~I., {et~al.} 2017, in
  Society of Photo-Optical Instrumentation Engineers (SPIE) Conference Series,
  Vol. 10401, Society of Photo-Optical Instrumentation Engineers (SPIE)
  Conference Series, 104010W

\bibitem[{{Conard} {et~al.}(2005){Conard}, {Azad}, {Boldt}, {Cheng}, {Cooper},
  {Darlington}, {Grey}, {Hayes}, {Hogue}, {Kosakowski}, {Magee}, {Morgan},
  {Rossano}, {Sampath}, {Schlemm}, \& {Weaver}}]{Conard2005}
{Conard}, S.~J., {Azad}, F., {Boldt}, J.~D., {et~al.} 2005, in {\procspie}
  (2005), Vol. 5906, {\it Astrobiology and Planetary Missions} (Eds. {Hoover},
  R.~B. et al.) 407-420, ed. R.~B. {Hoover}, G.~V. {Levin}, A.~Y. {Rozanov}, \&
  G.~R. {Gladstone}, 407--420

\bibitem[{{Cooray}(2016)}]{Cooray2016}
{Cooray}, A. 2016, Royal Society Open Science, 3, 150555,
  \dodoi{10.1098/rsos.150555}

\bibitem[{{Cooray} {et~al.}(2004){Cooray}, {Bock}, {Keatin}, {Lange}, \&
  {Matsumoto}}]{Cooray2004}
{Cooray}, A., {Bock}, J.~J., {Keatin}, B., {Lange}, A.~E., \& {Matsumoto}, T.
  2004, \apj, 606, 611, \dodoi{10.1086/383137}

\bibitem[{{Cooray} {et~al.}(2009){Cooray}, {Amblard}, {Beichman}, {Benford},
  {Bernstein}, {Bock}, {Brodwin}, {Bromm}, {Cen}, {Chary}, {Devlin}, {Dolch},
  {Dole}, {Dwek}, {Elbaz}, {' Fall}, {Fazio}, {Ferguson}, {Furlanetto},
  {Gardner}, {Giavalisco}, {Gilmore}, {Gnedin}, {Gonzalez}, {Haiman},
  {Kelsall}, {Komatsu}, {Lagache}, {Levenson}, {Loeb}, {Badau}, {Mather},
  {Matsumoto}, {Mattila}, {Moseley}, {Moustakas}, {Oh}, {Petro}, {Primack},
  {Reach}, {Renbarger}, {Shapiro}, {Stern}, {Sullivan}, {Venkatesan}, {Werner},
  {Windhorst}, {Wright}, \& {Zemcov}}]{Cooray2009}
{Cooray}, A., {Amblard}, A., {Beichman}, C., {et~al.} 2009, in astro2010: The
  Astronomy and Astrophysics Decadal Survey

\bibitem[{Cuzzi \& Estrada(1998)}]{Cuzzi_1998}
Cuzzi, J.~N., \& Estrada, P.~R. 1998, Icarus, 132, 1

\bibitem[{{Dixon} {et~al.}(2006){Dixon}, {Sankrit}, \& {Otte}}]{Dixon2006}
{Dixon}, W.~V.~D., {Sankrit}, R., \& {Otte}, B. 2006, \apj, 647, 328,
  \dodoi{10.1086/505168}

\bibitem[{{Dong} {et~al.}(2007){Dong}, {Udalski}, {Gould}, {Reach}, {Christie},
  {Boden}, {Bennett}, {Fazio}, {Griest}, {Szyma{\'n}ski}, {Kubiak},
  {Soszy{\'n}ski}, {Pietrzy{\'n}ski}, {Szewczyk}, {Wyrzykowski}, {Ulaczyk},
  {Wieckowski}, {Paczy{\'n}ski}, {DePoy}, {Pogge}, {Preston}, {Thompson}, \&
  {Patten}}]{Dong2007}
{Dong}, S., {Udalski}, A., {Gould}, A., {et~al.} 2007, \apj, 664, 862,
  \dodoi{10.1086/518536}

\bibitem[{{Drout} {et~al.}(2017){Drout}, {Piro}, {Shappee}, {Kilpatrick},
  {Simon}, {Contreras}, {Coulter}, {Foley}, {Siebert}, {Morrell}, {Boutsia},
  {Di Mille}, {Holoien}, {Kasen}, {Kollmeier}, {Madore}, {Monson},
  {Murguia-Berthier}, {Pan}, {Prochaska}, {Ramirez-Ruiz}, {Rest}, {Adams},
  {Alatalo}, {Ba{\~n}ados}, {Baughman}, {Beers}, {Bernstein}, {Bitsakis},
  {Campillay}, {Hansen}, {Higgs}, {Ji}, {Maravelias}, {Marshall}, {Moni Bidin},
  {Prieto}, {Rasmussen}, {Rojas-Bravo}, {Strom}, {Ulloa},
  {Vargas-Gonz{\'a}lez}, {Wan}, \& {Whitten}}]{Drout2017}
{Drout}, M.~R., {Piro}, A.~L., {Shappee}, B.~J., {et~al.} 2017, ArXiv e-prints.
\newblock \doarXiv{1710.05443}

\bibitem[{Durisen {et~al.}(1989)Durisen, Cramer, Murphy, Cuzzi, Mullikin, \&
  Cederbloom}]{Durisen_1989}
Durisen, R.~H., Cramer, N.~L., Murphy, B.~W., {et~al.} 1989, Icarus, 80, 136

\bibitem[{{Edelstein} {et~al.}(2000){Edelstein}, {Bowyer}, \&
  {Lampton}}]{Edelstein2000}
{Edelstein}, J., {Bowyer}, S., \& {Lampton}, M. 2000, \apj, 539, 187,
  \dodoi{10.1086/309192}

\bibitem[{{Elbert} {et~al.}(2018){Elbert}, {Bullock}, \&
  {Kaplinghat}}]{Elbert2018}
{Elbert}, O.~D., {Bullock}, J.~S., \& {Kaplinghat}, M. 2018, \mnras, 473, 1186,
  \dodoi{10.1093/mnras/stx1959}

\bibitem[{Estrada {et~al.}(2015)Estrada, Durisen, Cuzzi, \&
  Morgan}]{Estrada_2015}
Estrada, P.~R., Durisen, R.~H., Cuzzi, J.~N., \& Morgan, D.~A. 2015, Icarus,
  252, 415

\bibitem[{Feuchtgruber {et~al.}(1997)Feuchtgruber, Lellouch, de~Graauw,
  B{\'e}zard, Encrenaz, \& Griffin}]{Feuchtgruber_1997}
Feuchtgruber, H., Lellouch, E., de~Graauw, T., {et~al.} 1997, Nature, 389, 159

\bibitem[{{Fountain} {et~al.}(2008){Fountain}, {Kusnierkiewicz}, {Hersman},
  {Herder}, {Coughlin}, {Gibson}, {Clancy}, {Deboy}, {Hill}, {Kinnison},
  {Mehoke}, {Ottman}, {Rogers}, {Stern}, {Stratton}, {Vernon}, \&
  {Williams}}]{Fountain2008}
{Fountain}, G.~H., {Kusnierkiewicz}, D.~Y., {Hersman}, C.~B., {et~al.} 2008,
  \ssr, 140, 23, \dodoi{10.1007/s11214-008-9374-8}

\bibitem[{Frankland {et~al.}(2016)Frankland, James, {Carrillo-S\'anchez},
  Mangan, Willacy, Poppe, \& Plane}]{Frankland_2016}
Frankland, V.~L., James, A.~D., {Carrillo-S\'anchez}, J.~D., {et~al.} 2016,
  Icarus, 278, 88

\bibitem[{{Gehrels} {et~al.}(2016){Gehrels}, {Cannizzo}, {Kanner}, {Kasliwal},
  {Nissanke}, \& {Singer}}]{Gehrels2016}
{Gehrels}, N., {Cannizzo}, J.~K., {Kanner}, J., {et~al.} 2016, \apj, 820, 136,
  \dodoi{10.3847/0004-637X/820/2/136}

\bibitem[{{Gilliland} {et~al.}(2011){Gilliland}, {Chaplin}, {Dunham},
  {Argabright}, {Borucki}, {Basri}, {Bryson}, {Buzasi}, {Caldwell}, {Elsworth},
  {Jenkins}, {Koch}, {Kolodziejczak}, {Miglio}, {van Cleve}, {Walkowicz}, \&
  {Welsh}}]{Gilliland2011}
{Gilliland}, R.~L., {Chaplin}, W.~J., {Dunham}, E.~W., {et~al.} 2011, \apjs,
  197, 6, \dodoi{10.1088/0067-0049/197/1/6}

\bibitem[{{Gladstone} {et~al.}(2013){Gladstone}, {Stern}, \&
  {Pryor}}]{Gladstone2013}
{Gladstone}, G.~R., {Stern}, S.~A., \& {Pryor}, W.~R. 2013, {New Horizons
  Cruise Observations of Lyman-{$\alpha$} Emissions from the Interplanetary
  Medium}, ed. E.~{Qu{\'e}merais}, M.~{Snow}, \& R.-M. {Bonnet} (Springer, New
  York), 177

\bibitem[{{Gong} {et~al.}(2016){Gong}, {Cooray}, {Mitchell-Wynne}, {Chen},
  {Zemcov}, \& {Smidt}}]{Gong2016}
{Gong}, Y., {Cooray}, A., {Mitchell-Wynne}, K., {et~al.} 2016, \apj, 825, 104,
  \dodoi{10.3847/0004-637X/825/2/104}

\bibitem[{{Gordon} {et~al.}(1998){Gordon}, {Witt}, \& {Friedmann}}]{Gordon1998}
{Gordon}, K.~D., {Witt}, A.~N., \& {Friedmann}, B.~C. 1998, \apj, 498, 522,
  \dodoi{10.1086/305571}

\bibitem[{{Gorjian} {et~al.}(2000){Gorjian}, {Wright}, \&
  {Chary}}]{Gorjian2000}
{Gorjian}, V., {Wright}, E.~L., \& {Chary}, R.~R. 2000, \apj, 536, 550,
  \dodoi{10.1086/308974}

\bibitem[{{Gould}(1992)}]{Gould1992}
{Gould}, A. 1992, \apj, 392, 442, \dodoi{10.1086/171443}

\bibitem[{{Gould} {et~al.}(2009){Gould}, {Udalski}, {Monard}, {Horne}, {Dong},
  {Miyake}, {Sahu}, {Bennett}, {Wyrzykowski}, {Soszy{\'n}ski}, {Szyma{\'n}ski},
  {Kubiak}, {Pietrzy{\'n}ski}, {Szewczyk}, {Ulaczyk}, {OGLE Collaboration},
  {Allen}, {Christie}, {DePoy}, {Gaudi}, {Han}, {Lee}, {McCormick}, {Natusch},
  {Park}, {Pogge}, {{$\mu$}FUN Collaboration}, {Allan}, {Bode}, {Bramich},
  {Burgdorf}, {Dominik}, {Fraser}, {Kerins}, {Mottram}, {Snodgrass}, {Steele},
  {Street}, {Tsapras}, {RoboNet Collaboration}, {Abe}, {Bond}, {Botzler},
  {Fukui}, {Furusawa}, {Hearnshaw}, {Itow}, {Kamiya}, {Kilmartin}, {Korpela},
  {Lin}, {Ling}, {Masuda}, {Matsubara}, {Muraki}, {Nagaya}, {Ohnishi},
  {Okumura}, {Perrott}, {Rattenbury}, {Saito}, {Sako}, {Skuljan}, {Sullivan},
  {Sumi}, {Sweatman}, {Tristram}, {Yock}, {MOA Collaboration}, {Albrow},
  {Beaulieu}, {Coutures}, {Calitz}, {Caldwell}, {Fouque}, {Martin}, {Williams},
  \& {PLANET Collaboration}}]{Gould2009}
{Gould}, A., {Udalski}, A., {Monard}, B., {et~al.} 2009, \apjl, 698, L147,
  \dodoi{10.1088/0004-637X/698/2/L147}

\bibitem[{Grigorieva {et~al.}(2007)Grigorieva, Th{\'e}bault, Artymowicz, \&
  Brandeker}]{Grigorieva_2007}
Grigorieva, A., Th{\'e}bault, P., Artymowicz, P., \& Brandeker, A. 2007,
  Astron. Astrophys., 475, 755

\bibitem[{{Gr{\"u}n} {et~al.}(1995{\natexlab{a}}){Gr{\"u}n}, {Baguhl},
  {Divine}, {Fechtig}, {Hamilton}, {Hanner}, {Kissel}, {Lindblad}, {Linkert},
  {Linkert}, {Mann}, {McDonnell}, {Morfill}, {Polanskey}, {Riemann}, {Schwehm},
  {Siddique}, {Staubach}, \& {Zook}}]{Grun_Uly_1995}
{Gr{\"u}n}, E., {Baguhl}, M., {Divine}, N., {et~al.} 1995{\natexlab{a}},
  \planss, 43, 971, \dodoi{10.1016/0032-0633(94)00233-H}

\bibitem[{{Gr{\"u}n} {et~al.}(1995{\natexlab{b}}){Gr{\"u}n}, {Baguhl},
  {Divine}, {Fechtig}, {Hamilton}, {Hanner}, {Kissel}, {Lindblad}, {Linkert},
  {Linkert}, {Mann}, {McDonnell}, {Morfill}, {Polanskey}, {Riemann}, {Schwehm},
  {Siddique}, {Staubach}, \& {Zook}}]{Grun_Gal_1995}
---. 1995{\natexlab{b}}, \planss, 43, 953, \dodoi{10.1016/0032-0633(94)00234-I}

\bibitem[{Gurnett {et~al.}(1997)Gurnett, Ansher, Kurth, \&
  Granroth}]{Gurnett_1997}
Gurnett, D.~A., Ansher, J.~A., Kurth, W.~S., \& Granroth, L.~J. 1997, Geophys.
  Res. Lett., 24, 3125

\bibitem[{Gustafson(1994)}]{Gustafson_1994}
Gustafson, B. A.~S. 1994, Annu. Rev. Earth Planet. Sci., 22, 553

\bibitem[{Hahn {et~al.}(2002)Hahn, Zook, Cooper, \& Sunkara}]{Hahn_2002}
Hahn, J.~M., Zook, H.~A., Cooper, B., \& Sunkara, B. 2002, Icarus, 158, 360

\bibitem[{{Hamden} {et~al.}(2013){Hamden}, {Schiminovich}, \&
  {Seibert}}]{Hamden2013}
{Hamden}, E.~T., {Schiminovich}, D., \& {Seibert}, M. 2013, \apj, 779, 180,
  \dodoi{10.1088/0004-637X/779/2/180}

\bibitem[{Han {et~al.}(2011)Han, Poppe, Piquette, Gr{\"u}n, \&
  Hor{\'a}nyi}]{Han_2011}
Han, D., Poppe, A.~R., Piquette, M., Gr{\"u}n, E., \& Hor{\'a}nyi, M. 2011,
  Geophys. Res. Lett., 38, L24102

\bibitem[{{Hanner} {et~al.}(1974){Hanner}, {Weinberg}, {DeShields}, {Green}, \&
  {Toller}}]{Hanner1974}
{Hanner}, M.~S., {Weinberg}, J.~L., {DeShields}, II, L.~M., {Green}, B.~A., \&
  {Toller}, G.~N. 1974, \jgr, 79, 3671, \dodoi{10.1029/JA079i025p03671}

\bibitem[{{Hauser} \& {Dwek}(2001)}]{Hauser2001}
{Hauser}, M.~G., \& {Dwek}, E. 2001, \araa, 39, 249,
  \dodoi{10.1146/annurev.astro.39.1.249}

\bibitem[{Hedman {et~al.}(2009)Hedman, Murray, Cooper, Tiscareno, Beurle,
  Evans, \& Burns}]{Hedman_2009}
Hedman, M.~M., Murray, C.~D., Cooper, N.~J., {et~al.} 2009, Icarus, 199, 378

\bibitem[{{Heller}(2014)}]{Heller2014}
{Heller}, R. 2014, \apj, 787, 14, \dodoi{10.1088/0004-637X/787/1/14}

\bibitem[{{Heller}(2017)}]{Heller2017}
---. 2017, ArXiv e-prints.
\newblock \doarXiv{1701.04706}

\bibitem[{{Heller} {et~al.}(2016{\natexlab{a}}){Heller}, {Hippke}, \&
  {Jackson}}]{Heller2016a}
{Heller}, R., {Hippke}, M., \& {Jackson}, B. 2016{\natexlab{a}}, \apj, 820, 88,
  \dodoi{10.3847/0004-637X/820/2/88}

\bibitem[{{Heller} {et~al.}(2016{\natexlab{b}}){Heller}, {Hippke}, {Placek},
  {Angerhausen}, \& {Agol}}]{Heller2016b}
{Heller}, R., {Hippke}, M., {Placek}, B., {Angerhausen}, D., \& {Agol}, E.
  2016{\natexlab{b}}, \aap, 591, A67, \dodoi{10.1051/0004-6361/201628573}

\bibitem[{{Henry} {et~al.}(2018){Henry}, {Murthy}, \& {Overduin}}]{RCH2018}
{Henry}, R., {Murthy}, J., \& {Overduin}, J. 2018, ArXiv e-prints,
  arXiv:1805.09658.
\newblock \doarXiv{1805.09658}

\bibitem[{{Henry}(1991)}]{Henry1991}
{Henry}, R.~C. 1991, \araa, 29, 89, \dodoi{10.1146/annurev.aa.29.090191.000513}

\bibitem[{{Henry} {et~al.}(2015){Henry}, {Murthy}, {Overduin}, \&
  {Tyler}}]{Henry2015}
{Henry}, R.~C., {Murthy}, J., {Overduin}, J., \& {Tyler}, J. 2015, \apj, 798,
  14, \dodoi{10.1088/0004-637X/798/1/14}

\bibitem[{{H.E.S.S.~Collaboration} {et~al.}(2013){H.E.S.S.~Collaboration},
  {Abramowski}, {Acero}, {Aharonian}, {Akhperjanian}, {Anton}, {Balenderan},
  {Balzer}, {Barnacka}, {Becherini}, {Becker Tjus}, {Bernl{\"o}hr}, {Birsin},
  {Biteau}, {Bochow}, {Boisson}, {Bolmont}, {Bordas}, {Brucker}, {Brun},
  {Brun}, {Bulik}, {Carrigan}, {Casanova}, {Cerruti}, {Chadwick},
  {Charbonnier}, {Chaves}, {Cheesebrough}, {Cologna}, {Conrad}, {Couturier},
  {Dalton}, {Daniel}, {Davids}, {Degrange}, {Deil}, {deWilt}, {Dickinson},
  {Djannati-Ata{\"i}}, {Domainko}, {O'C.~Drury}, {Dubus}, {Dutson}, {Dyks},
  {Dyrda}, {Egberts}, {Eger}, {Espigat}, {Fallon}, {Farnier}, {Fegan},
  {Feinstein}, {Fernandes}, {Fernandez}, {Fiasson}, {Fontaine}, {F{\"o}rster},
  {F{\"u}{\ss}ling}, {Gajdus}, {Gallant}, {Garrigoux}, {Gast}, {Giebels},
  {Glicenstein}, {Gl{\"u}ck}, {G{\"o}ring}, {Grondin}, {H{\"a}ffner}, {Hague},
  {Hahn}, {Hampf}, {Harris}, {Heinz}, {Heinzelmann}, {Henri}, {Hermann},
  {Hillert}, {Hinton}, {Hofmann}, {Hofverberg}, {Holler}, {Horns},
  {Jacholkowska}, {Jahn}, {Jamrozy}, {Jung}, {Kastendieck}, {Katarzy{\'n}ski},
  {Katz}, {Kaufmann}, {Kh{\'e}lifi}, {Klochkov}, {Klu{\'z}niak}, {Kneiske},
  {Komin}, {Kosack}, {Kossakowski}, {Krayzel}, {Laffon}, {Lamanna}, {Lenain},
  {Lennarz}, {Lohse}, {Lopatin}, {Lu}, {Marandon}, {Marcowith}, {Masbou},
  {Maurin}, {Maxted}, {Mayer}, {McComb}, {Medina}, {M{\'e}hault}, {Menzler},
  {Moderski}, {Mohamed}, {Moulin}, {Naumann}, {Naumann-Godo}, {de Naurois},
  {Nedbal}, {Nguyen}, {Niemiec}, {Nolan}, {Ohm}, {de O{\~n}a Wilhelmi},
  {Opitz}, {Ostrowski}, {Oya}, {Panter}, {Parsons}, {Paz Arribas}, {Pekeur},
  {Pelletier}, {Perez}, {Petrucci}, {Peyaud}, {Pita}, {P{\"u}hlhofer}, {Punch},
  {Quirrenbach}, {Raue}, {Reimer}, {Reimer}, {Renaud}, {de los Reyes},
  {Rieger}, {Ripken}, {Rob}, {Rosier-Lees}, {Rowell}, {Rudak}, {Rulten},
  {Sahakian}, {Sanchez}, {Santangelo}, {Schlickeiser}, {Schulz}, {Schwanke},
  {Schwarzburg}, {Schwemmer}, {Sheidaei}, {Skilton}, {Sol}, {Spengler},
  {Stawarz}, {Steenkamp}, {Stegmann}, {Stinzing}, {Stycz}, {Sushch}, {Szostek},
  {Tavernet}, {Terrier}, {Tluczykont}, {Valerius}, {van Eldik}, {Vasileiadis},
  {Venter}, {Viana}, {Vincent}, {V{\"o}lk}, {Volpe}, {Vorobiov}, {Vorster},
  {Wagner}, {Ward}, {White}, {Wierzcholska}, {Wouters}, {Zacharias}, {Zajczyk},
  {Zdziarski}, {Zech}, \& {Zechlin}}]{HESS2013}
{H.E.S.S.~Collaboration}, {Abramowski}, A., {Acero}, F., {et~al.} 2013, \aap,
  550, A4, \dodoi{10.1051/0004-6361/201220355}

\bibitem[{{Holberg}(1986)}]{Holberg1986}
{Holberg}, J.~B. 1986, \apj, 311, 969, \dodoi{10.1086/164834}

\bibitem[{{Holberg} \& {Barber}(1985)}]{Holberg1985}
{Holberg}, J.~B., \& {Barber}, H.~B. 1985, \apj, 292, 16,
  \dodoi{10.1086/163128}

\bibitem[{{Hor{\'a}nyi} {et~al.}(2008){Hor{\'a}nyi}, {Hoxie}, {James}, {Poppe},
  {Bryant}, {Grogan}, {Lamprecht}, {Mack}, {Bagenal}, {Batiste}, {Bunch},
  {Chanthawanich}, {Christensen}, {Colgan}, {Dunn}, {Drake}, {Fernandez},
  {Finley}, {Holland}, {Jenkins}, {Krauss}, {Krauss}, {Krauss}, {Lankton},
  {Mitchell}, {Neeland}, {Reese}, {Rash}, {Tate}, {Vaudrin}, \&
  {Westfall}}]{Horanyi2008}
{Hor{\'a}nyi}, M., {Hoxie}, V., {James}, D., {et~al.} 2008, \ssr, 140, 387,
  \dodoi{10.1007/s11214-007-9250-y}

\bibitem[{{Howell}(2017)}]{Howell2017}
{Howell}, D.~A. 2017, {Superluminous Supernovae}, ed. A.~W. {Alsabti} \&
  P.~{Murdin} (Springer International Publishing AG, New York), 431

\bibitem[{{Howett} {et~al.}(2017){Howett}, {Parker}, {Olkin}, {Reuter},
  {Ennico}, {Grundy}, {Graps}, {Harrison}, {Throop}, {Buie}, {Lovering},
  {Porter}, {Weaver}, {Young}, {Stern}, {Beyer}, {Binzel}, {Buratti}, {Cheng},
  {Cook}, {Cruikshank}, {Dalle Ore}, {Earle}, {Jennings}, {Linscott},
  {Lunsford}, {Parker}, {Phillippe}, {Protopapa}, {Quirico}, {Schenk},
  {Schmitt}, {Singer}, {Spencer}, {Stansberry}, {Tsang}, {Weigle}, \&
  {Verbiscer}}]{Howett2017}
{Howett}, C.~J.~A., {Parker}, A.~H., {Olkin}, C.~B., {et~al.} 2017, \icarus,
  287, 140, \dodoi{10.1016/j.icarus.2016.12.007}

\bibitem[{{Huang} {et~al.}(2018){Huang}, {Shporer}, {Dragomir}, {Fausnaugh},
  {Levine}, {Morgan}, {Nguyen}, {Ricker}, {Wall}, {Woods}, \&
  {Vanderspek}}]{Huang2018}
{Huang}, C.~X., {Shporer}, A., {Dragomir}, D., {et~al.} 2018, ArXiv e-prints.
\newblock \doarXiv{1807.11129}

\bibitem[{Humes(1980)}]{Humes_1980}
Humes, D.~H. 1980, J. Geophys. Res., 85, 5841

\bibitem[{{Hunter}(2007)}]{Hunter2007}
{Hunter}, J.~D. 2007, Computing In Science \& Engineering, 9, 90,
  \dodoi{10.1109/MCSE.2007.55}

\bibitem[{Ipatov {et~al.}(2008)Ipatov, Kutyrev, Madsen, Mather, Moseley, \&
  Reynolds}]{Ipatov_2008}
Ipatov, S.~I., Kutyrev, A.~S., Madsen, G.~J., {et~al.} 2008, Icarus, 194, 769

\bibitem[{J{\"a}ger {et~al.}(2003)J{\"a}ger, Dorschner, Mutschke, Posch, \&
  Henning}]{Jager_2003}
J{\"a}ger, C., Dorschner, J., Mutschke, H., Posch, T., \& Henning, T. 2003,
  Astron. Astrophys., 408, 193

\bibitem[{{Jenkins} {et~al.}(2010){Jenkins}, {Caldwell}, {Chandrasekaran},
  {Twicken}, {Bryson}, {Quintana}, {Clarke}, {Li}, {Allen}, {Tenenbaum}, {Wu},
  {Klaus}, {Van Cleve}, {Dotson}, {Haas}, {Gilliland}, {Koch}, \&
  {Borucki}}]{Jenkins2010}
{Jenkins}, J.~M., {Caldwell}, D.~A., {Chandrasekaran}, H., {et~al.} 2010,
  \apjl, 713, L120, \dodoi{10.1088/2041-8205/713/2/L120}

\bibitem[{Jenniskens(1993)}]{Jenniskens_1993}
Jenniskens, P. 1993, Astron. Astrophys., 274, 653

\bibitem[{{Jones} {et~al.}(2001){Jones}, {Oliphant}, {Peterson},
  {et~al.}}]{SciPy}
{Jones}, E., {Oliphant}, T., {Peterson}, P., {et~al.} 2001, {SciPy}: Open
  source scientific tools for {Python}.
\newblock \url{http://www.scipy.org/}

\bibitem[{{Keenan} {et~al.}(2010){Keenan}, {Barger}, {Cowie}, \&
  {Wang}}]{Keenan2010}
{Keenan}, R.~C., {Barger}, A.~J., {Cowie}, L.~L., \& {Wang}, W.-H. 2010, \apj,
  723, 40, \dodoi{10.1088/0004-637X/723/1/40}

\bibitem[{Kelley {et~al.}(2013)Kelley, {Fern\'andez}, Licandro, Lisse, Reach,
  A'Hearn, Bauer, Campins, Fitzsimmons, Groussin, Lamy, Lowry, Meech,
  Pittichov\'a, Snodgrass, Toth, \& Weaver}]{Kelley_2013}
Kelley, M.~S., {Fern\'andez}, Y.~R., Licandro, J., {et~al.} 2013, Icarus, 225,
  475

\bibitem[{{Kipping} {et~al.}(2010){Kipping}, {Fossey}, {Campanella},
  {Schneider}, \& {Tinetti}}]{Kipping2010}
{Kipping}, D.~M., {Fossey}, S.~J., {Campanella}, G., {Schneider}, J., \&
  {Tinetti}, G. 2010, in Astronomical Society of the Pacific Conference Series,
  Vol. 430, Pathways Towards Habitable Planets, ed. V.~{Coud{\'e} du Foresto},
  D.~M. {Gelino}, \& I.~{Ribas}, 139

\bibitem[{{Leinert} {et~al.}(1998){Leinert}, {Bowyer}, {Haikala}, {Hanner},
  {Hauser}, {Levasseur-Regourd}, {Mann}, {Mattila}, {Reach}, {Schlosser},
  {Staude}, {Toller}, {Weiland}, {Weinberg}, \& {Witt}}]{Leinert1998}
{Leinert}, C., {Bowyer}, S., {Haikala}, L.~K., {et~al.} 1998, \aaps, 127, 1,
  \dodoi{10.1051/aas:1998105}

\bibitem[{{Levenson} {et~al.}(2007){Levenson}, {Wright}, \&
  {Johnson}}]{Levenson2007}
{Levenson}, L.~R., {Wright}, E.~L., \& {Johnson}, B.~D. 2007, \apj, 666, 34,
  \dodoi{10.1086/520112}

\bibitem[{Liou {et~al.}(1996)Liou, Zook, \& Dermott}]{Liou_1996}
Liou, J.-C., Zook, H.~A., \& Dermott, S.~F. 1996, Icarus, 124, 429

\bibitem[{Liou {et~al.}(1995)Liou, Zook, \& Jackson}]{Liou_1995}
Liou, J.-C., Zook, H.~A., \& Jackson, A.~A. 1995, Icarus, 116, 186

\bibitem[{{Madau} \& {Pozzetti}(2000)}]{Madau2000}
{Madau}, P., \& {Pozzetti}, L. 2000, \mnras, 312, L9,
  \dodoi{10.1046/j.1365-8711.2000.03268.x}

\bibitem[{{Mather} \& {Beichman}(1996)}]{Mather1996}
{Mather}, J.~C., \& {Beichman}, C.~A. 1996, in American Institute of Physics
  Conference Series, Vol. 348, American Institute of Physics Conference Series,
  ed. E.~{Dwek}, 271--277

\bibitem[{{Matsumoto} {et~al.}(2018){Matsumoto}, {Tsumura}, {Matsuoka}, \&
  {Pyo}}]{Matsumoto2018}
{Matsumoto}, T., {Tsumura}, K., {Matsuoka}, Y., \& {Pyo}, J. 2018, \aj, 156,
  86, \dodoi{10.3847/1538-3881/aad0f0}

\bibitem[{{Matsumoto} {et~al.}(2005){Matsumoto}, {Matsuura}, {Murakami},
  {Tanaka}, {Freund}, {Lim}, {Cohen}, {Kawada}, \& {Noda}}]{Matsumoto2005}
{Matsumoto}, T., {Matsuura}, S., {Murakami}, H., {et~al.} 2005, \apj, 626, 31,
  \dodoi{10.1086/429383}

\bibitem[{{Matsuoka} {et~al.}(2011){Matsuoka}, {Ienaka}, {Kawara}, \&
  {Oyabu}}]{Matsuoka2011}
{Matsuoka}, Y., {Ienaka}, N., {Kawara}, K., \& {Oyabu}, S. 2011, \apj, 736,
  119, \dodoi{10.1088/0004-637X/736/2/119}

\bibitem[{Matsuura {et~al.}(2014)Matsuura, Yano, Yonetoku, Funase, Mori,
  Shirasawa, \& Group}]{Matsuura2014}
Matsuura, S., Yano, H., Yonetoku, D., {et~al.} 2014, Transactions of the Japan
  Society for Aeronautical and Space Sciences, Aerospace Technology Japan, 12,
  {Tr\_1}

\bibitem[{{Matsuura} {et~al.}(2017){Matsuura}, {Arai}, {Bock}, {Cooray},
  {Korngut}, {Kim}, {Lee}, {Lee}, {Levenson}, {Matsumoto}, {Onishi},
  {Shirahata}, {Tsumura}, {Wada}, \& {Zemcov}}]{Matsuura2017}
{Matsuura}, S., {Arai}, T., {Bock}, J.~J., {et~al.} 2017, \apj, 839, 7,
  \dodoi{10.3847/1538-4357/aa6843}

\bibitem[{{Mattila}(2003)}]{Mattila2003}
{Mattila}, K. 2003, \apj, 591, 119, \dodoi{10.1086/375182}

\bibitem[{{Mattila}(2006)}]{Mattila2006}
---. 2006, \mnras, 372, 1253, \dodoi{10.1111/j.1365-2966.2006.10934.x}

\bibitem[{{Mattila} {et~al.}(2012){Mattila}, {Lehtinen}, {V{\"a}is{\"a}nen},
  {von Appen-Schnur}, \& {Leinert}}]{Mattila2012}
{Mattila}, K., {Lehtinen}, K., {V{\"a}is{\"a}nen}, P., {von Appen-Schnur}, G.,
  \& {Leinert}, C. 2012, in IAU Symposium, Vol. 284, {\it The Spectral Energy
  Distribution of Galaxies - SED 2011} (Eds. {Tuffs}, R.~J. and {Popescu},
  C.~C.) 429-436, ed. R.~J. {Tuffs} \& C.~C. {Popescu}, 429--436

\bibitem[{{Mattila} {et~al.}(2017){Mattila}, {V{\"a}is{\"a}nen}, {Lehtinen},
  {von Appen-Schnur}, \& {Leinert}}]{Mattila2017}
{Mattila}, K., {V{\"a}is{\"a}nen}, P., {Lehtinen}, K., {von Appen-Schnur}, G.,
  \& {Leinert}, C. 2017, \mnras, 470, 2152, \dodoi{10.1093/mnras/stx1296}

\bibitem[{{Metzger}(2017)}]{Metzger2017}
{Metzger}, B.~D. 2017, ArXiv e-prints.
\newblock \doarXiv{1710.05931}

\bibitem[{{Morgan} {et~al.}(2005){Morgan}, {Conard}, {Weaver}, {Barnouin-Jha},
  {Cheng}, {Taylor}, {Cooper}, {Barkhouser}, {Boucarut}, {Darlington}, {Grey},
  {Kuznetsov}, {Madison}, {Quijada}, {Sahnow}, \& {Stock}}]{Morgan2005}
{Morgan}, F., {Conard}, S.~J., {Weaver}, H.~A., {et~al.} 2005, in \procspie,
  Vol. 5906, {\it Astrobiology and Planetary Missions} (Eds. {Hoover}, R.~B. et
  al.) 421-432, ed. R.~B. {Hoover}, G.~V. {Levin}, A.~Y. {Rozanov}, \& G.~R.
  {Gladstone}, 421--432

\bibitem[{Moses(1992)}]{Moses_1992}
Moses, J.~I. 1992, Icarus, 99, 368

\bibitem[{Moses {et~al.}(2000)Moses, Lellouch, B{\'e}zard, Gladstone,
  Feuchtgruber, \& Allen}]{Moses_2000b}
Moses, J.~I., Lellouch, E., B{\'e}zard, B., {et~al.} 2000, Icarus, 145, 166

\bibitem[{{Moses} \& {Poppe}(2017)}]{Moses_2017}
{Moses}, J.~I., \& {Poppe}, A.~R. 2017, \icarus, 297, 33,
  \dodoi{10.1016/j.icarus.2017.06.002}

\bibitem[{{Mroz} {et~al.}(2017){Mroz}, {Ryu}, {Skowron}, {Udalski}, {Gould},
  {Szymanski}, {Soszynski}, {Poleski}, {Pietrukowicz}, {Kozlowski}, {Pawlak},
  {Ulaczyk}, {Albrow}, {Chung}, {Jung}, {Han}, {Hwang}, {Shin}, {Yee}, {Zhu},
  {Cha}, {Kim}, {Kim}, {Kim}, {Lee}, {Lee}, {Lee}, {Park}, \&
  {Pogge}}]{Mroz2017}
{Mroz}, P., {Ryu}, Y.-H., {Skowron}, J., {et~al.} 2017, ArXiv e-prints.
\newblock \doarXiv{1712.01042}

\bibitem[{{Muraki} {et~al.}(2011){Muraki}, {Han}, {Bennett}, {Suzuki},
  {Monard}, {Street}, {Jorgensen}, {Kundurthy}, {Skowron}, {Becker}, {Albrow},
  {Fouqu{\'e}}, {Heyrovsk{\'y}}, {Barry}, {Beaulieu}, {Wellnitz}, {Bond},
  {Sumi}, {Dong}, {Gaudi}, {Bramich}, {Dominik}, {Abe}, {Botzler}, {Freeman},
  {Fukui}, {Furusawa}, {Hayashi}, {Hearnshaw}, {Hosaka}, {Itow}, {Kamiya},
  {Korpela}, {Kilmartin}, {Lin}, {Ling}, {Makita}, {Masuda}, {Matsubara},
  {Miyake}, {Nishimoto}, {Ohnishi}, {Perrott}, {Rattenbury}, {Saito},
  {Skuljan}, {Sullivan}, {Sweatman}, {Tristram}, {Wada}, {Yock}, {MOA
  Collaboration}, {Christie}, {DePoy}, {Gorbikov}, {Gould}, {Kaspi}, {Lee},
  {Mallia}, {Maoz}, {McCormick}, {Moorhouse}, {Natusch}, {Park}, {Pogge},
  {Polishook}, {Shporer}, {Thornley}, {Yee}, {{$\mu$}FUN Collaboration},
  {Allan}, {Browne}, {Horne}, {Kains}, {Snodgrass}, {Steele}, {Tsapras},
  {RoboNet Collaboration}, {Batista}, {Bennett}, {Brillant}, {Caldwell},
  {Cassan}, {Cole}, {Corrales}, {Coutures}, {Dieters}, {Dominis Prester},
  {Donatowicz}, {Greenhill}, {Kubas}, {Marquette}, {Martin}, {Menzies}, {Sahu},
  {Waldman}, {Williams}, {Zub}, {PLANET Collaboration}, {Bourhrous},
  {Matsuoka}, {Nagayama}, {Oi}, {Randriamanakoto}, {IRSF Observers}, {Bozza},
  {Burgdorf}, {Calchi Novati}, {Dreizler}, {Finet}, {Glitrup}, {Harps{\o}e},
  {Hinse}, {Hundertmark}, {Liebig}, {Maier}, {Mancini}, {Mathiasen}, {Rahvar},
  {Ricci}, {Scarpetta}, {Skottfelt}, {Surdej}, {Southworth}, {Wambsganss},
  {Zimmer}, {MiNDSTEp Consortium}, {Udalski}, {Poleski}, {Wyrzykowski},
  {Ulaczyk}, {Szyma{\'n}ski}, {Kubiak}, {Pietrzy{\'n}ski}, {Soszy{\'n}ski}, \&
  {OGLE Collaboration}}]{Muraki2011}
{Muraki}, Y., {Han}, C., {Bennett}, D.~P., {et~al.} 2011, \apj, 741, 22,
  \dodoi{10.1088/0004-637X/741/1/22}

\bibitem[{{Murthy}(2009)}]{Murthy2009}
{Murthy}, J. 2009, \apss, 320, 21, \dodoi{10.1007/s10509-008-9855-y}

\bibitem[{{Murthy}(2016)}]{Murthy2016}
---. 2016, \mnras, 459, 1710, \dodoi{10.1093/mnras/stw755}

\bibitem[{{Murthy} {et~al.}(1999){Murthy}, {Hall}, {Earl}, {Henry}, \&
  {Holberg}}]{Murthy1999}
{Murthy}, J., {Hall}, D., {Earl}, M., {Henry}, R.~C., \& {Holberg}, J.~B. 1999,
  \apj, 522, 904, \dodoi{10.1086/307652}

\bibitem[{{Murthy} {et~al.}(1991){Murthy}, {Henry}, \& {Holberg}}]{Murthy1991}
{Murthy}, J., {Henry}, R.~C., \& {Holberg}, J.~B. 1991, \apj, 383, 198,
  \dodoi{10.1086/170776}

\bibitem[{{Murthy} {et~al.}(2001){Murthy}, {Henry}, {Shelton}, \&
  {Holberg}}]{Murthy2001}
{Murthy}, J., {Henry}, R.~C., {Shelton}, R.~L., \& {Holberg}, J.~B. 2001,
  \apjl, 557, L47, \dodoi{10.1086/323041}

\bibitem[{{Murthy} {et~al.}(1993){Murthy}, {Im}, {Henry}, \&
  {Holberg}}]{Murthyvoy1993}
{Murthy}, J., {Im}, M., {Henry}, R.~C., \& {Holberg}, J.~B. 1993, \apj, 419,
  739, \dodoi{10.1086/173524}

\bibitem[{{Nesvorn{\'y}} {et~al.}(2011){Nesvorn{\'y}}, {Vokrouhlick{\'y}},
  {Pokorn{\'y}}, \& {Janches}}]{Nesvorny_2011}
{Nesvorn{\'y}}, D., {Vokrouhlick{\'y}}, D., {Pokorn{\'y}}, P., \& {Janches}, D.
  2011, \apj, 743, 37, \dodoi{10.1088/0004-637X/743/1/37}

\bibitem[{{Nicholl} {et~al.}(2017){Nicholl}, {Berger}, {Margutti}, {Blanchard},
  {Guillochon}, {Leja}, \& {Chornock}}]{Nicholl2017}
{Nicholl}, M., {Berger}, E., {Margutti}, R., {et~al.} 2017, \apjl, 845, L8,
  \dodoi{10.3847/2041-8213/aa82b1}

\bibitem[{{Nissanke} {et~al.}(2013){Nissanke}, {Kasliwal}, \&
  {Georgieva}}]{Nissanke2013}
{Nissanke}, S., {Kasliwal}, M., \& {Georgieva}, A. 2013, \apj, 767, 124,
  \dodoi{10.1088/0004-637X/767/2/124}

\bibitem[{{Noble} {et~al.}(2009){Noble}, {Conard}, {Weaver}, {Hayes}, \&
  {Cheng}}]{Noble2009}
{Noble}, M.~W., {Conard}, S.~J., {Weaver}, H.~A., {Hayes}, J.~R., \& {Cheng},
  A.~F. 2009, in \procspie, Vol. 7441, Instruments and Methods for Astrobiology
  and Planetary Missions XII, 74410Y

\bibitem[{{Oliphant}(2006)}]{NumPy}
{Oliphant}, T.~E. 2006, A guide to NumPy (Trelgol Publishing, USA)

\bibitem[{{Olkin} {et~al.}(2006){Olkin}, {Reuter}, {Lunsford}, {Binzel}, \&
  {Stern}}]{Olkin2006}
{Olkin}, C.~B., {Reuter}, D., {Lunsford}, A., {Binzel}, R.~P., \& {Stern},
  S.~A. 2006, in Bulletin of the American Astronomical Society, Vol.~38,
  AAS/Division for Planetary Sciences Meeting Abstracts \#38, 597

\bibitem[{{Pan} {et~al.}(2017){Pan}, {Kilpatrick}, {Simon}, {Xhakaj},
  {Boutsia}, {Coulter}, {Drout}, {Foley}, {Kasen}, {Morrell},
  {Murguia-Berthier}, {Osip}, {Piro}, {Prochaska}, {Ramirez-Ruiz}, {Rest},
  {Rojas-Bravo}, {Shappee}, \& {Siebert}}]{Pan2017}
{Pan}, Y.-C., {Kilpatrick}, C.~D., {Simon}, J.~D., {et~al.} 2017, \apjl, 848,
  L30, \dodoi{10.3847/2041-8213/aa9116}

\bibitem[{{Perez} \& {Granger}(2007)}]{Perez2007}
{Perez}, F., \& {Granger}, B.~E. 2007, Computing in Science Engineering, 9, 21,
  \dodoi{10.1109/MCSE.2007.53}

\bibitem[{{Pian} {et~al.}(2017){Pian}, {D'Avanzo}, {Benetti}, {Branchesi},
  {Brocato}, {Campana}, {Cappellaro}, {Covino}, {D'Elia}, {Fynbo}, {Getman},
  {Ghirlanda}, {Ghisellini}, {Grado}, {Greco}, {Hjorth}, {Kouveliotou},
  {Levan}, {Limatola}, {Malesani}, {Mazzali}, {Melandri}, {M{\o}ller},
  {Nicastro}, {Palazzi}, {Piranomonte}, {Rossi}, {Salafia}, {Selsing},
  {Stratta}, {Tanaka}, {Tanvir}, {Tomasella}, {Watson}, {Yang}, {Amati},
  {Antonelli}, {Ascenzi}, {Bernardini}, {Bo{\"e}r}, {Bufano}, {Bulgarelli},
  {Capaccioli}, {Casella}, {Castro-Tirado}, {Chassande-Mottin}, {Ciolfi},
  {Copperwheat}, {Dadina}, {De Cesare}, {di Paola}, {Fan}, {Gendre},
  {Giuffrida}, {Giunta}, {Hunt}, {Israel}, {Jin}, {Kasliwal}, {Klose}, {Lisi},
  {Longo}, {Maiorano}, {Mapelli}, {Masetti}, {Nava}, {Patricelli}, {Perley},
  {Pescalli}, {Piran}, {Possenti}, {Pulone}, {Razzano}, {Salvaterra},
  {Schipani}, {Spera}, {Stamerra}, {Stella}, {Tagliaferri}, {Testa}, {Troja},
  {Turatto}, {Vergani}, \& {Vergani}}]{Pian2017}
{Pian}, E., {D'Avanzo}, P., {Benetti}, S., {et~al.} 2017, \nat, 551, 67,
  \dodoi{10.1038/nature24298}

\bibitem[{{Poleski}(2016)}]{Poleski2016}
{Poleski}, R. 2016, \mnras, 455, 3656, \dodoi{10.1093/mnras/stv2569}

\bibitem[{Poppe \& Hor{\'a}nyi(2011)}]{Poppe_2011b}
Poppe, A., \& Hor{\'a}nyi, M. 2011, Planet. Space Sci., 59, 1647

\bibitem[{{Poppe} {et~al.}(2010){Poppe}, {James}, {Jacobsmeyer}, \&
  {Hor{\'a}nyi}}]{Poppe_2010a}
{Poppe}, A., {James}, D., {Jacobsmeyer}, B., \& {Hor{\'a}nyi}, M. 2010, \grl,
  37, L11101, \dodoi{10.1029/2010GL043300}

\bibitem[{Poppe(2016)}]{Poppe_2016}
Poppe, A.~R. 2016, Icarus, 264, 369

\bibitem[{Quinten {et~al.}(2002)Quinten, Kreibig, Henning, \&
  Mutschke}]{Quinten_2002}
Quinten, M., Kreibig, U., Henning, T., \& Mutschke, H. 2002, Appl. Optics, 41,
  7102

\bibitem[{{Reuter} {et~al.}(2008){Reuter}, {Stern}, {Scherrer}, {Jennings},
  {Baer}, {Hanley}, {Hardaway}, {Lunsford}, {McMuldroch}, {Moore}, {Olkin},
  {Parizek}, {Reitsma}, {Sabatke}, {Spencer}, {Stone}, {Throop}, {van Cleve},
  {Weigle}, \& {Young}}]{Reuter2008}
{Reuter}, D.~C., {Stern}, S.~A., {Scherrer}, J., {et~al.} 2008, \ssr, 140, 129,
  \dodoi{10.1007/s11214-008-9375-7}

\bibitem[{{Rice}(2014)}]{Rice2014}
{Rice}, K. 2014, Challenges, 5, 296, \dodoi{10.3390/challe5020296}

\bibitem[{Rogers {et~al.}(2006)Rogers, Schwinger, Kaidy, Strikwerda, Casini,
  Landi, Bettarini, \& Lorenzini}]{Rogers2006}
Rogers, G., Schwinger, M., Kaidy, J., {et~al.} 2006, Autonomous Star Tracker
  Performance for the New Horizons Mission (American Institute of Aeronautics
  and Astronautics).
\newblock \url{https://doi.org/10.2514/6.2006-6383}

\bibitem[{{Rybicki} \& {Lightman}(1986)}]{Rybicki1986}
{Rybicki}, G.~B., \& {Lightman}, A.~P. 1986, {Radiative Processes in
  Astrophysics} (Wiley-VCH, Weinheim)

\bibitem[{{Sano} {et~al.}(2015){Sano}, {Kawara}, {Matsuura}, {Kataza}, {Arai},
  \& {Matsuoka}}]{Sano2015}
{Sano}, K., {Kawara}, K., {Matsuura}, S., {et~al.} 2015, \apj, 811, 77,
  \dodoi{10.1088/0004-637X/811/2/77}

\bibitem[{{Singer} {et~al.}(2016){Singer}, {Chen}, {Holz}, {Farr}, {Price},
  {Raymond}, {Cenko}, {Gehrels}, {Cannizzo}, {Kasliwal}, {Nissanke},
  {Coughlin}, {Farr}, {Urban}, {Vitale}, {Veitch}, {Graff}, {Berry},
  {Mohapatra}, \& {Mandel}}]{Singer2016}
{Singer}, L.~P., {Chen}, H.-Y., {Holz}, D.~E., {et~al.} 2016, \apjl, 829, L15,
  \dodoi{10.3847/2041-8205/829/1/L15}

\bibitem[{Stern(1996)}]{Stern_1996}
Stern, S.~A. 1996, Astron. Astrophys., 310, 999

\bibitem[{{Stern} {et~al.}(2018){Stern}, {Weaver}, {Spencer}, \&
  {Elliott}}]{Stern2018}
{Stern}, S.~A., {Weaver}, H.~A., {Spencer}, J.~R., \& {Elliott}, H.~A. 2018,
  \ssr, 214, 77

\bibitem[{{Stern} {et~al.}(2008){Stern}, {Slater}, {Scherrer}, {Stone},
  {Dirks}, {Versteeg}, {Davis}, {Gladstone}, {Parker}, {Young}, \&
  {Siegmund}}]{Stern2008}
{Stern}, S.~A., {Slater}, D.~C., {Scherrer}, J., {et~al.} 2008, \ssr, 140, 155,
  \dodoi{10.1007/s11214-008-9407-3}

\bibitem[{Stone {et~al.}(2015)Stone, Alkalai, Friedman, Arora, Arya, Barnes,
  Brashears, Brown, Cauley, Cesarone, Dyson, Garber, Goldsmith, Jemison,
  Johnson, Liewer, Lubin, Maccone, Males, McDonough, Ralph L.~McNutt, Mewaldt,
  Michael, Montgomery, Opher, Provornikova, Rankin, Redfield, Shao, Shotwell,
  Strange, Svitek, Swain, Turyshev, Werner, \& Zank}]{Stone2015}
Stone, E., Alkalai, L., Friedman, L., {et~al.} 2015, {Science and Enabling
  Technologies for the Exploration of the Interstellar Medium}, Tech. rep.,
  Keck Institute for Space Studies, Pasadena

\bibitem[{{Street} {et~al.}(2016){Street}, {Udalski}, {Calchi Novati},
  {Hundertmark}, {Zhu}, {Gould}, {Yee}, {Tsapras}, {Bennett}, {RoboNet
  Project}, {Consortium}, {J{\o}rgensen}, {Dominik}, {Andersen}, {Bachelet},
  {Bozza}, {Bramich}, {Burgdorf}, {Cassan}, {Ciceri}, {D'Ago}, {Dong}, {Evans},
  {Gu}, {Harkonnen}, {Hinse}, {Horne}, {Figuera Jaimes}, {Kains}, {Kerins},
  {Korhonen}, {Kuffmeier}, {Mancini}, {Menzies}, {Mao}, {Peixinho}, {Popovas},
  {Rabus}, {Rahvar}, {Ranc}, {Tronsgaard Rasmussen}, {Scarpetta}, {Schmidt},
  {Skottfelt}, {Snodgrass}, {Southworth}, {Steele}, {Surdej}, {Unda-Sanzana},
  {Verma}, {von Essen}, {Wambsganss}, {Wang}, {Wertz}, {OGLE Project},
  {Poleski}, {Pawlak}, {Szyma{\'n}ski}, {Skowron}, {Mr{\'o}z}, {Koz{\l}owski},
  {Wyrzykowski}, {Pietrukowicz}, {Pietrzy{\'n}ski}, {Soszy{\'n}ski}, {Ulaczyk},
  {Spitzer Team}, {Beichman}, {Bryden}, {Carey}, {Gaudi}, {Henderson}, {Pogge},
  {Shvartzvald}, {MOA Collaboration}, {Abe}, {Asakura}, {Bhattacharya}, {Bond},
  {Donachie}, {Freeman}, {Fukui}, {Hirao}, {Inayama}, {Itow}, {Koshimoto},
  {Li}, {Ling}, {Masuda}, {Matsubara}, {Muraki}, {Nagakane}, {Nishioka},
  {Ohnishi}, {Oyokawa}, {Rattenbury}, {Saito}, {Sharan}, {Sullivan}, {Sumi},
  {Suzuki}, {Tristram}, {Wakiyama}, {Yonehara}, {KMTNet Modeling Team}, {Han},
  {Choi}, {Park}, {Jung}, \& {Shin}}]{Street2016}
{Street}, R.~A., {Udalski}, A., {Calchi Novati}, S., {et~al.} 2016, \apj, 819,
  93, \dodoi{10.3847/0004-637X/819/2/93}

\bibitem[{{Sullivan} {et~al.}(2015){Sullivan}, {Winn}, {Berta-Thompson},
  {Charbonneau}, {Deming}, {Dressing}, {Latham}, {Levine}, {McCullough},
  {Morton}, {Ricker}, {Vanderspek}, \& {Woods}}]{Sullivan2015}
{Sullivan}, P.~W., {Winn}, J.~N., {Berta-Thompson}, Z.~K., {et~al.} 2015, \apj,
  809, 77, \dodoi{10.1088/0004-637X/809/1/77}

\bibitem[{{Sumi} \& {Penny}(2016)}]{SumiPenny2016}
{Sumi}, T., \& {Penny}, M.~T. 2016, \apj, 827, 139,
  \dodoi{10.3847/0004-637X/827/2/139}

\bibitem[{Szalay {et~al.}(2013)Szalay, Piquette, \& Hor{\'a}nyi}]{Szalay_2013}
Szalay, J.~R., Piquette, M., \& Hor{\'a}nyi, M. 2013, Earth Planets Space, 65,
  1145

\bibitem[{{Toller} {et~al.}(1987){Toller}, {Tanabe}, \&
  {Weinberg}}]{Toller1987}
{Toller}, G., {Tanabe}, H., \& {Weinberg}, J.~L. 1987, \aap, 188, 24

\bibitem[{{Toller}(1983)}]{Toller1983}
{Toller}, G.~N. 1983, \apjl, 266, L79, \dodoi{10.1086/183982}

\bibitem[{{Totani} {et~al.}(2001){Totani}, {Yoshii}, {Iwamuro}, {Maihara}, \&
  {Motohara}}]{Totani2001}
{Totani}, T., {Yoshii}, Y., {Iwamuro}, F., {Maihara}, T., \& {Motohara}, K.
  2001, \apjl, 550, L137, \dodoi{10.1086/319646}

\bibitem[{{Tsapras} {et~al.}(2016){Tsapras}, {Hundertmark}, {Wyrzykowski},
  {Horne}, {Udalski}, {Snodgrass}, {Street}, {Bramich}, {Dominik}, {Bozza},
  {Figuera Jaimes}, {Kains}, {Skowron}, {Szyma{\'n}ski}, {Pietrzy{\'n}ski},
  {Soszy{\'n}ski}, {Ulaczyk}, {Koz{\l}owski}, {Pietrukowicz}, \&
  {Poleski}}]{Tsapras2016}
{Tsapras}, Y., {Hundertmark}, M., {Wyrzykowski}, {\L}., {et~al.} 2016, \mnras,
  457, 1320, \dodoi{10.1093/mnras/stw023}

\bibitem[{{Tsumura} {et~al.}(2013){Tsumura}, {Matsumoto}, {Matsuura}, {Sakon},
  \& {Wada}}]{Tsumura2013}
{Tsumura}, K., {Matsumoto}, T., {Matsuura}, S., {Sakon}, I., \& {Wada}, T.
  2013, \pasj, 65, 121, \dodoi{10.1093/pasj/65.6.121}

\bibitem[{{Tyson}(1995)}]{Tyson1995}
{Tyson}, J.~A. 1995, in {\it Extragalactic Background Radiation Meeting} (Ed.
  Calzetti, D. et al.) 103-133, ed. D.~{Calzetti}, M.~{Livio}, \& P.~{Madau},
  103--133

\bibitem[{{Valenti} {et~al.}(2017){Valenti}, {Sand}, {Yang}, {Cappellaro},
  {Tartaglia}, {Corsi}, {Jha}, {Reichart}, {Haislip}, \&
  {Kouprianov}}]{Valenti2017}
{Valenti}, S., {Sand}, D.~J., {Yang}, S., {et~al.} 2017, \apjl, 848, L24,
  \dodoi{10.3847/2041-8213/aa8edf}

\bibitem[{{Vanderburg} \& {Johnson}(2014)}]{Vanderburg2014}
{Vanderburg}, A., \& {Johnson}, J.~A. 2014, \pasp, 126, 948,
  \dodoi{10.1086/678764}

\bibitem[{Verbiscer {et~al.}(2009)Verbiscer, Skrutskie, \&
  Hamilton}]{Verbiscer_2009}
Verbiscer, A.~J., Skrutskie, M.~F., \& Hamilton, D.~P. 2009, Nature, 461, 1098

\bibitem[{{Villanueva} {et~al.}(2018){Villanueva}, {Dragomir}, \&
  {Gaudi}}]{Villanueva2018}
{Villanueva}, Jr., S., {Dragomir}, D., \& {Gaudi}, B.~S. 2018, ArXiv e-prints.
\newblock \doarXiv{1805.00956}

\bibitem[{{Vitense} {et~al.}(2012){Vitense}, {Krivov}, {Kobayashi}, \&
  {L{\"o}hne}}]{Vitense_2012}
{Vitense}, C., {Krivov}, A.~V., {Kobayashi}, H., \& {L{\"o}hne}, T. 2012, \aap,
  540, A30, \dodoi{10.1051/0004-6361/201118551}

\bibitem[{{Vitense} {et~al.}(2010){Vitense}, {Krivov}, \&
  {L{\"o}hne}}]{Vitense_2010}
{Vitense}, C., {Krivov}, A.~V., \& {L{\"o}hne}, T. 2010, \aap, 520, A32,
  \dodoi{10.1051/0004-6361/201014208}

\bibitem[{{Warren}(1984)}]{Warren_1984}
{Warren}, S.~G. 1984, \ao, 23, 1206, \dodoi{10.1364/AO.23.001206}

\bibitem[{{Weaver} {et~al.}(2008){Weaver}, {Gibson}, {Tapley}, {Young}, \&
  {Stern}}]{Weaver2008}
{Weaver}, H.~A., {Gibson}, W.~C., {Tapley}, M.~B., {Young}, L.~A., \& {Stern},
  S.~A. 2008, \ssr, 140, 75, \dodoi{10.1007/s11214-008-9376-6}

\bibitem[{{Weinberg} {et~al.}(1974){Weinberg}, {Hanner}, {Beeson}, {DeShields},
  \& {Green}}]{Weinberg1974}
{Weinberg}, J.~L., {Hanner}, M.~S., {Beeson}, D.~E., {DeShields}, II, L.~M., \&
  {Green}, B.~A. 1974, \jgr, 79, 3665, \dodoi{10.1029/JA079i025p03665}

\bibitem[{{Wright}(2001)}]{Wright2001}
{Wright}, E.~L. 2001, \apj, 553, 538, \dodoi{10.1086/320942}

\bibitem[{{Wright}(2004)}]{Wright2004}
---. 2004, \nar, 48, 465, \dodoi{10.1016/j.newar.2003.12.054}

\bibitem[{{Wyrzykowski} {et~al.}(2011){Wyrzykowski}, {Skowron}, {Koz{\l}owski},
  {Udalski}, {Szyma{\'n}ski}, {Kubiak}, {Pietrzy{\'n}ski}, {Soszy{\'n}ski},
  {Szewczyk}, {Ulaczyk}, {Poleski}, \& {Tisserand}}]{Wyrzykowski2011}
{Wyrzykowski}, L., {Skowron}, J., {Koz{\l}owski}, S., {et~al.} 2011, \mnras,
  416, 2949, \dodoi{10.1111/j.1365-2966.2011.19243.x}

\bibitem[{{Yee} {et~al.}(2015){Yee}, {Gould}, {Beichman}, {Calchi Novati},
  {Carey}, {Gaudi}, {Henderson}, {Nataf}, {Penny}, {Shvartzvald}, \&
  {Zhu}}]{Yee2015}
{Yee}, J.~C., {Gould}, A., {Beichman}, C., {et~al.} 2015, \apj, 810, 155,
  \dodoi{10.1088/0004-637X/810/2/155}

\bibitem[{{Zemcov} {et~al.}(2017){Zemcov}, {Immel}, {Nguyen}, {Cooray},
  {Lisse}, \& {Poppe}}]{Zemcov2017}
{Zemcov}, M., {Immel}, P., {Nguyen}, C., {et~al.} 2017, Nature Communications,
  8, 15003, \dodoi{10.1038/ncomms15003}

\bibitem[{{Zemcov} {et~al.}(2014){Zemcov}, {Smidt}, {Arai}, {Bock}, {Cooray},
  {Gong}, {Kim}, {Korngut}, {Lam}, {Lee}, {Matsumoto}, {Matsuura}, {Nam},
  {Roudier}, {Tsumura}, \& {Wada}}]{Zemcov2014}
{Zemcov}, M., {Smidt}, J., {Arai}, T., {et~al.} 2014, Science, 346, 732,
  \dodoi{10.1126/science.1258168}

\bibitem[{{Zhu} {et~al.}(2017{\natexlab{a}}){Zhu}, {Udalski}, {Huang}, {Calchi
  Novati}, {Sumi}, {Poleski}, {Skowron}, {Mr{\'o}z}, {Szyma{\'n}ski},
  {Soszy{\'n}ski}, {Pietrukowicz}, {Koz{\l}owski}, {Ulaczyk}, {Pawlak}, {OGLE
  Collaboration}, {Beichman}, {Bryden}, {Carey}, {Gaudi}, {Gould}, {Henderson},
  {Shvartzvald}, {Yee}, {Spitzer Team}, {Bond}, {Bennett}, {Suzuki},
  {Rattenbury}, {Koshimoto}, {Abe}, {Asakura}, {Barry}, {Bhattacharya},
  {Donachie}, {Evans}, {Fukui}, {Hirao}, {Itow}, {Kawasaki}, {Li}, {Ling},
  {Masuda}, {Matsubara}, {Miyazaki}, {Munakata}, {Muraki}, {Nagakane},
  {Ohnishi}, {Ranc}, {Saito}, {Sharan}, {Sullivan}, {Tristram}, {Yamada},
  {Yonehara}, \& {MOA Collaboration}}]{Zhu2017a}
{Zhu}, W., {Udalski}, A., {Huang}, C.~X., {et~al.} 2017{\natexlab{a}}, \apjl,
  849, L31, \dodoi{10.3847/2041-8213/aa93fa}

\bibitem[{{Zhu} {et~al.}(2017{\natexlab{b}}){Zhu}, {Huang}, {Udalski},
  {Soares-Furtado}, {Poleski}, {Skowron}, {Mr{\'o}z}, {Szyma{\'n}ski},
  {Soszy{\'n}ski}, {Pietrukowicz}, {Koz{\L}owski}, {Ulaczyk}, \&
  {Pawlak}}]{Zhu2017b}
{Zhu}, W., {Huang}, C.~X., {Udalski}, A., {et~al.} 2017{\natexlab{b}}, \pasp,
  129, 104501, \dodoi{10.1088/1538-3873/aa7dd7}

\end{thebibliography}

\end{document}